\documentclass[review]{elsarticle}

\usepackage{graphicx}%
\usepackage{multirow}%
\usepackage{amsmath,amssymb,amsfonts}%
\usepackage{amsthm}%
\usepackage{mathrsfs}%
\usepackage[title]{appendix}%
\usepackage{xcolor}%
\usepackage{textcomp}%
\usepackage{manyfoot}%
\usepackage{booktabs}%
\usepackage{algorithm}%
\usepackage{algorithmicx}%
\usepackage{algpseudocode}%
\usepackage{listings}%

\usepackage{amssymb}
\usepackage{amsmath}
\usepackage{amsthm}
\usepackage{xr}
\usepackage{lineno,hyperref}
\usepackage{cleveref}
\usepackage{epsfig}
\usepackage{arydshln}
\usepackage{float}
\usepackage{verbatim}
\usepackage{comment}
\newcommand{\comm}[1]{}
\usepackage{color}
\usepackage{xcolor}
\usepackage{booktabs}
\usepackage{threeparttable}
\newcommand{\MM}[1]{{\color{black} #1}}

\newcommand{\SN}[1]{{\color{black} #1}}

\usepackage{lineno}

\begin{document}
	
	\begin{frontmatter}
		
		
		
		
		
		\author[addr1,addr2]{Bartosz Bucholc}
		
		\author[addr1]{Marcin Ma\'zdziarz}
		
		\author[addr1]{Szymon Nosewicz\corref{mycorrespondingauthor}}
		\ead{snosew@ippt.pan.pl}
		
		\address[addr1]{Institute of Fundamental Technological Research Polish Academy of Sciences, Pawi\'nskiego 5B, 02-106 Warsaw, Poland}
		\cortext[mycorrespondingauthor]{Corresponding author}
		
		\address[addr2]{{\L}ukasiewicz Research Network Institute of Microelectronics and Photonics,
			Lotnik\'ow 32/46, 02-668 Warsaw, Poland}
		
		
		\title{Evaluation of upper-scale mechanical parameters of particulate aluminum-based nanocomposites from molecular simulations}

		\begin{abstract}
		{This paper focuses on the numerical and analytical determination of the mechanical parameters of aluminum-matrix nanocomposites using atomistic-scale simulations within a multiscale modeling framework. Molecular statics simulations were employed to evaluate the tensile and compressive behavior of Al-based composites reinforced with nanosized alumina particles. Two classes of atomistic models were investigated. First, a unit-cell model containing a single Al$_2$O$_3$ monocrystal embedded in an aluminum matrix was used to examine the influence of the volume fraction of the ceramic inclusion on the effective mechanical properties. Second, a polycrystalline RVE model comprising an alumina particle was analyzed to assess the effect of the number of Al grains on effective properties. For each case, the mechanical response was characterized by stress–strain curves, enabling the identification of upper-scale material parameters associated with the elastic regime (bulk, shear and Young's modulus) and the damage stage (ultimate strength, ultimate strain, and failure strain). The isotropized elastic constants evaluated from molecular simulations were compared with classical Voigt-Reuss-Hill bounds to assess their applicability in the elastic regime. Finally, atomistic-scale simulation outputs provided insight into the dominant deformation/damage mechanisms (brittle versus ductile) and fracture modes (transgranular versus intergranular).}

		\end{abstract}

		\comm{
	\begin{graphicalabstract}
	\end{graphicalabstract}
	
	\begin{highlights}
		\item Research highlight 1
		\item Research highlight 2
	\end{highlights}

}



\end{frontmatter}
	
	
	\newpage
	\section{Introduction}
	\label{sec:Int}
	Due to their exceptional properties, particulate composites are among the most extensively studied classes of engineering materials. They exhibit excellent mechanical performance, particularly a high strength-to-weight ratio, superior resistance to wear and fracture, and favorable thermal properties \cite{MENG202214}. As a result, composite materials are widely used in advanced technological fields such as aerospace, automotive engineering, electronics, and many others. Of equal importance is the concurrent development of advanced fabrication techniques, including microwave sintering \cite{Alem17022020}, spark/pulse plasma sintering \cite{Zygmuntowicz2024}, and additive manufacturing \cite{Goyal2025}. These methods enable precise control over microstructure, allowing it to be tailored to specific application requirements, thereby significantly enhancing the flexibility of the material design process.
	
	Particular attention has been devoted to composites based on lightweight metal matrices, such as aluminum, which are commonly reinforced with ceramic particles, e.g., oxides \cite{Kok2005}, borides \cite{Kumar2016}, and carbides \cite{NIKITIN2026247}. This matrix–reinforcement configuration enables a synergistic effect between the lightweight metallic matrix, which provides structural integrity and ductility, and the hard ceramic particles, resulting in enhanced strength, wear resistance, and overall durability. The high technological relevance of such materials, combined with increasing industrial demand, continues to drive intensive research efforts. In recent years, these studies have increasingly focused on advanced fabrication and processing techniques \cite{SAMAL2020131}, the effects of hybrid reinforcement strategies \cite{BabuR2024}, metal-ceramic interface design \cite{ZHOU202663}, as well as nanoreinforcements \cite{Golmohammadi2025}. While nanometer-scale reinforcements offer significant potential for further property enhancement, they also pose substantial challenges for predominantly experimental investigations, as many nanoscale phenomena are difficult or even impossible to observe and analyze with conventional characterization techniques, such as microscopy. Consequently, computational approaches, particularly molecular dynamics (MD) simulations, are increasingly employed to investigate composite materials and the fundamental mechanisms that govern their performance.
	
	MD is a computational simulation method used to model atoms and particles and describe their behavior by employing mathematical models and classical Newtonian mechanics, enabling the calculation of particle trajectories as a function of time. It is widely applied to the investigation of mechanical \cite{MAZDZIARZ2024109953}, thermal \cite{Kumar2019}, and transport properties \cite{Chaparro2024}, as well as to the analysis of deformation mechanisms and defect evolution \cite{Zhang2021}. MD has been employed in studies of a broad range of materials, including polymers \cite{LI2019348}, metals \cite{ZepedaRuiz2017}, ceramics \cite{Hong2023}, and structural composites \cite{LAU2018282}. These investigations have provided valuable insights into thermal conductivity, atomic-scale transport phenomena, structural stability, and have supported the optimization of reinforcement strategies and the improvement of material durability.
	
	Although MD simulations remain constrained by the limited system sizes that can be realistically modeled and computed within reasonable time and computational cost, they are increasingly employed in the study of metal matrix composites (MMCs) \cite{Zhao2024}. This approach enables a detailed investigation and understanding of how nanoreinforcements contribute to the strengthening of materials. In particular, MD provides deep insight into dislocation motion \cite{Li2017}, the Orowan mechanism \cite{Huo2018}, and load transfer phenomena \cite{Zhao2024} – processes that are extremely challenging to capture using experimental techniques.
	Moreover, interfacial phenomena occurring at the boundary between the matrix and the reinforcing particle play a critical role in these mechanisms. MD simulations allow for the analysis of interfacial bonding characteristics, including bonding strength, relative orientation, and chemical bonding \cite{SAZGAR2015200} as well as crack initiation and propagation \cite{ZHOU201627}, and their influence on strength, ductility, and overall mechanical performance.
	
	Despite their significant technological importance and favorable properties, the number of studies employing MD methods to investigate {Al–Al$_2$O$_3$} materials remains relatively limited. Research efforts have primarily focused on interfacial phenomena, modeling the mechanical properties of the metal–ceramic interface, and simulating MMCs featuring one or more embedded particles.
	
	Mei et al. \cite{MEI2015538} investigated energetically preferred {Al–Al$_2$O$_3$} interfacial configurations through an analysis of the work of adhesion. Their results revealed the formation of an interfacial aluminum suboxide layer between Al and Al$_2$O$_3$, which is in good agreement with experimental observations. Other interfacial phenomena have been investigated by Ma et al. \cite{Ma2011}, who examined nitrogen-related effects in nanostructured Al-based composites produced via cryomilling in liquid nitrogen. Experimental observations of enhanced strength and thermal stability in these materials were attributed to N absorption at the metal-ceramic interface. Owing to the inherent experimental challenges associated with observing such processes at the nanoscale, MD simulations were employed to investigate the underlying mechanisms. The simulations demonstrated that not only atomic N but also molecular N$_2$ can be absorbed at the interface during cryomilling. Furthermore, the presence of defects, particularly vacancies, was identified as a key factor facilitating N absorption into the Al lattice, which is a prerequisite for the subsequent formation of the AlN phase. Another important interfacial aspect, particularly relevant for composites fabricated with the involvement of a liquid phase, is related to diffusion phenomena. Tahani et al. \cite{ma16124324} determined both average and cross ternary interdiffusion coefficients for a composite system in which an AlSi matrix was reinforced with Al$_2$O$_3$ particles. One of the key findings of their study was the formation of an interdiffusion zone, whose thickness was shown to be proportional to both the annealing temperature and time.
	
	The reinforcement efficiency of composite materials strongly depends on the quality of the bond between the reinforcing particles and the metallic matrix, which justifies the need for detailed investigation and understanding of the factors governing interfacial integrity. In this context, Dandekar et al. \cite{DANDEKAR2011350} employed MD simulations to perform a detailed interfacial characterization of an {Al–Al$_2$O$_3$} composite system. The parameterized traction–separation law developed in their study was found to be consistent with continuum-based cohesive zone models (CZMs), and the simulated interfacial fracture energy values showed good agreement with experimental data. Furthermore, the authors also examined the influence of porosity on the plastic deformation behavior of the composite. Sazgar et al. \cite{SAZGAR2015200} employed a reactive force field (ReaxFF) potential within MD simulations to investigate the bond strength between Al and Al$_2$O$_3$ interfaces subjected to tensile and shear loading. This approach enabled a detailed assessment of the effects of crystallographic orientation and the nature of atomic layers within the interfacial region on deformation behavior and fracture strength. In a subsequent study published in 2017 \cite{SAZGAR2017116}, the same research group adopted a broader, hierarchical modeling strategy. They developed an interfacial cohesive zone model (CZM) based on traction–separation relationships derived from MD simulations of {Al–Al$_2$O$_3$} interfaces at different temperatures.

 \SN{Recent atomistic studies have further demonstrated the importance of nanoscale
structure and processing conditions in Al- and Al$_2$O$_3$-based systems.
Khoei et al.~\cite{Khoei2024} investigated the thermo-mechanical behavior of oxide-coated
Al nanopowders using reactive molecular dynamics, showing that their effective
response depends strongly on temperature, density, and the presence of the
surface oxide layer. In a related study, the mechanical response of oxidized
Al nanopowders under triaxial compression was analyzed, demonstrating that an
increase in the Al$_2$O$_3$ shell thickness increases hardness while reducing
ductility \cite{Khoei2022}. More recently, reactive molecular dynamics simulations of
Al$_2$O$_3$ nanopowders under warm compaction demonstrated significant effects
of nanoparticle size, temperature, and confining pressure on densification
behavior \cite{Khoei2024WarmCompaction}. These results underline the importance of particle-scale
structure, interfaces, and loading conditions in determining the effective
behavior of Al/Al$_2$O$_3$ nanosystems and provide additional motivation for
the atomistic analysis adopted in the present study.}
	
	As demonstrated by the studies discussed above, MD simulations have been successfully applied to investigate the strength of {Al–Al$_2$O$_3$} interfacial bonding and to analyze a range of interfacial phenomena and properties. However, relatively few studies have employed MD to simulate the mechanical behavior of composite materials as a whole, extending beyond the interface itself to include a representative portion of the bulk material, in which reinforcing nanoparticles are embedded within the metallic matrix. In this context, the work of Mei et al. \cite{Mei2013} is particularly noteworthy, as it examined the influence of the volume fraction of reinforcement particles embedded in Al single-crystal and bicrystal matrices on the ultimate strength and elastic modulus.
	
\MM{In the present work}, an atomistic approach is employed to systematically investigate the mechanical behavior of {Al–Al$_2$O$_3$} nanocomposite models under tensile and compressive loading conditions. The analysis considers the influence of the reinforcement volume fraction and microstructural complexity, including the presence of grain boundaries, on \MM{elastic properties}, as well as deformation and damage mechanisms. Moreover, analytical estimates based on Voigt-Reuss-Hill bounds were used to benchmark the atomistic results for the \MM{elastic moduli}.

    \SN{
The paper is organized as follows. Section~\ref{sec:Fotm} introduces the
motivation for the adopted multiscale framework, the atomistic unit-cell and
polycrystalline RVE configurations, and the computational methodology.
Section~\ref{sec:Res} presents the mechanical response of the investigated {Al-Al$_2$O$_3$}
systems, including the effective elastic parameters and tensile and compressive
behavior of the unit-cell and RVE models. Section~\ref{sec:Disc} discusses the atomistic
results in relation to classical homogenization estimates and available
experimental and computational literature. Finally, Section~\ref{sec:Concl} summarizes the
main findings, limitations, and implications of the present atomistic
parameter-identification approach for subsequent higher-scale modeling.
}
	
\section{Fundamentals of the method}
\label{sec:Fotm}
    
     \SN{
This section introduces the theoretical and computational framework adopted for
the atomistic characterization of {Al-Al$_2$O$_3$} nanostructures. First, the
motivation for the multiscale approach and the role of atomistic simulations in
the identification of effective mechanical parameters are discussed, together
with the unit-cell and polycrystalline RVE concepts used in the present study.
Subsequently, the computational methodology is described, including the
crystallographic structures and model configurations, calculation of effective
elastic properties, molecular statics procedure, and selection of the
interatomic potentials employed for Al, Al$_2$O$_3$, and their interactions.
}

	\subsection{Formulation and motivation of the problem}
    
\begin{figure}[!h] 
		\centering
		\begin{tabular}{cc}
			{\footnotesize (a)} & {\footnotesize (b)} \\
			\includegraphics[width=0.405\linewidth]{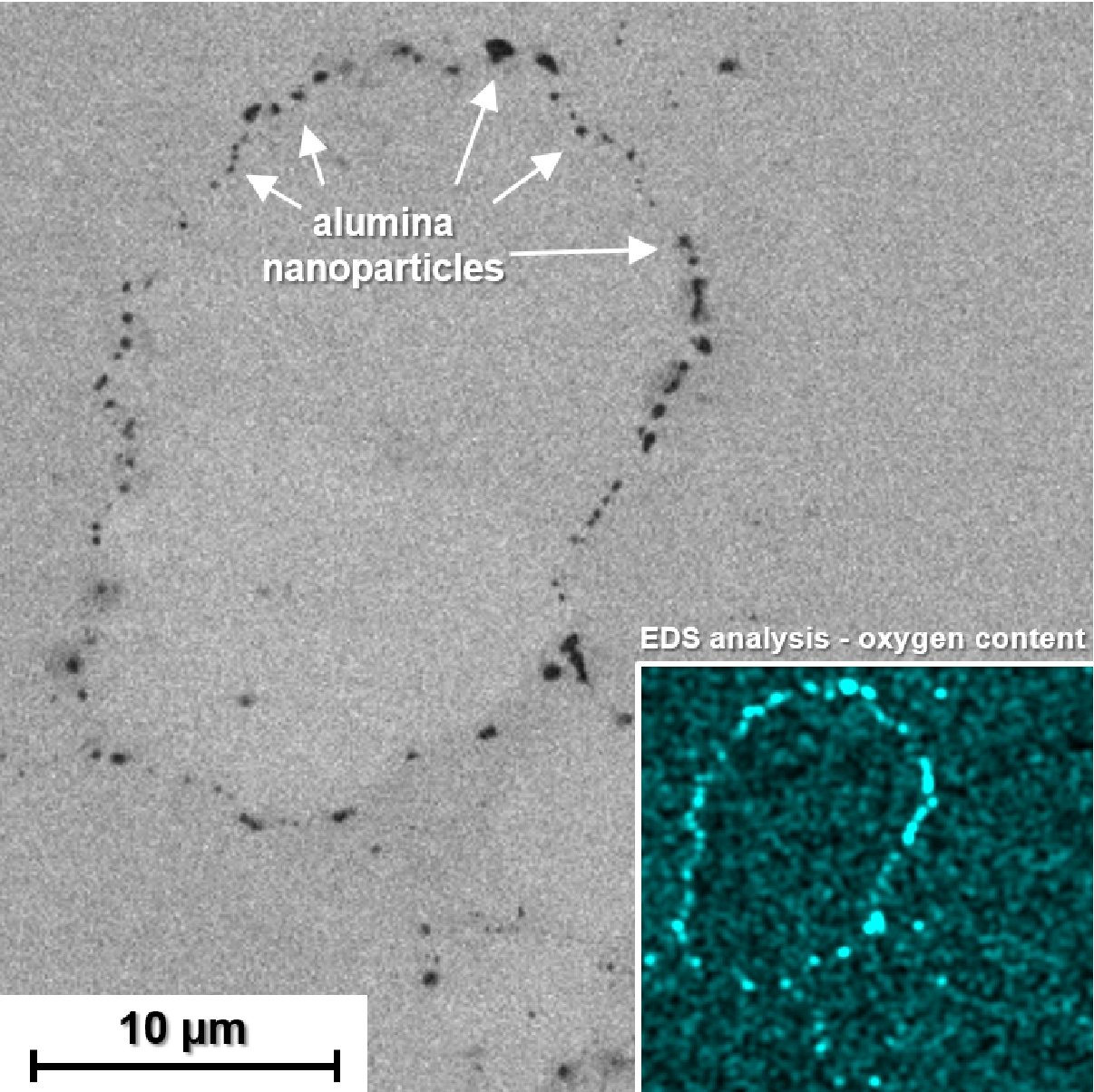} & \hspace{-0.1cm}
			\includegraphics[width=0.495\linewidth]{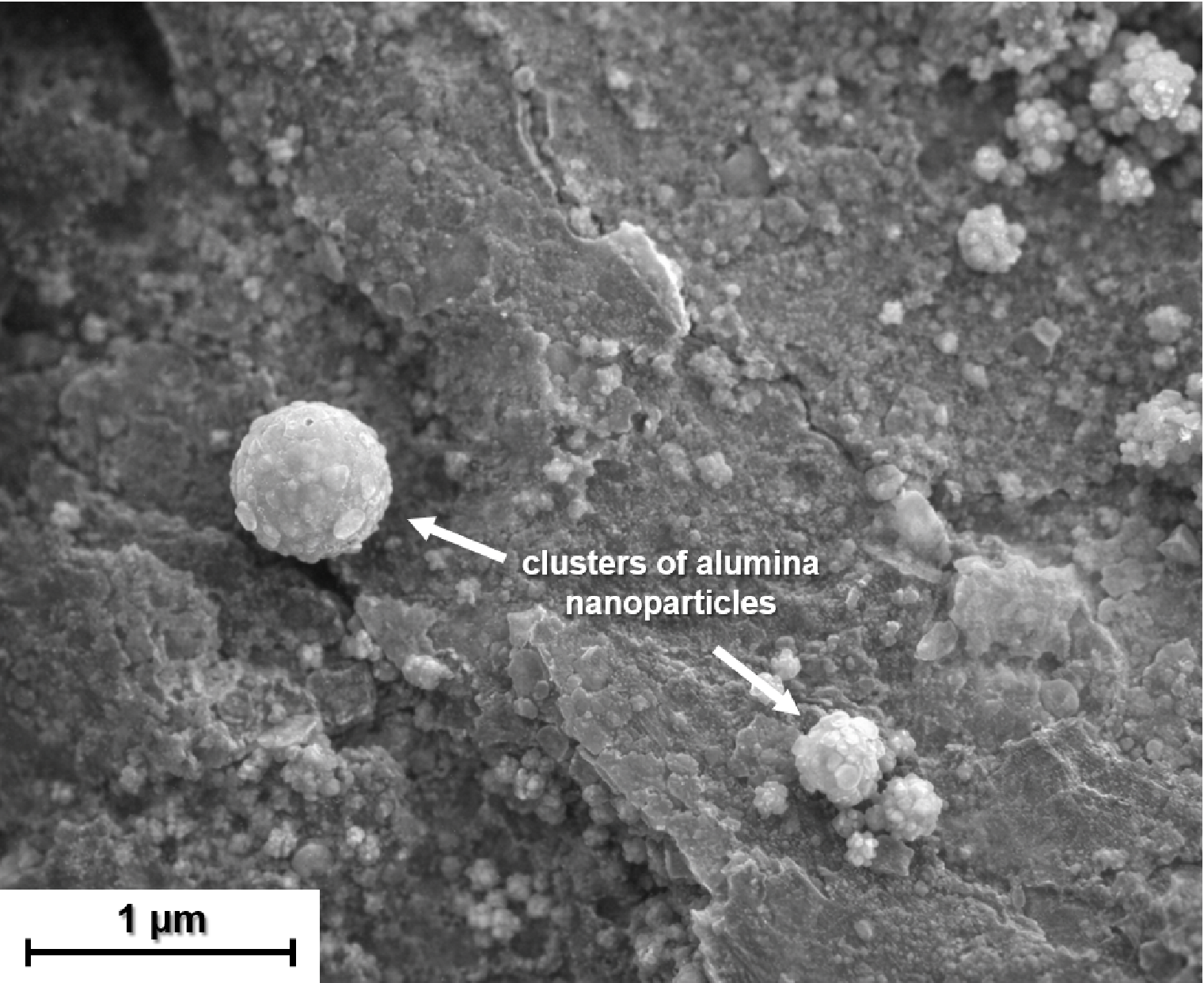}\\
		\end{tabular}
		\caption{SEM images of a representative metal-matrix composite with alumina nanoparticles (a) and the fracture surface of the {Al-Al$_2$O$_3$} nanocomposite (b).}
		\label{fig:SEM}
	\end{figure}

	The microstructure of the particle-reinforced nanocomposite exhibits pronounced heterogeneity, characterized by a non-uniform spatial distribution of reinforcing nanoparticles and the coexistence of isolated particles and agglomerates (Fig.~\ref{fig:SEM}a). A significant fraction of the nanoparticles is preferentially located along grain boundaries, where clustering (Fig.~\ref{fig:SEM}b) intensifies local stiffness contrast and may act as stress-concentration sites, influencing the overall mechanical response of the composite \cite{SHIN2017217}.
	
	In nanocomposite modeling, identifying deformation and damage parameters is challenging due to the inherently multiscale nature of the governing mechanisms \cite{ZENG2008191}. The macroscopic mechanical response results from processes spanning from atomistic interactions and interfacial bonding at the nanoscale to collective deformation and damage evolution at the microscale. Micromechanical models are typically formulated in terms of parameters such as interfacial strength, cohesive stiffness, and damage evolution laws \cite{NOSEWICZ2025104390}, which are difficult or impossible to measure experimentally. In situ experimental characterization of nanoscale deformation and damage processes remains highly limited.
	
	On the one hand, atomistic simulations may improve understanding of the interfacial conditions and mechanical properties of matrix and reinforcement polycrystals \cite{MAZDZIARZ2024109953}. On the other hand, atomistic modeling provides detailed insight into local bonding, defect evolution, and interfacial phenomena \cite{Pan2023, SAZGAR2015200}; however, it is inherently limited to small spatial and temporal scales, restricting simulations to highly localized material regions. Consequently, global material performance—such as macroscopic stiffness, strength, or failure response—cannot be directly assessed and must be inferred through scale-bridging assumptions that introduce additional uncertainty \cite{Guo2007Multiscale}.
	
\begin{figure}[!h] 
		\centering
		\includegraphics[width=0.999\linewidth]{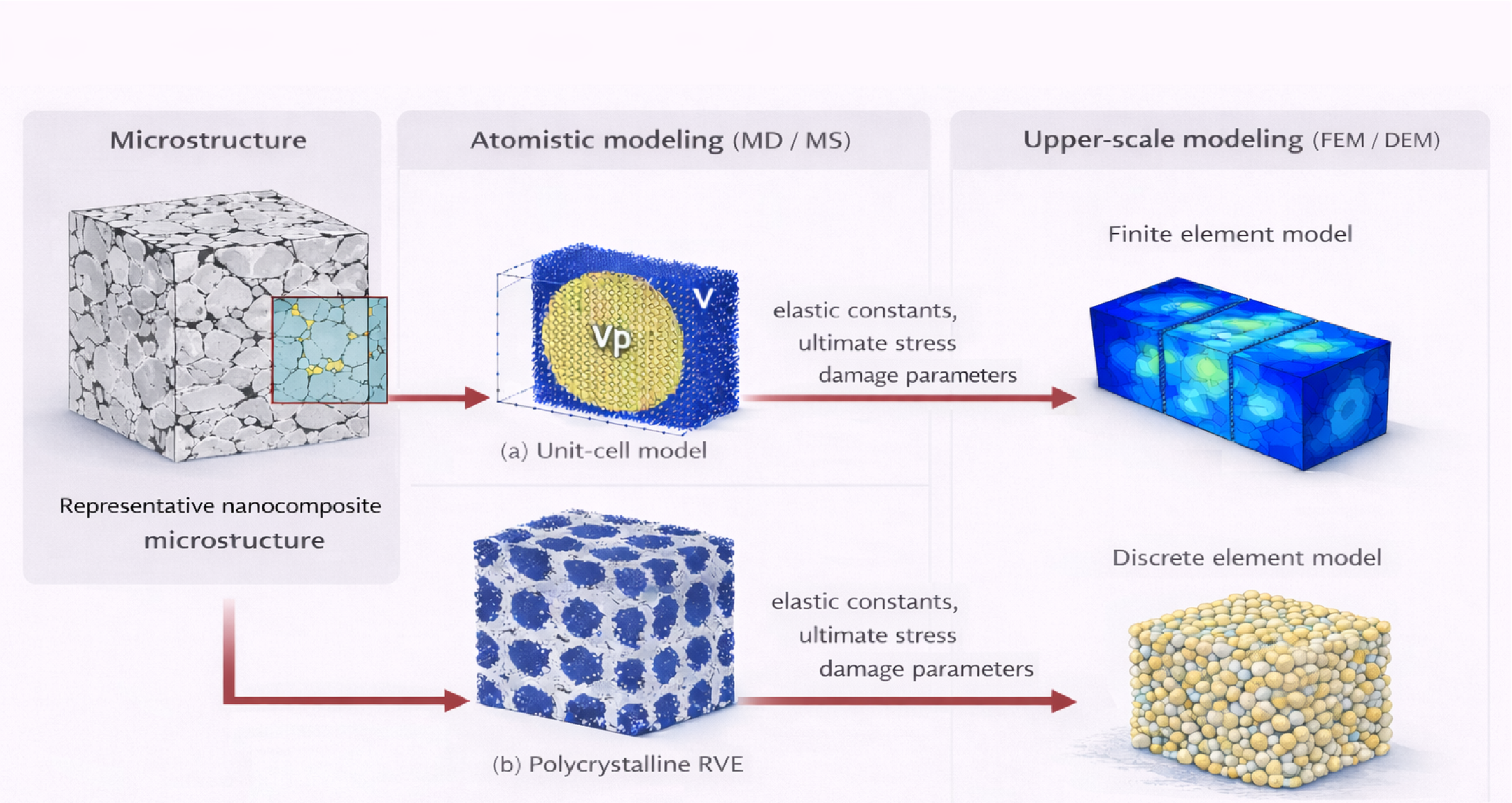}
		\caption{ \SN{
Conceptual scale-bridging framework illustrating the possible transfer
of atomistically identified mechanical parameters to subsequent higher-scale
FEM/DEM models. Only the atomistic parameter-identification stage is addressed
in the present study.
}}
		\label{fig:scope}
	\end{figure}

	In contrast to traditional single-scale approaches, multiscale analysis concurrently integrates models operating at different length and time scales, thereby combining the computational efficiency of macroscopic descriptions with the physical fidelity of microscopic models (Fig.~\ref{fig:scope}). This framework exploits the strengths of lower-scale methods, such as molecular simulations that resolve atomic-level mechanisms over short timescales, while coupling them with macroscopic simulations that predict material behavior over significantly longer spatial and temporal domains.
	
	This paper aims to determine atomistically derived elastic and damage-related
parameters suitable for subsequent higher-scale modeling. Two complementary
model classes were considered: a unit-cell model containing a single ceramic
reinforcement particle and a polycrystalline RVE containing up to 128 Al grains.
The investigated configurations were used to assess the effects of ceramic
particle volume fraction and matrix grain structure. The resulting effective
mechanical parameters can subsequently be transferred to higher-scale models,
such as finite-element or discrete-element formulations, following an appropriate
scale-transition procedure (Fig.~\ref{fig:scope}). 

    Here, the properties of each element in the discretized domain can be represented by an effective mechanical response of unit-cell/RVE models determined from molecular dynamics simulations. Assuming considerable heterogeneity in the composite nanostructure due to varying ceramic particle fraction and sizes, together with the presence of ceramic agglomerates at grain boundaries, such a multiscale modeling framework can serve as a suitable tool for representing the micro- and macro-mechanical responses of nanocomposites. In particular, it can be highly useful for composites exhibiting a highly non-uniform distribution of reinforcement within the matrix, including functionally graded metal-matrix composites \cite{TAYYEBI2022165078} and additively manufactured composites, where layer-by-layer fabrication often induces reinforcement-rich and reinforcement-poor zones due to melt-pool dynamics and particle migration \cite{HMEIDAT2020101385}. Finally, the numerical analysis of the polycrystal RVE models with various numbers of Al grains surrounding a single Al$_2$O$_3$ particle can be used to represent composite microstructures subjected to severe plastic deformation (SPD). Such processing can refine the metallic grain structure to the ultrafine regime and thereby enhance mechanical strength \cite{NosewiczHPTMultiscale2021}.

     \SN{

     In contrast to our previous work \cite{MAZDZIARZ2024109953}, which focused on the atomistic characterization of individual structural components of a NiAl--Al$_2$O$_3$ composite, including the matrix, ceramic reinforcement, grain-boundary representations, and isolated metal--ceramic interfaces, the present study considers complete Al--Al$_2$O$_3$ particle-containing composite configurations. Here, the Al matrix, Al$_2$O$_3$ reinforcement, and their interface are modeled simultaneously within monocrystalline unit-cell and polycrystalline RVE systems. This enables the systematic assessment of the effects of ceramic volume fraction, matrix grain structure, and loading direction on the effective elastic and damage-related response of the composite. The resulting parameters therefore represent effective properties of the integrated composite configurations rather than the properties of isolated structural components.
     
    The trends identified in the present study should be interpreted within the limits of the adopted idealized atomistic configurations. The unit-cell model contains a single spherical Al$_2$O$_3$ inclusion and does not account for particle-size distribution, clustering, residual porosity, variations in particle shape, or a statistical distribution of matrix--particle interface orientations.
    }

	\subsection{Computational methodology}
	\label{sec:Cm}
	
\begin{figure}[!h] 
		\centering
		\begin{tabular}{cc}
			{\footnotesize (a)} & {\footnotesize (b)}\\
			\includegraphics[width=0.460\linewidth]{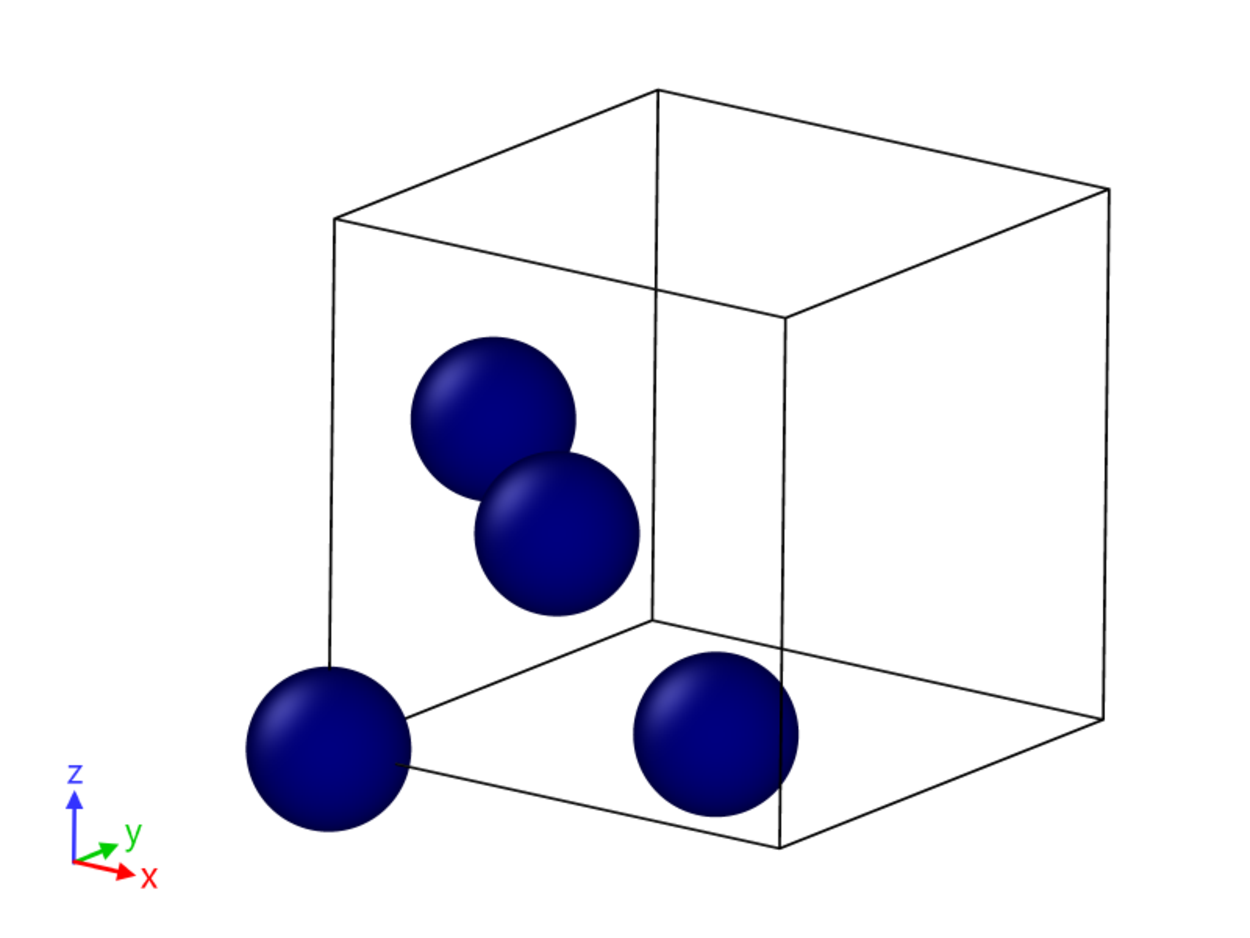} &
			\includegraphics[width=0.460\linewidth]{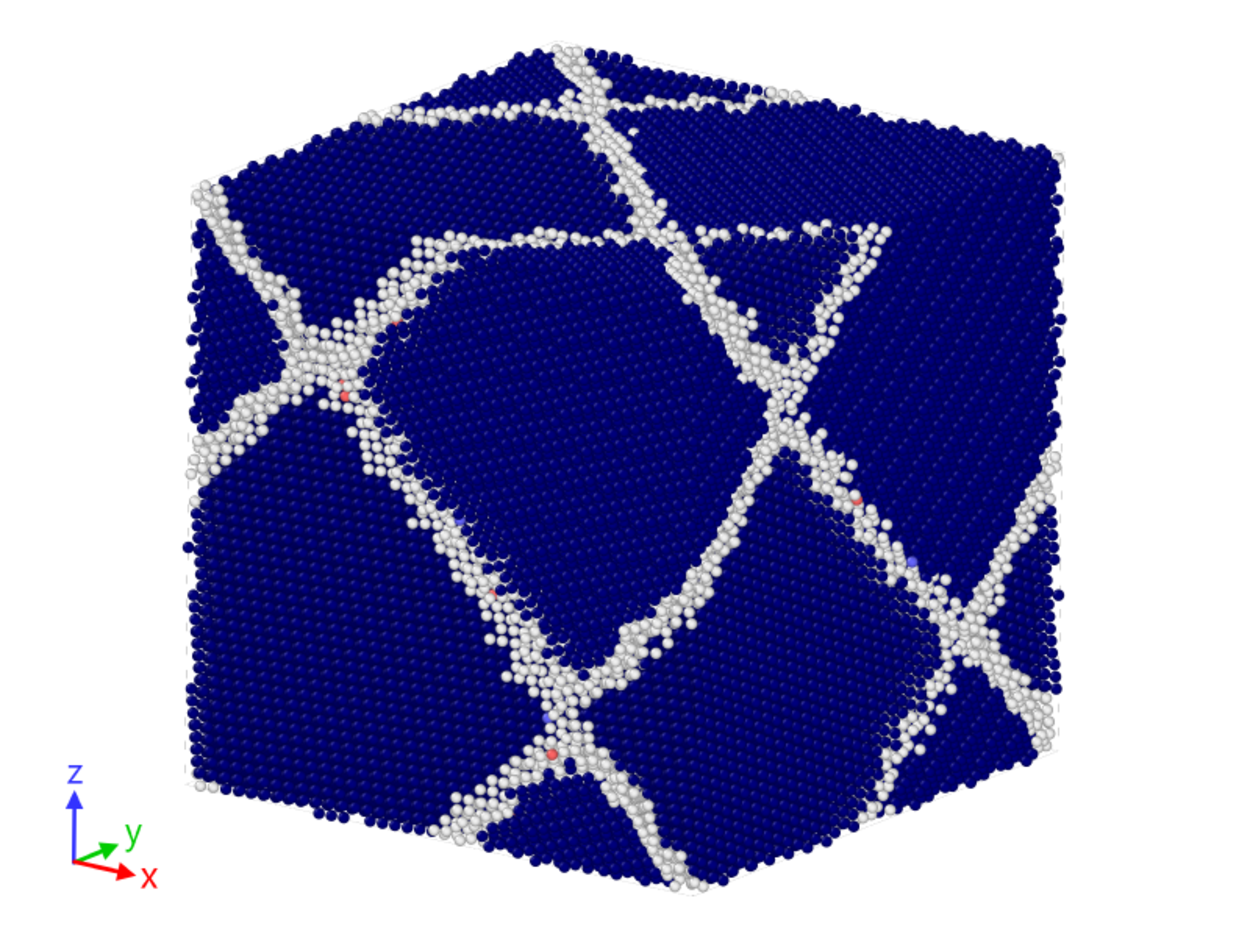} \\
			{\footnotesize (c)} & 	{\footnotesize (d)}\\
			\includegraphics[width=0.460\linewidth]{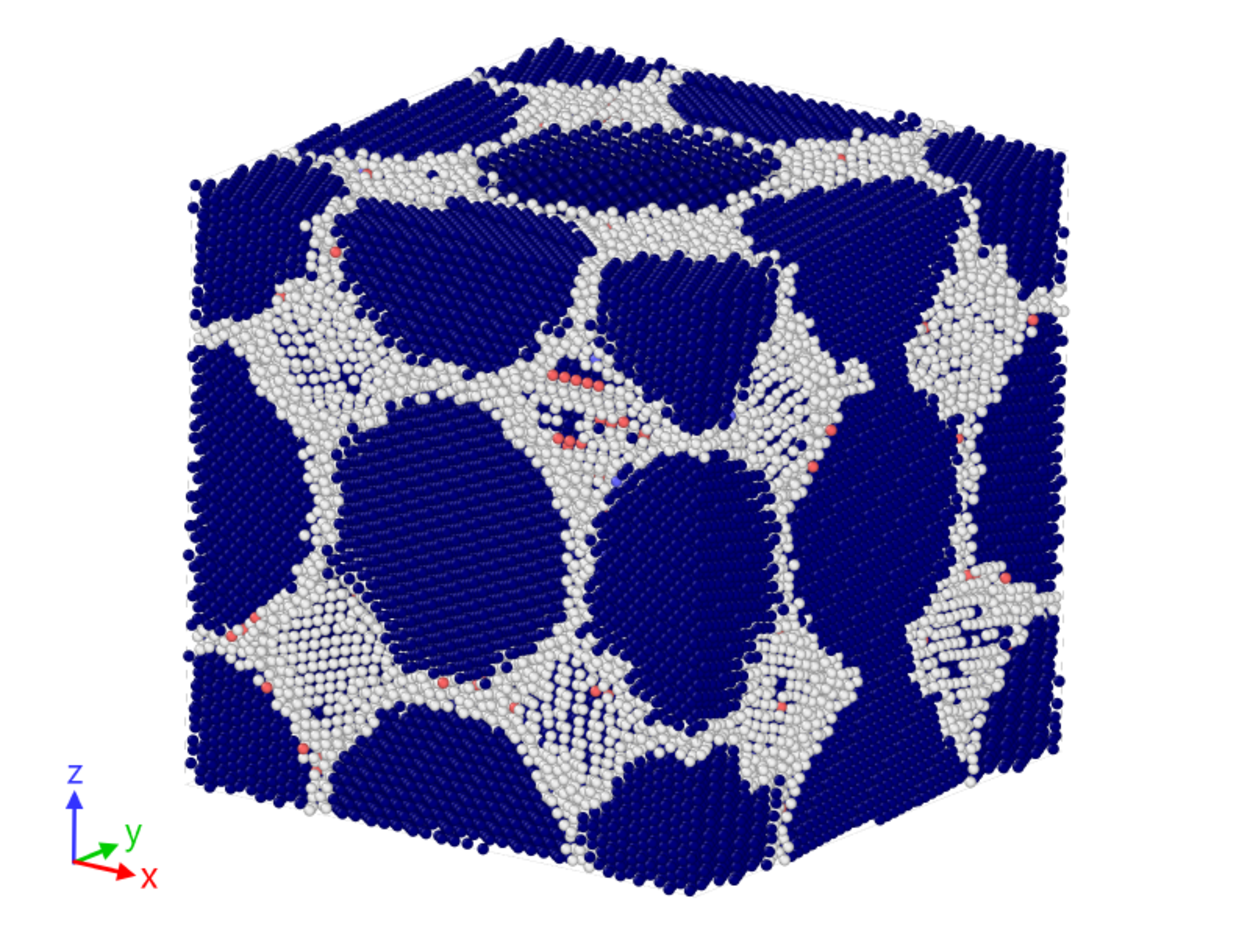} & 	
			\includegraphics[width=0.460\linewidth]{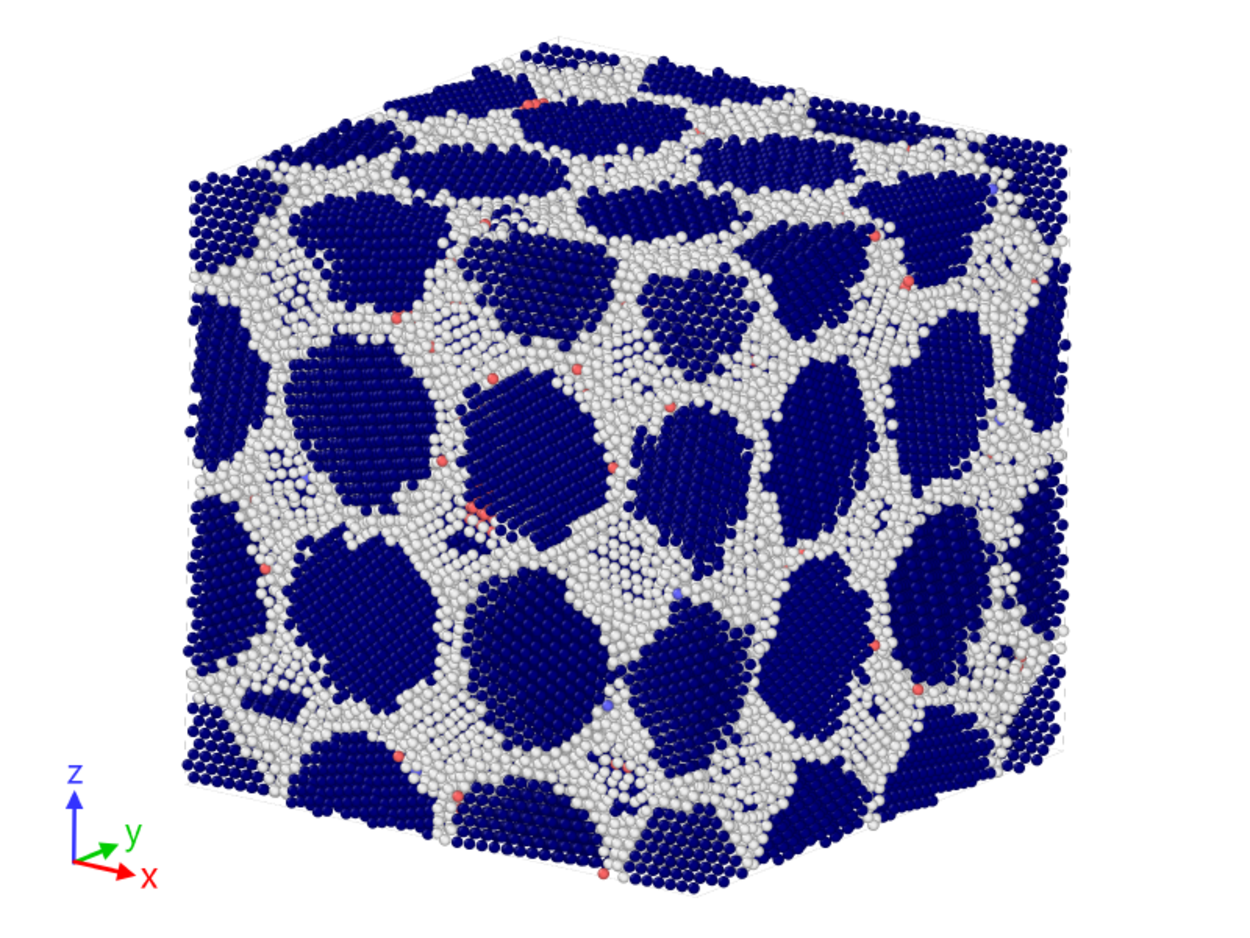} \\
			{\footnotesize (e)} \\
			\includegraphics[width=0.460\linewidth]{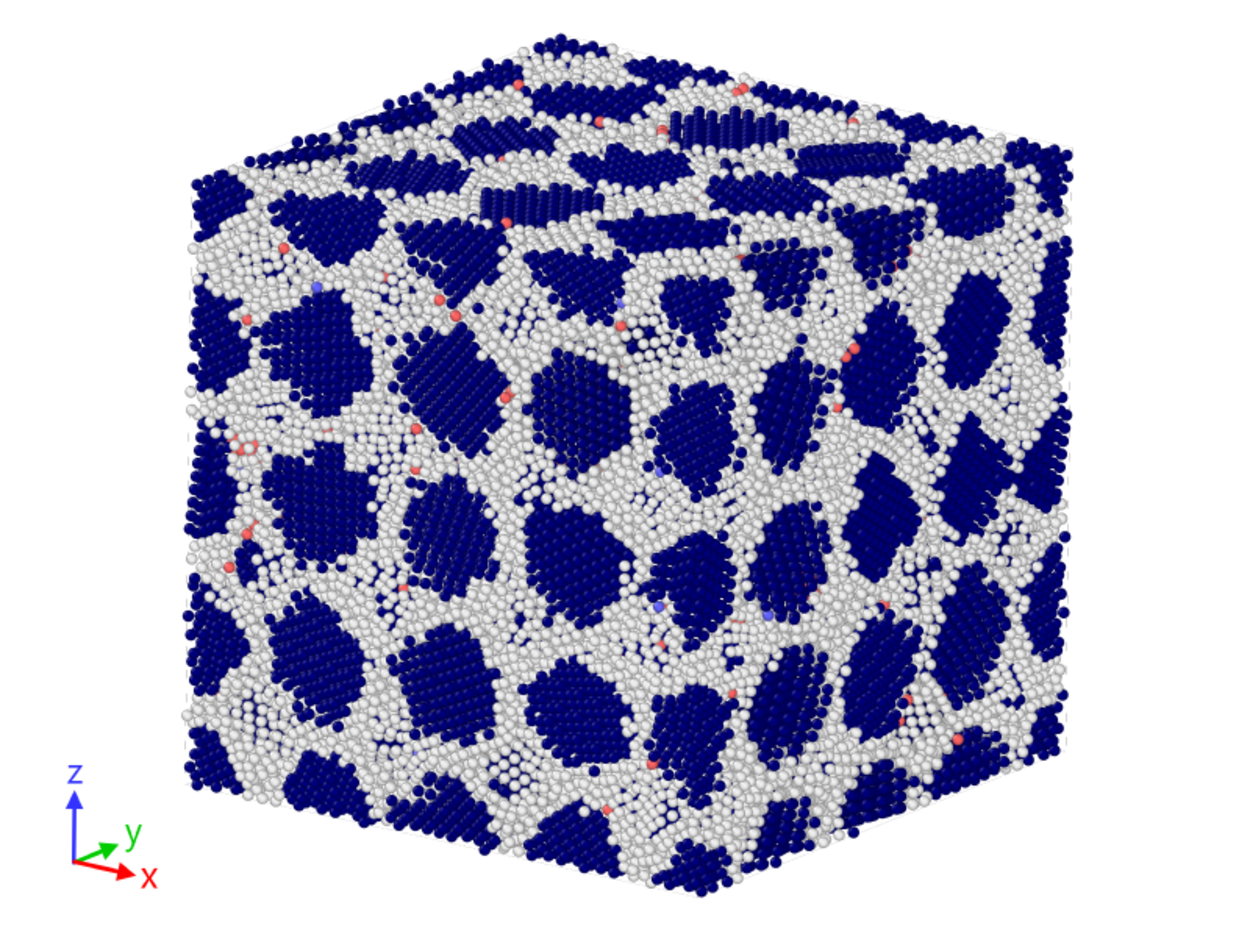} \\
		\end{tabular}
		\caption{Al: (a) FCC basic cell X=[100] Y=[010] Z=[001], (b) {polycrystal} with a size of $32 \times 32 \times 32$ unit cells and 2 grains, (c) {polycrystal} with a size of $32 \times 32 \times 32$ unit cells and 16 grains, (d) {polycrystal} with a size of $32 \times 32 \times 32$ unit cells and 54 grains, (e) {polycrystal} with a size of $32 \times 32 \times 32$ unit cells and 128 grains (The blue atoms represent FCC structures).}
		\label{fig:Al}
	\end{figure}

	Aluminum, a post-transition metal, crystallizes in the face-centered cubic (FCC) Fm$\bar{3}$m space group. For the conve\MM{n}tional unit cell where the crystallographic axes of the crystal lattice are oriented in the $X$=[100], $Y$=[010] and $Z$=[001] directions, see Fig.~\ref{fig:Al}a; lattice constants are $a=b=c=4.025${\,\AA}.
	
\begin{figure}[!h] 
		\centering
		\begin{tabular}{cc}
			{\footnotesize (a)} & {\footnotesize (b)} \\
			\includegraphics[width=0.470\linewidth]{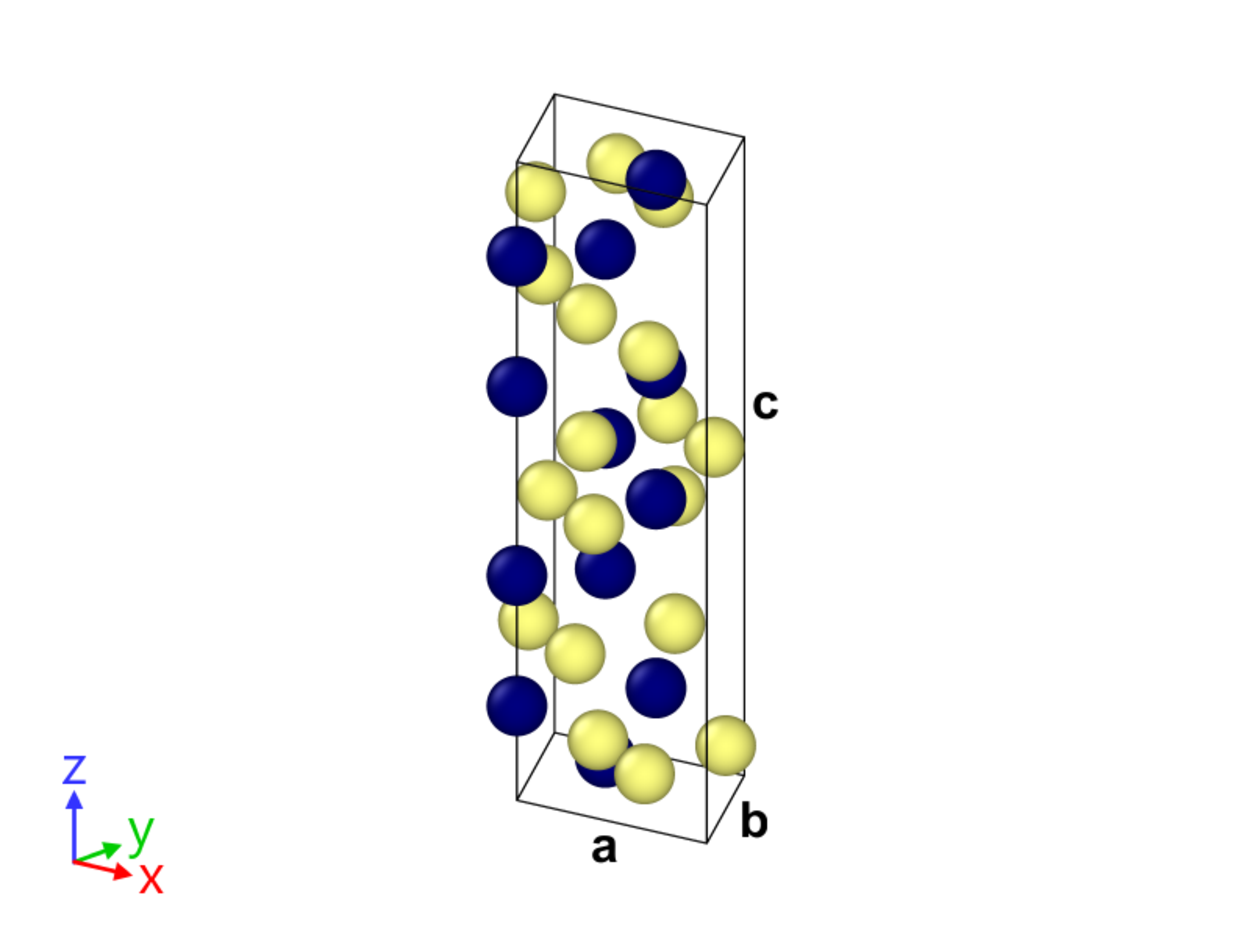} &
			\includegraphics[width=0.470\linewidth]{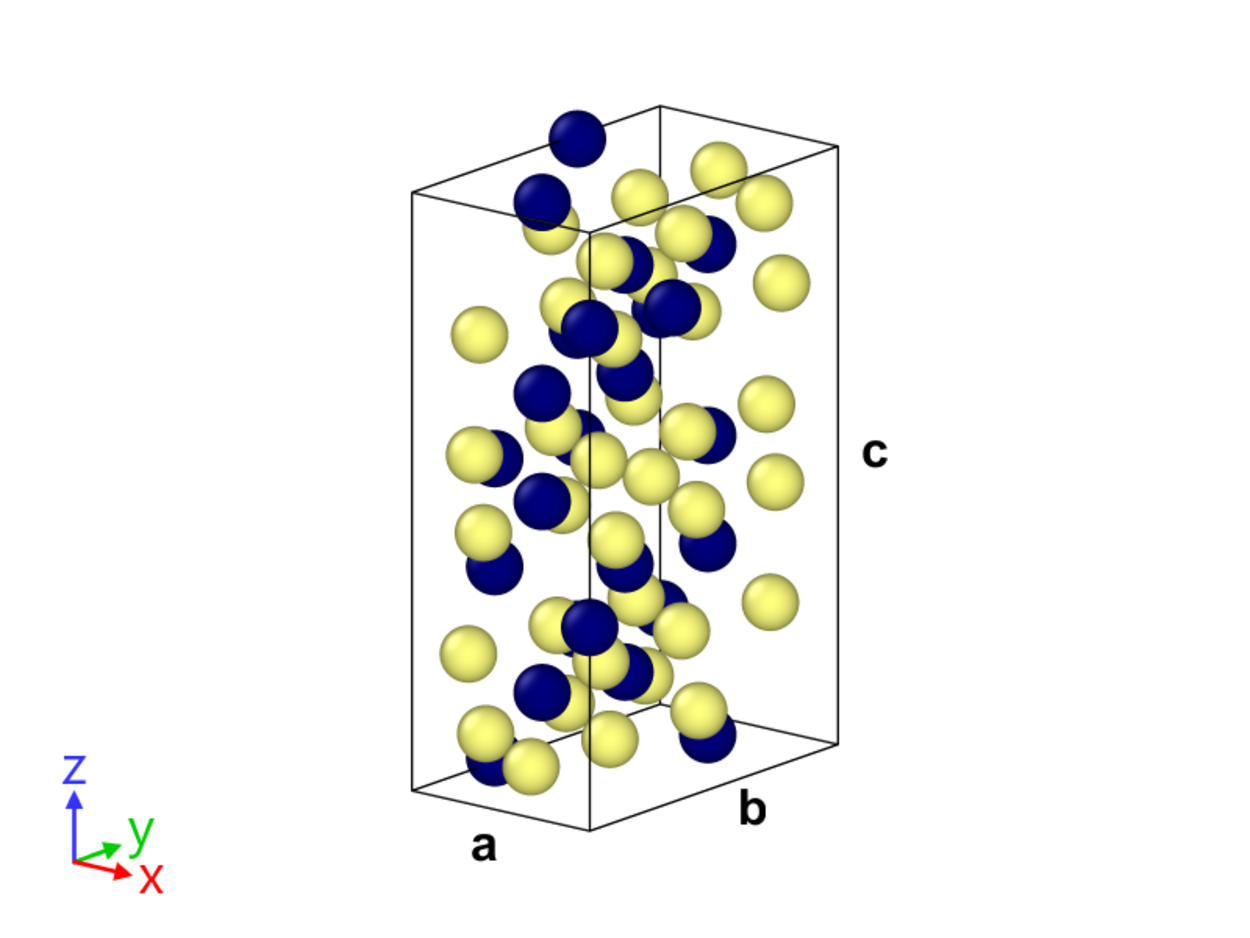}\\
		\end{tabular}
		\caption{Al$_2$O$_3$: (a) hexagonal and (b) orthorhombic basic cell (The yellow and blue atoms represent O and Al, respectively).}
		\label{fig:Al2O3_geo}
	\end{figure}

	Corundum is a crystalline form of alumina, i.e., aluminum oxide ($\alpha$-Al$_2$O$_3$)\MM{, crystallizing} in the trigonal R$\bar{3}$c space group. For the conve\MM{n}tional unit cell shown in Fig.~\ref{fig:Al2O3_geo}a, $a=b=4.758${\,\AA} and $c=12.99${\,\AA}. For the orthorhombic basic cell shown in Fig.~\ref{fig:Al2O3_geo}b, $a=4.758${\,\AA}, $b=8.24${\,\AA} and $c=12.99${\,\AA}.
	
\begin{figure}[!h] 
		\centering
		\begin{tabular}{cc}
			{\footnotesize (a)} & {\footnotesize (b)} \\
			\includegraphics[width=0.470\linewidth]{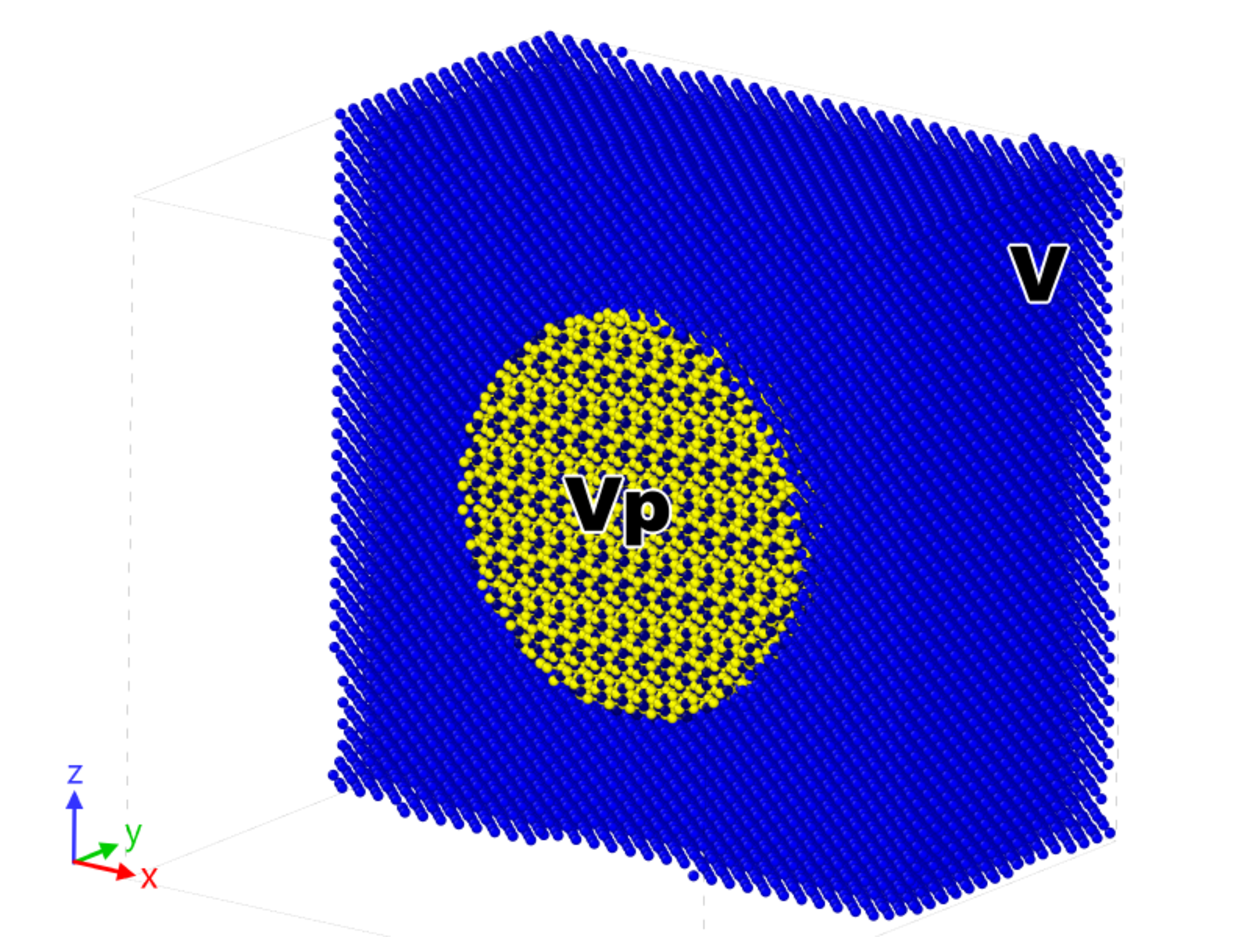} &
			\includegraphics[width=0.470\linewidth]{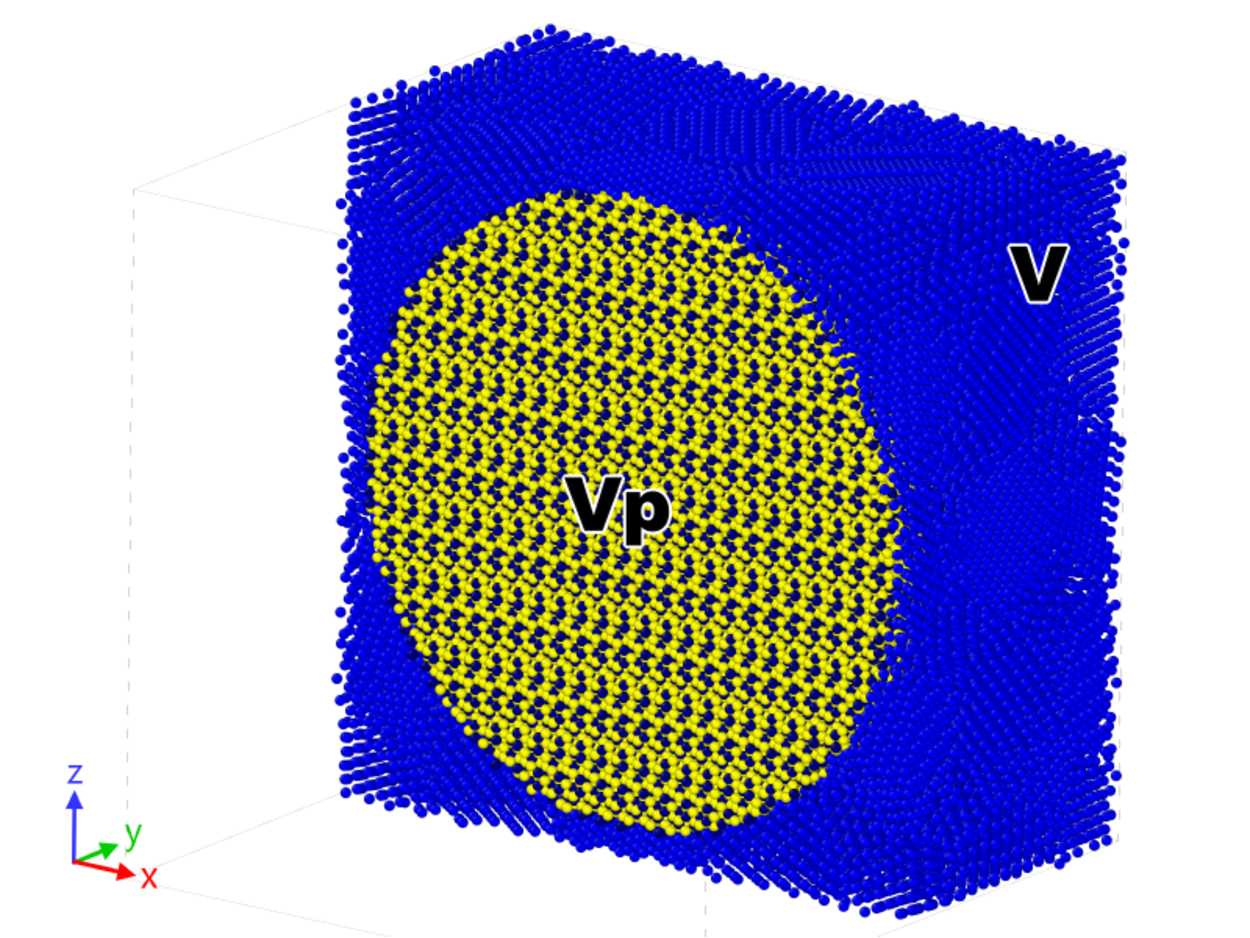}\\
		\end{tabular}
		\caption{Al-Al$_2$O$_3$ cross section: (a) monocrystal Al with a size of $32 \times 32 \times 32$ unit cells and 10\% spherical Al$_2$O$_3$ particle,  (b) monocrystal Al with a size of $32 \times 32 \times 32$ unit cells and 40\% spherical Al$_2$O$_3$ particle (The yellow and blue atoms represent O and Al, respectively).}
		\label{fig:AlAl2O3}
	\end{figure}

	\MM{In this study}, we analyze the mechanical behavior of aluminum-based composites reinforced with corundum nanoparticles and define the volume fraction of corundum (see Fig.~\ref{fig:AlAl2O3}) in the entire sample using a defined quantity called the volume ratio $\mathbf{V_R}$: 
	\begin{align}
		\mathbf{{V}_{R}} :  = \dfrac{\mathbf{V_p}}{\mathbf{V}},
		\label{eqn:VR}
	\end{align}
	where $\mathbf{V_p}$ is volume of corundum spherical nanoparticle and $\mathbf{V}$ is total volume of the cubic sample. \MM{If a value in the $\mathbf{V_R}$ text is expressed as [\%], it means that it has been multiplied by 100.}
	
\begin{figure}[!h] 
		\centering
		\begin{tabular}{cc}
			{\footnotesize (a)} & {\footnotesize (b)}\\
			\includegraphics[width=0.470\linewidth]{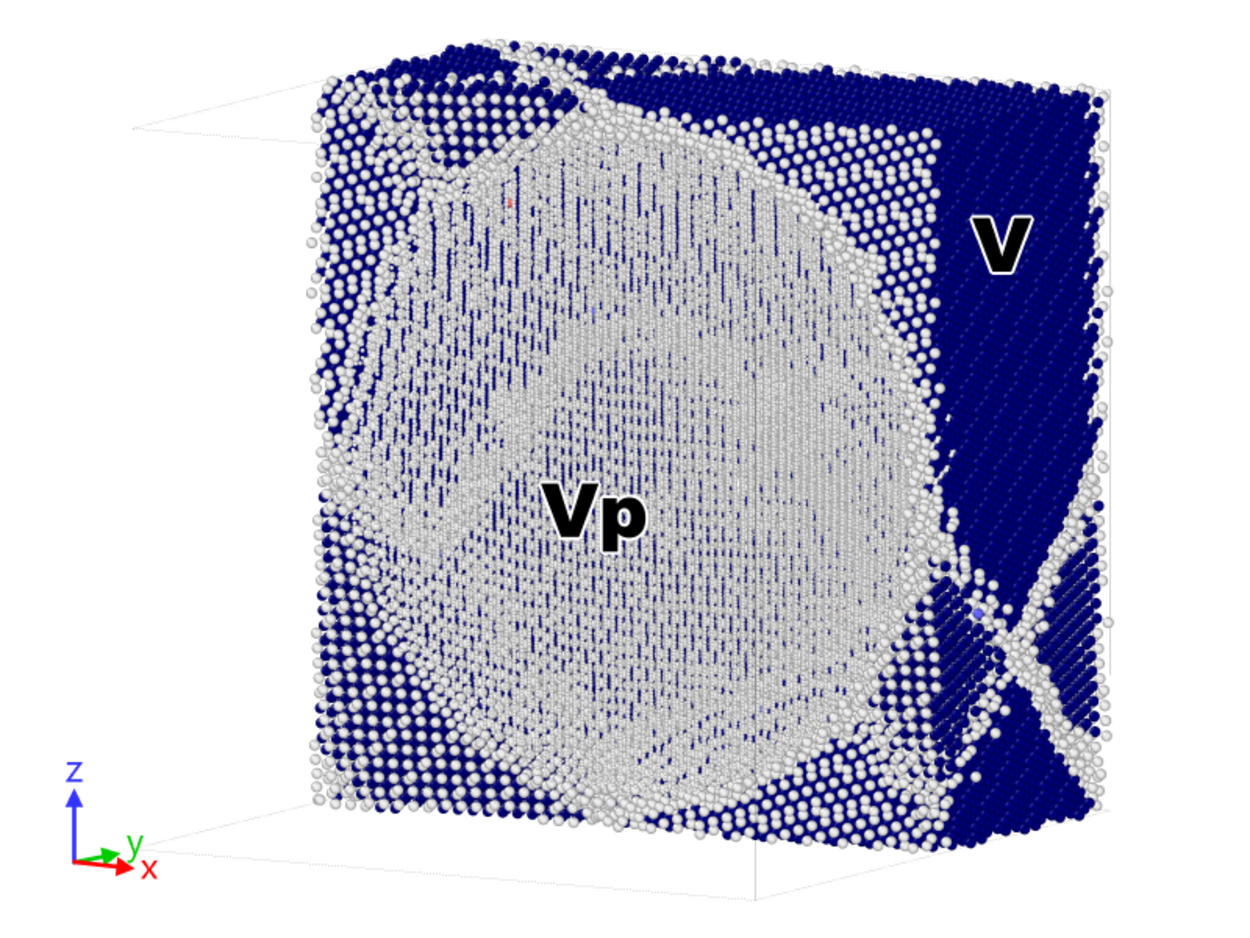} &
			\includegraphics[width=0.470\linewidth]{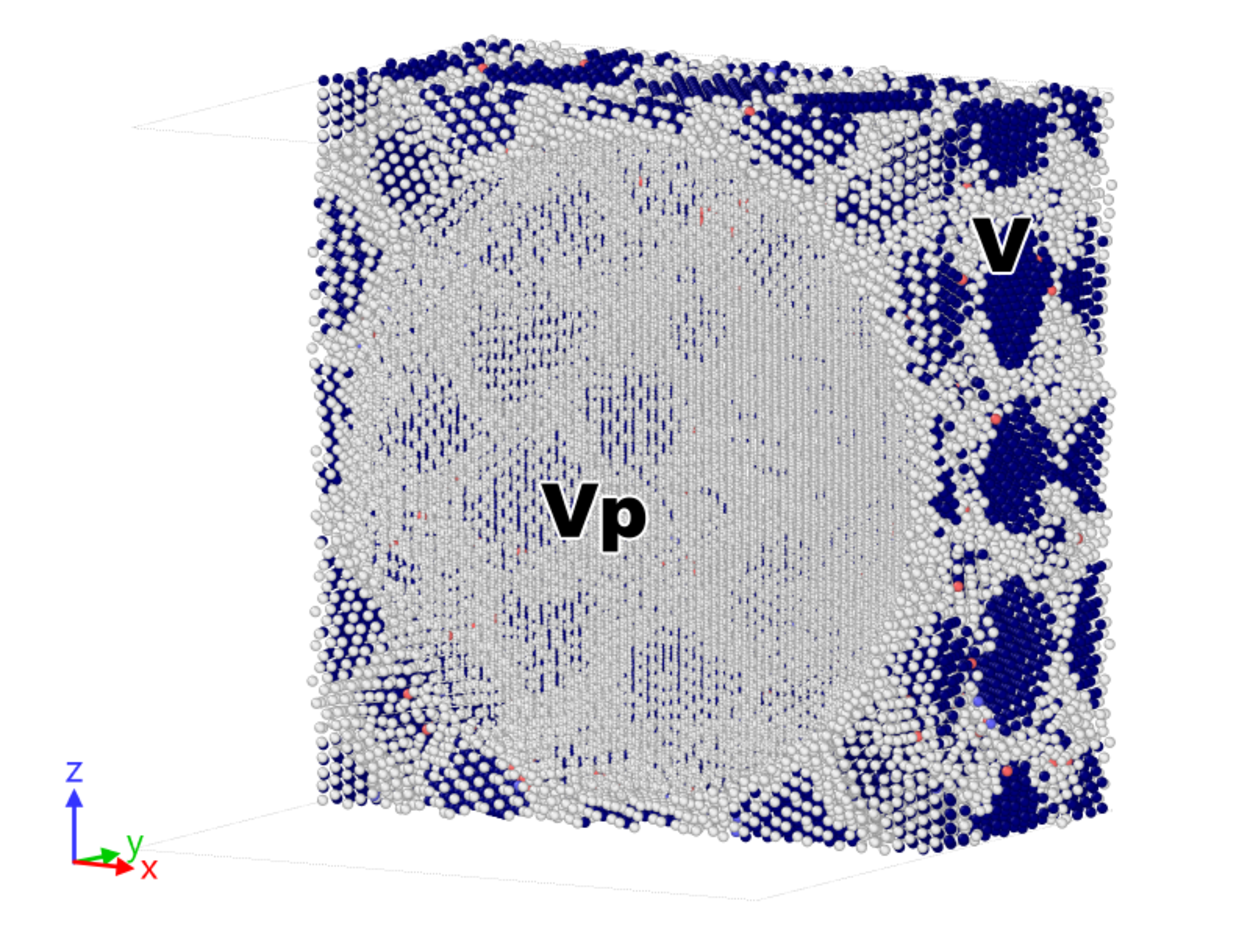} \\
		\end{tabular}
		\caption{{Al-Al$_2$O$_3$} cross section: (a) polycrystal Al with a size of $32 \times 32 \times 32$ unit cells, 2 grains and 40\% spherical Al$_2$O$_3$ particle,  (b) polycrystal Al with a size of $32 \times 32 \times 32$ unit cells, 128 grains and 40\% spherical Al$_2$O$_3$ particle (The blue atoms represent FCC structures).}
		\label{fig:Al-Al2O332Ng}
	\end{figure}

	All the analyzed cases are enumerated in Table~\ref{tab1} and visualized in sample Figs.~\ref{fig:Al}-\ref{fig:Al-Al2O332Ng}. The analyzed systems include five aluminum samples: one monocrystalline and four polycrystalline models containing 2–128 grains. Additional systems comprise monocrystalline corundum, monocrystalline Al samples of different sizes containing 40\% corundum, monocrystalline Al with a fixed size and variable corundum content, and polycrystalline Al with a fixed size, variable grain count, and 40\% corundum. For each of these 18 cases, the stiffness tensor \cite{Mazdziarz2025am} and \MM{stress-strain} curves were determined for tensile and compression tests in the X and Z directions. \MM{Due to the cubic symmetry of aluminum and the trigonal symmetry of corundum, as well as the construction of the samples under analysis, they exhibit identical properties in the X and Y directions.} \comm{Full details can be found in the \href{run:./Al_Al2O3_Supplementary_data.pdf}{Supplementary data}.}
	
	The homogenized isotropic bulk modulus (\textbf{$B$}), the shear modulus (\textbf{$G$}), Young's modulus (\textbf{$E$}), and Poisson's ratio (\textbf{$\nu$}) were calculated by means of a Voigt–Reuss–Hill (VRH) average \cite{Hill_1952, Kube2016, Mazdziarz2025am}, whereas the universal elastic anisotropy index (\textbf{$A^U$}) was determined in accordance with \cite{OstojaStarzewski}. 
	
	\begin{table}[h!]
		\centering
			\scriptsize
			\caption{Summary of the molecular model configurations, including domain size, number of Al grains, alumina nanoparticle volume fraction \,${V_p}/{V}$\,[\%] and the interatomic potentials employed.}\label{tab1}%
			\begin{tabular}{@{}ccccc@{}}
				\hline
				Sample & Size  & No. of grains & ${V_p}/{V}$ & Potential \\
				\hline
				\multicolumn{5}{c}{\hyperref[fig:Al]{Aluminum}} \\
				\hline
				\hyperref[supp-fig:Al]{Al\_32\_1g~EAM} & $32 \times 32 \times 32$  & 1    & - & EAM \\
				{Al\_32\_2g~EAM} & $32 \times 32 \times 32$  & 2    & - & EAM \\
				{Al\_32\_16g~EAM} & $32 \times 32 \times 32$ & 16   & - & EAM \\
				{Al\_32\_54g~EAM} & $32 \times 32 \times 32$ & 54   & - & EAM \\
				{Al\_32\_128g~EAM} & $32 \times 32 \times 32$ & 128 & - & EAM \\
				
				\hline
				\multicolumn{5}{c}{\hyperref[fig:Al2O3_geo]{Alumina}} \\
				\hline
				\hyperref[supp-fig:Al2O3]{Al$_2$O$_3$~COMB3} & $1 \times 1 \times 1$  & 1    & - & COMB3 \\
				
				\hline
				\multicolumn{5}{c}{\hyperref[fig:AlAl2O3]{Al-Al$_2$O$_3$}} \\
				\hline
				\hyperref[supp-fig:Al-Al2O3_N]{Al-Al$_2$O$_3$\_4\_40~EAM+COMB3} & $4 \times 4 \times 4$    & 1    & 40 & EAM+COMB3 \\
				{Al-Al$_2$O$_3$\_8\_40~EAM+COMB3} & $8 \times 8 \times 8$    & 1    & 40 & EAM+COMB3 \\
				{Al-Al$_2$O$_3$\_16\_40~EAM+COMB3} & $16 \times 16 \times 16$  & 1    & 40 & EAM+COMB3 \\
				{Al-Al$_2$O$_3$\_32\_40~EAM+COMB3} & $32 \times 32 \times 32$  & 1    & 40 & EAM+COMB3 \\
				\hdashline
				\hyperref[supp-fig:Al-Al2O3_32N0]{Al-Al$_2$O$_3$\_32\_10~EAM+COMB3} & $32 \times 32 \times 32$  & 1    & 10 & EAM+COMB3 \\
				{Al-Al$_2$O$_3$\_32\_20~EAM+COMB3} & $32 \times 32 \times 32$  & 1    & 20 & EAM+COMB3 \\
				{Al-Al$_2$O$_3$\_32\_30~EAM+COMB3} & $32 \times 32 \times 32$  & 1    & 30 & EAM+COMB3 \\
				{Al-Al$_2$O$_3$\_32\_40~EAM+COMB3} & $32 \times 32 \times 32$  & 1    & 40 & EAM+COMB3 \\
				\hdashline
				\hyperref[supp-fig:Al-Al2O3_32_Ng]{Al-Al$_2$O$_3$\_32\_2g\_40~EAM+COMB3} & $32 \times 32 \times 32$  & 2    & 40 & EAM+COMB3 \\
				{Al-Al$_2$O$_3$\_32\_16g\_40~EAM+COMB3} & $32 \times 32 \times 32$  & 16    & 40 & EAM+COMB3 \\
				{Al-Al$_2$O$_3$\_32\_54g\_40~EAM+COMB3} & $32 \times 32 \times 32$  & 54    & 40 & EAM+COMB3 \\
				{Al-Al$_2$O$_3$\_32\_128g\_40~EAM+COMB3} & $32 \times 32 \times 32$  & 128    & 40 & EAM+COMB3 \\
				\hline
			\end{tabular}
	\end{table}
	
	All the atomistic calculations were performed at a temperature of 0\,K through energy minimization using the molecular statics (MS) approach \comm{\cite{Maz2011, Mazdziarz2015}} and the Large-scale Atomic/Molecular Massively Parallel Simulator (\href{https://www.lammps.org/}{LAMMPS}) \cite{Plimpton1995}. 
        \MM{A shortcoming of MS is that it cannot directly account for the effect of temperature on the behavior of the material being analyzed. To account for the effect of temperature, molecular dynamics simulations should be performed using an appropriate thermostat. Determining the mechanical properties of a material under analysis, such as the stiffness tensor or stress-strain curves, requires the application of strain to the sample in molecular calculations. An inherent feature of molecular dynamics simulations is that strain is typically imposed dynamically by rescaling the simulation box coordinates at a defined strain rate. For computational reasons, at the nanoscale these strain rates are very high and often range from approximately 10$^7$\,s$^{-1}$ to 10$^{10}$\,s$^{-1}$. Unfortunately, the mechanical properties depend strongly on strain rates. For nanocrystalline copper \cite{RIDA2020109294}, for example, when the strain rate changes by three orders of magnitude, the calculated Young's modulus changes by nearly a factor of 2, and the maximum stress on the stress-strain curves changes by more than a factor of 2, making it rather difficult to expect the results to stabilize for lower rates.}
	Molecular statics can be regarded as a special, limiting case of molecular dynamics (MD), \MM{though it is likely more reliable for determining mechanical properties. Both terms are used interchangeably in the text, but it should be understood that these are, in fact, static calculations.}
	
	The aluminum polycrystals shown in Figs.~\ref{fig:Al}b-e were generated using a methodology taken from \cite{Kowalczyk18, Kowalczyk19, MAZDZIARZ2024109953}. All samples used in the computations were generated with the \href{https://atomsk.univ-lille.fr/}{Atomsk} \cite{Hirel2015212} code and visualized at the atomistic level with the \href{https://www.ovito.org/}{OVITO} \cite{Ovito2010} program. OVITO was also used to quantify local lattice disorder, cohesive energy per atom ($E_c$/atom), and the centrosymmetry parameter (CSP). If the generated corundum particles were not fully stoichiometric, which implies a lack of charge neutrality, they were manually modified to meet this requirement. \MM{Full data on the computational samples, simulation parameters, and complete results can be found in the \href{run:./Al_Al2O3_Supplementary_data.pdf}{Supplementary data}.}
	
	The choice of interatomic potential in molecular simulations is of fundamental importance for the quality of the calculations.
	The embedded-atom method (EAM) potentials are the natural choice for metals, and here, for Al, the following potential \cite{Winey2009} was used.
	Modeling the behavior of ionic compounds is significantly more expensive than modeling the behavior of metals. This is related to the long-range interactions between the point charges present there and the necessity for charge equilibration (QEq) in more advanced potentials.   
	  Here, for pure Al$_2$O$_3$, as well as for interactions between Al and Al$_2$O$_3$, the charge-optimized many-body (COMB) potential \cite{Choudhary2015, Kumar2015} was used. \comm{In this case, the balance between accuracy and computational costs is particularly evident \cite{Gramatte2023}.}
	We decided to compare the cost of determining the stiffness tensor for the same sample size for aluminum (EAM potential) and for corundum (different potentials available in LAMMPS), i.e., 60 atoms. While the normalized cost for EAM \cite{Winey2009} is 1, it is 60 for Vashishta \cite{Vashishta2008}, 300 for COMB, 1,200 for Reactive Force Field (ReaxFF)  \cite{Sen2014}, 1,600 for SM \cite{Streitz1994}, and 190,000 for Second-Moment tight-Binding QEq (SMTB-Q) \cite{SALLES2016181}. Interestingly, the Vashishta potential performs well with pure, ideal corundum. Unfortunately, however, it does not balance the charges and cannot describe the interaction between Al and Al$_2$O$_3$. 
	\MM{An alternative to the COMB potential for modeling the aluminum-corundum system could possibly be the ReaxFF potential. However, for a system chemically similar to ours, i.e., alumina formed on top of aluminum, it was noted in \cite{LUU2022128342} that ReaxFF is not suitable for modeling the mechanical properties of aluminum and alumina during nanoindentation.}  
	Although COMB can also be used for pure aluminum, EAM is more accurate and significantly cheaper \cite{Choudhary2014}.

     \SN{The identification of damage and fracture mechanisms was based on the combined interpretation of the mechanical response and atomistic structural evolution. Brittle- and ductile-type damage were distinguished primarily from the post-peak stress--strain response, with a pronounced abrupt loss of load-carrying capacity associated with brittle-type behavior and a more gradual post-peak stress decrease associated with progressive, ductile-type damage. The location of fracture in the final deformed unit-cell and RVE configurations was subsequently used to distinguish interface-dominated, transgranular, and intergranular damage paths. Cohesive-energy distributions provided complementary information on regions of bond weakening and energetic disturbance, whereas CSP was used as a qualitative measure of local lattice disorder.}

   \SN{Direct experimental characterization of mechanical response and deformation mechanisms at the atomistic scale remains highly challenging. Molecular simulations are frequently employed as a stand-alone approach to investigate nanoscale mechanical behavior without direct experimental reproduction of the simulated configuration. This strategy has been adopted for {Al--Al$_2$O$_3$} interfaces and nanocomposites \cite{SAZGAR2015200, Mei2013}, as well as for the deformation and failure of alumina at the nanoscale \cite{Le2022TensileAl2O3, Lee2024AmorphousAl2O3, Xu2023}. In these studies, the atomistic simulations were not accompanied by dedicated experiments reproducing and directly validating the simulated mechanical response.

Such an approach is particularly relevant for nanoscale atomistic models. The largest system considered here is approximately 12.9\,nm in size ($32^3$ Al unit cells). Direct experimental reproduction at this scale remains highly challenging despite advances in in situ TEM- and MEMS-based testing, mainly because of limitations in specimen preparation, loading, force measurement, and atomic-scale observation \cite{Kang2017InSitu, Li2023InSituTEM}. Therefore, the present models should be regarded as controlled nanoscale systems for identifying intrinsic mechanical trends rather than direct replicas of macroscopic specimens.}

	\section{Results}
    \label{sec:Res}

\SN{This section presents the mechanical response obtained from the atomistic
simulations. The results are organized into two complementary groups. First,
the unit-cell models are used to examine the effect of Al$_2$O$_3$ volume
fraction on the effective elastic response and on the tensile and compressive
behavior of the composite. Second, the polycrystalline RVE models are analyzed
to assess the influence of matrix grain structure on the corresponding
effective properties and damage response.}

	\subsection{Unit-cell model}
	\comm{\subsubsection{The effect of ceramic content}}
	
\begin{figure}[!h] 
		\centering
		\begin{tabular}{cc}
			{\footnotesize (a)} & {\footnotesize (b)} \\
			\includegraphics[width=0.47\linewidth]{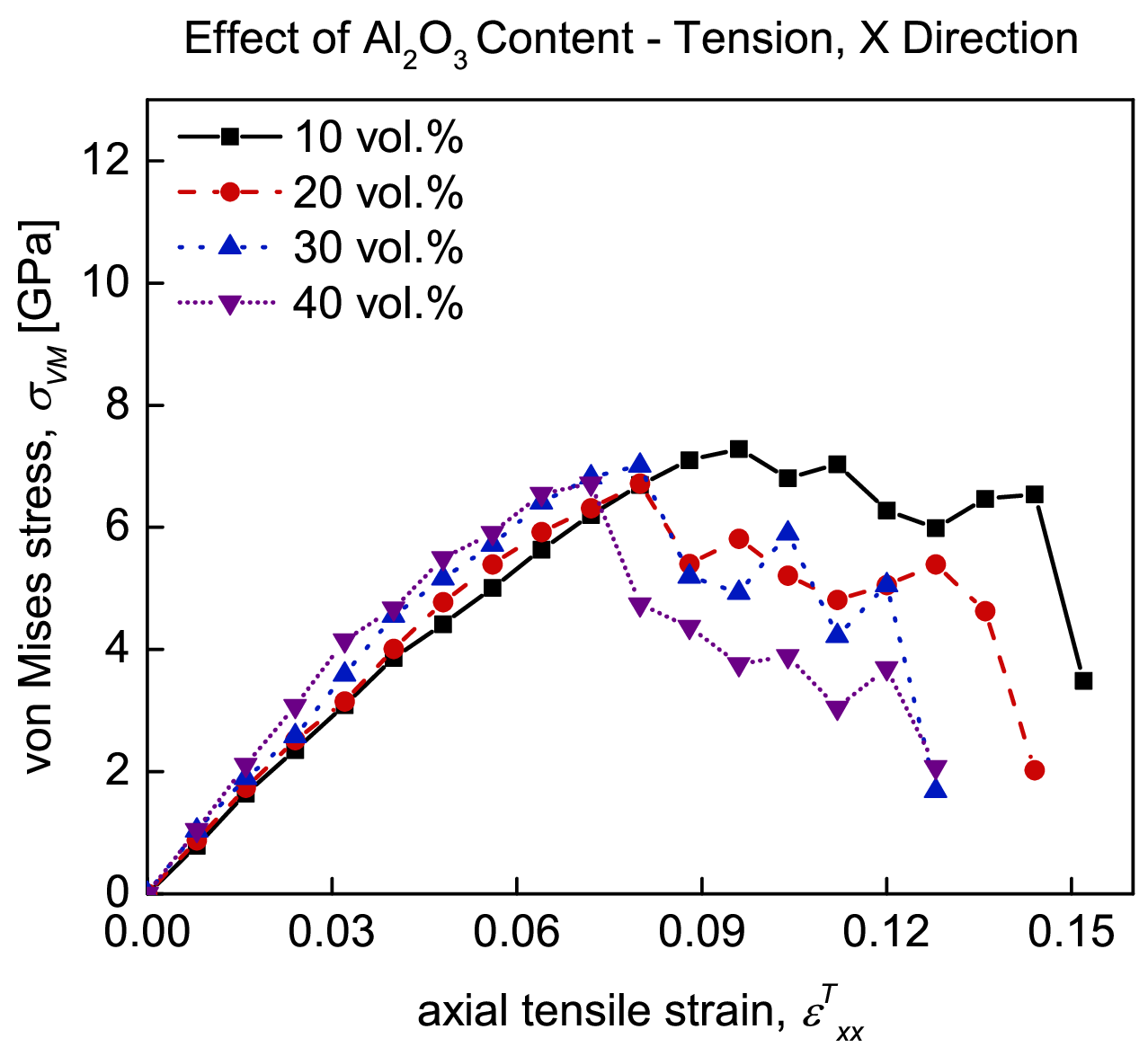} &
			\includegraphics[width=0.47\linewidth]{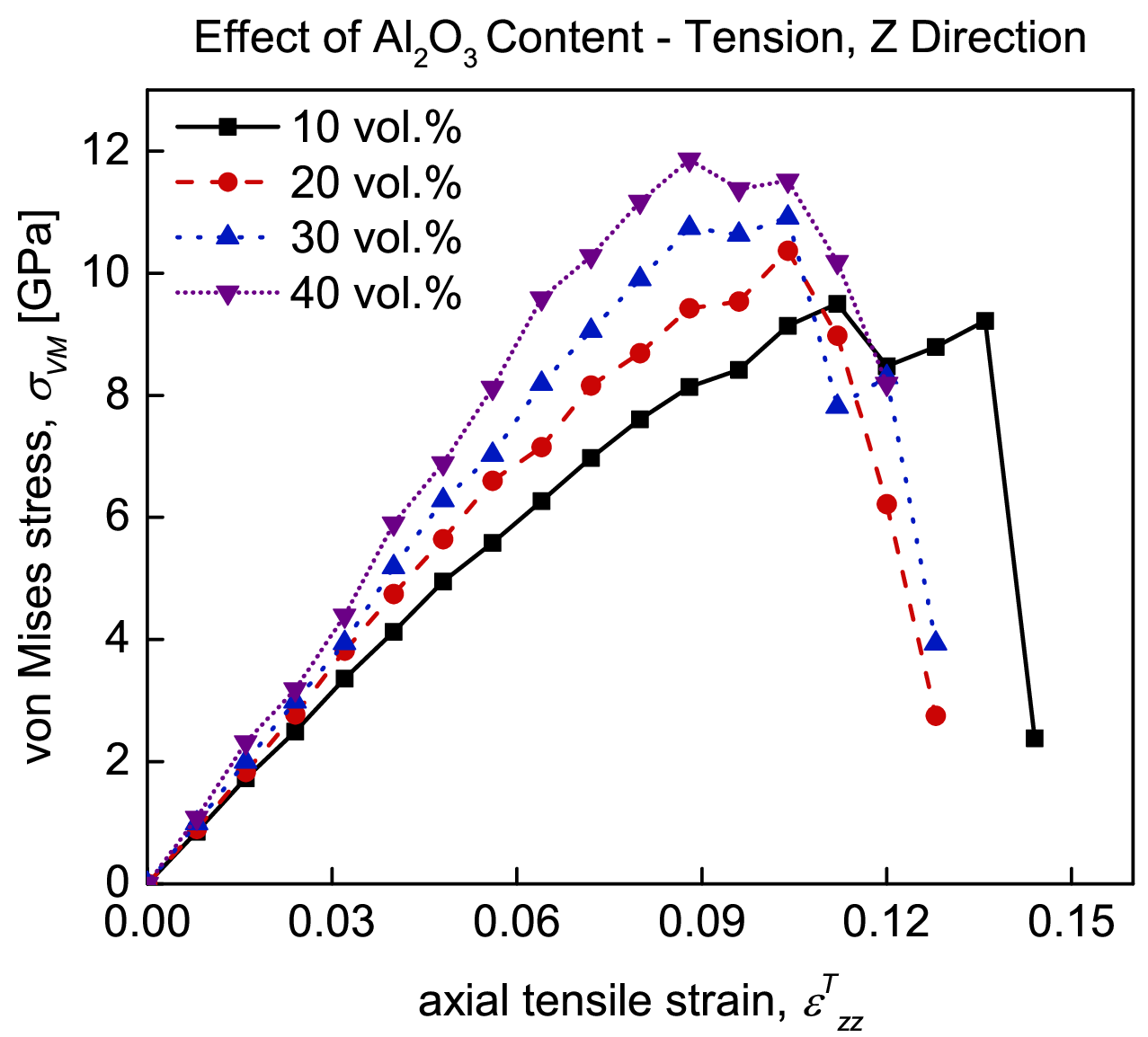}\\
		\end{tabular}
		\caption{The stress-strain evolution of {Al-Al$_2$O$_3$} unit-cell samples with different ceramic content subjected to uniaxial tension in: (a) X and (b) Z direction.}
		\label{fig:Tensile_DiffContAlumina_Mono}
	\end{figure}

	Figure~\ref{fig:Tensile_DiffContAlumina_Mono} presents von Mises tensile stress–strain curves for the {Al-Al$_2$O$_3$} unit-cell model with increasing ceramic volume fraction (10–40\%) loaded in the X (a) and Z (b) directions. For both loading directions, increasing Al$_2$O$_3$ content results in a steeper initial slope of the stress–strain curves (especially for the Z direction), indicating a higher effective stiffness of the model. This reflects the growing contribution of the stiff ceramic phase to load transfer in the early deformation regime. In the X direction (Fig.~\ref{fig:Tensile_DiffContAlumina_Mono}a), the maximum tensile stress is reached at moderate strains ($\approx$ 0.07–0.1). The composite with 10\% Al$_2$O$_3$ exhibits the highest peak stress, whereas higher ceramic contents (30–40\%) yield lower peak stresses. This suggests that although stiffness increases with ceramic fraction, excessive reinforcement promotes earlier damage initiation and limits the attainable macroscopic strength.
	
	\begin{table}[!h]
		\centering
		\caption{Summary of ultimate strength ($\sigma^{US}$), ultimate strain ($\varepsilon^{US}$) and failure strain ($\varepsilon^{FS}$) of {Al-Al$_2$O$_3$} unit-cell models with different ceramic volume fractions obtained from atomistic uniaxial tensile simulations in the X and Z directions. $\varepsilon^{FS}$ was defined as the strain corresponding to the last data point preceding the first stress drop of at least 40\% between two consecutive simulation points.} \vspace{3mm}
		\label{tab:content_strength_vm}
		\small
		\setlength{\tabcolsep}{12pt}
		\renewcommand{\arraystretch}{0.8}
		
		\begin{tabular}{c ccc ccc}
			\toprule
			& \multicolumn{3}{c}{x direction} & \multicolumn{3}{c}{z direction}\\
			\midrule
			Al$_2$O$_3$ (vol.\%) &$\sigma^{US}$ & $\varepsilon^{US}_{xx}$ & $\varepsilon^{FS}_{xx}$ &
			$\sigma^{US}$ & $\varepsilon^{US}_{zz}$ & $\varepsilon^{FS}_{zz}$ \\
			\midrule
			10 & 7.29 & 0.096 & 0.144 & 9.50 & 0.112 & 0.136 \\
			20 & 6.72 & 0.080 & 0.136 & 10.37 & 0.104 & 0.120 \\
			30 & 7.01 & 0.080 & 0.120 & 10.91 & 0.104 & 0.120 \\
			40 & 6.72 & 0.072 & 0.120 & 11.87 & 0.088 & 0.120 \\
			\bottomrule
		\end{tabular}
		\vspace{1mm}
	\end{table}
	
	In contrast, in the Z direction, the peak stress increases with ceramic content, with 30–40\% Al$_2$O$_3$ achieving the highest maxima ($\approx$ 12\,GPa). This indicates a relatively strong directional effect, likely associated with the anisotropy of alumina monocrystals. In both directions, higher ceramic fractions systematically reduce the strain at which peak stress occurs. Composites with 30–40\% Al$_2$O$_3$ fail at noticeably lower strains than those with 10\%, highlighting a clear stiffness–ductility trade-off. The reduction in ductility is more pronounced in the X direction, which can be seen in Table~\ref{tab:content_strength_vm}.
	
	All curves in Fig.~\ref{fig:Tensile_DiffContAlumina_Mono}a exhibit a gradual post-peak stress decay rather than an abrupt drop, indicating progressive damage accumulation. Higher ceramic content leads to an earlier onset of softening and lower residual stress levels at large strains. Such failure behavior can be characterized by ductile-type damage. In the Z direction, the drop in stress is more pronounced and can be treated as brittle-type damage. This confirms that the effective mechanical response is governed not only by ceramic volume fraction but also by the orientation of the matrix/reinforcing phase and load-bearing paths within the unit cell. 
	
\begin{figure}[!h] 
		\centering
		\begin{tabular}{ccc}
			{\footnotesize (a)} & {\footnotesize (b)} & {\footnotesize (c)} \\
			\includegraphics[width=0.310\linewidth]{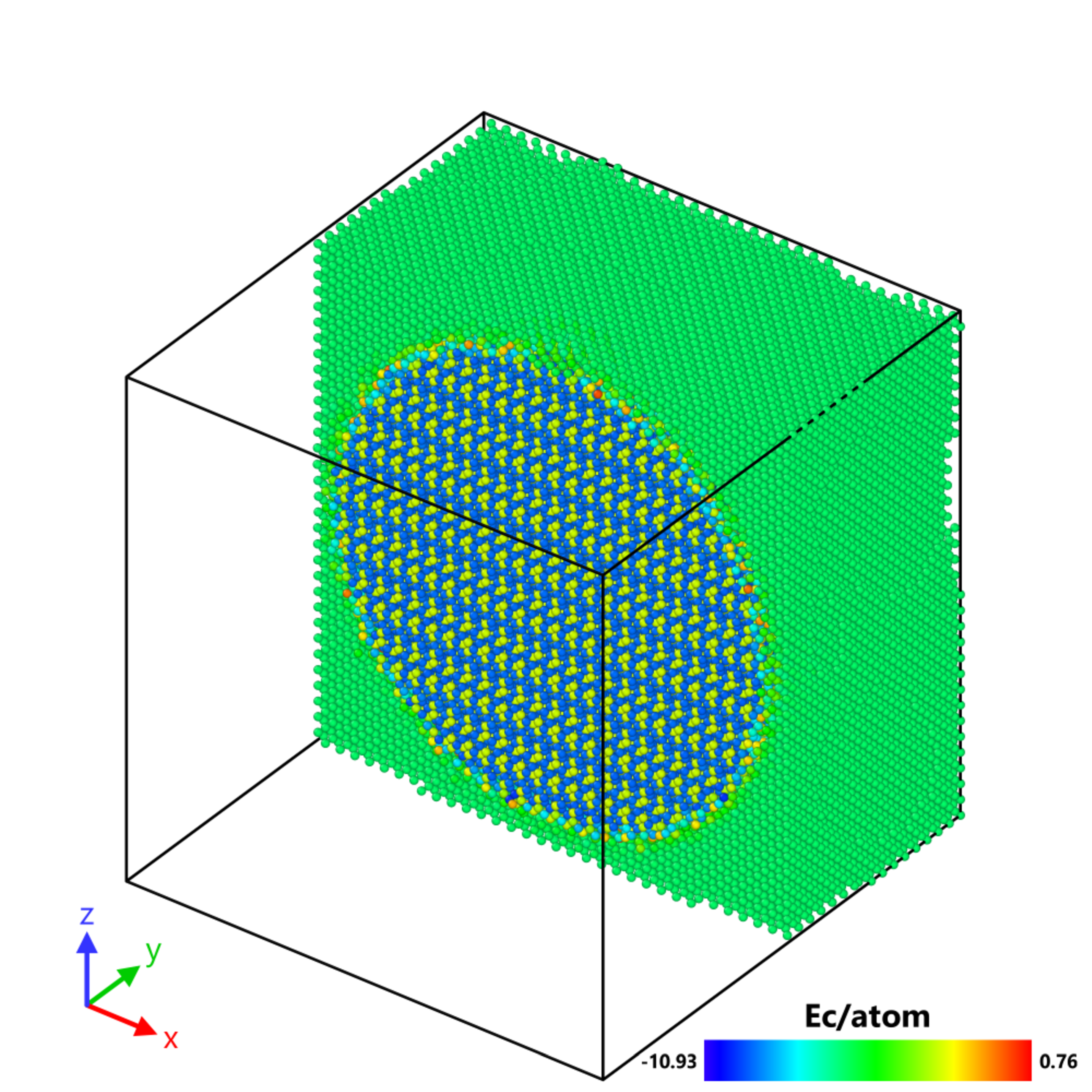} &
			\includegraphics[width=0.310\linewidth]{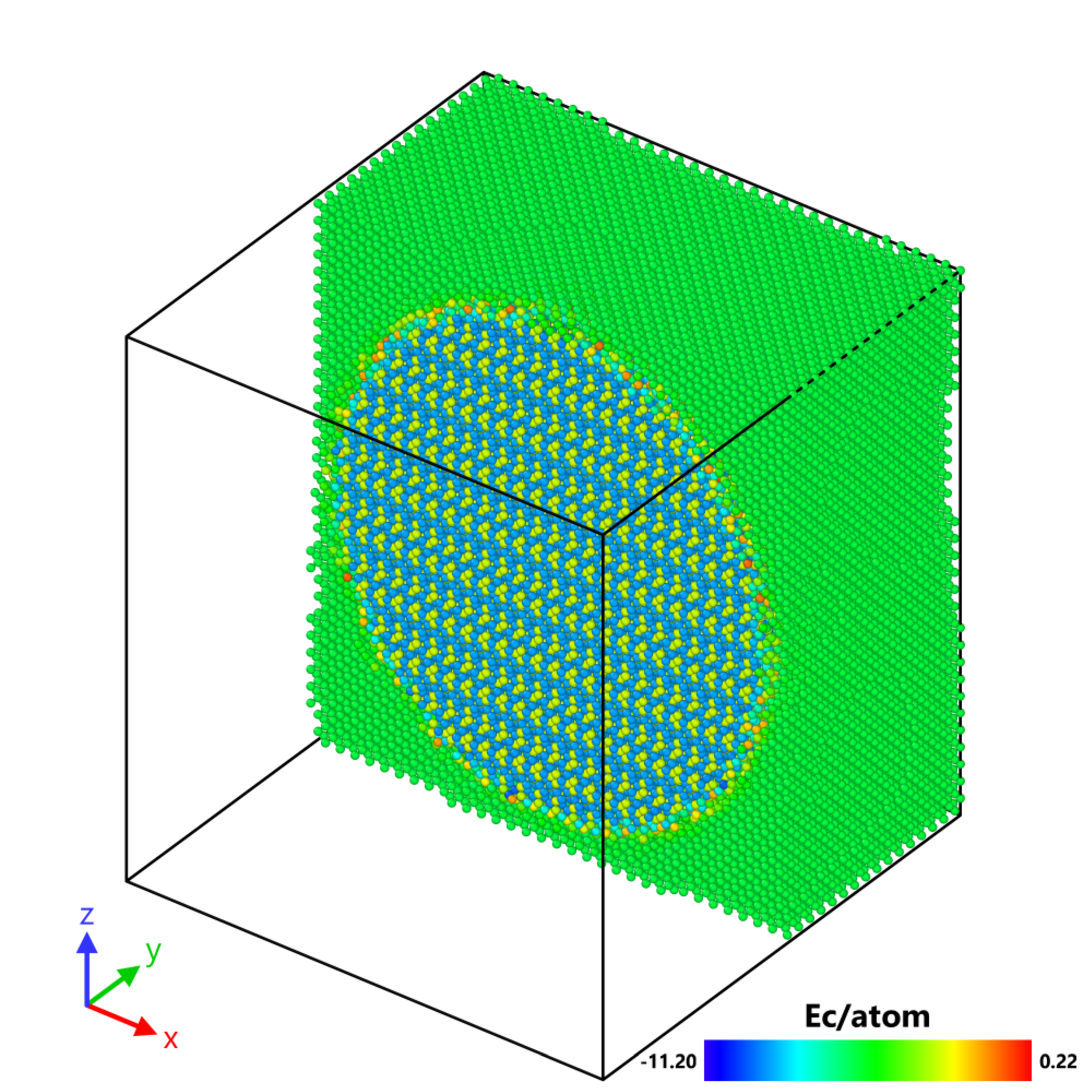} &
			\includegraphics[width=0.310\linewidth]{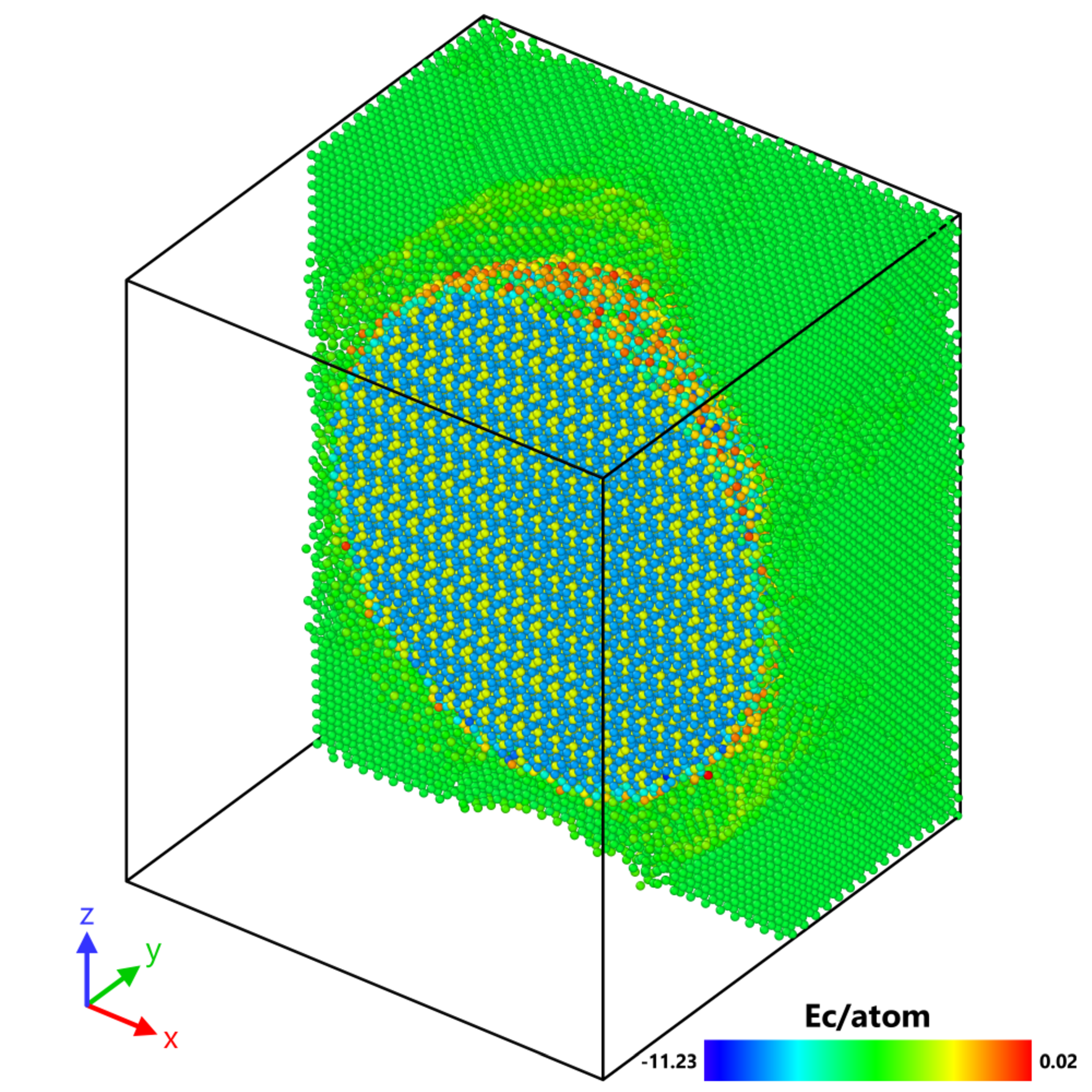}\\
			{\footnotesize (d)} & {\footnotesize (e)} & {\footnotesize (f)} \\
			\includegraphics[width=0.310\linewidth]{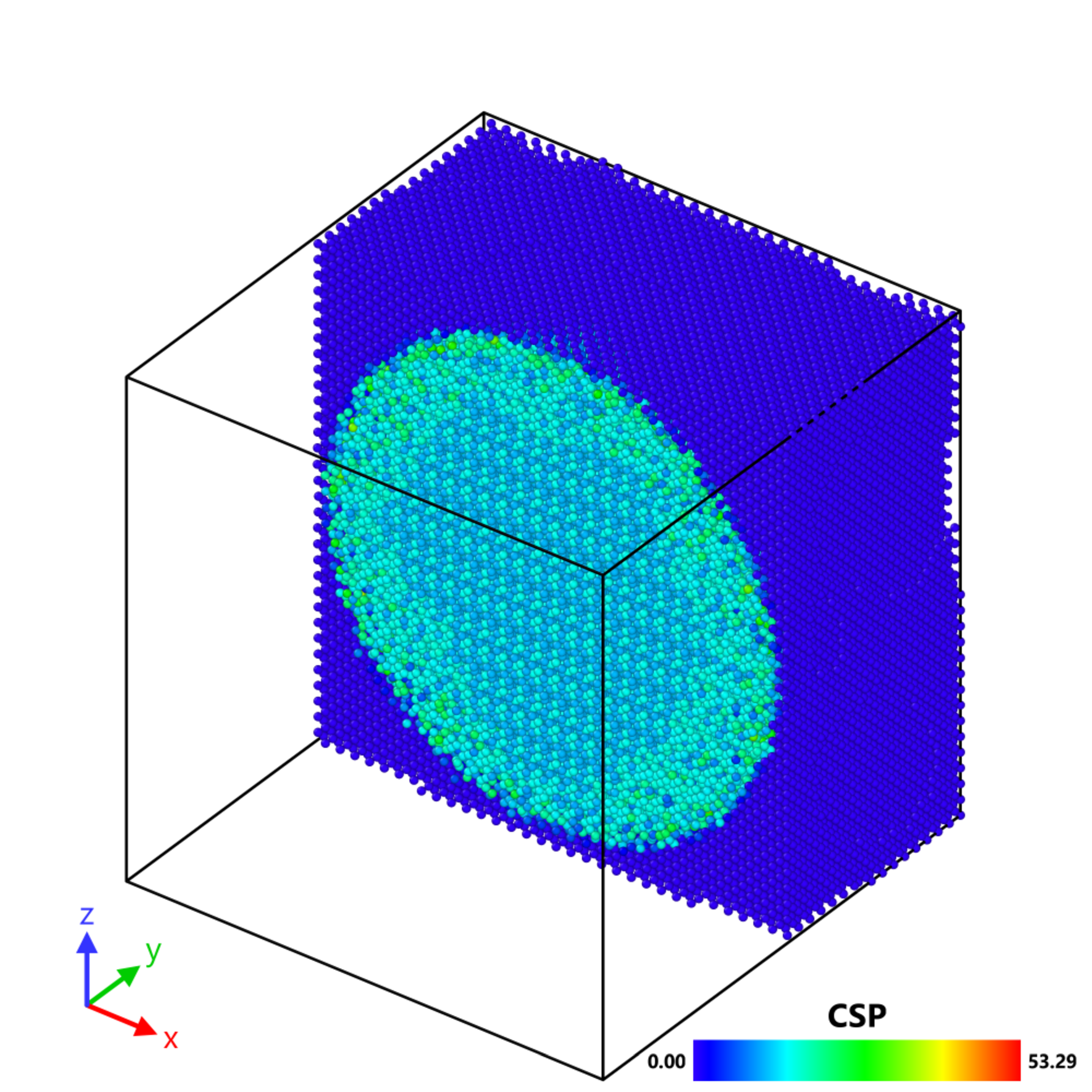} &
			\includegraphics[width=0.310\linewidth]{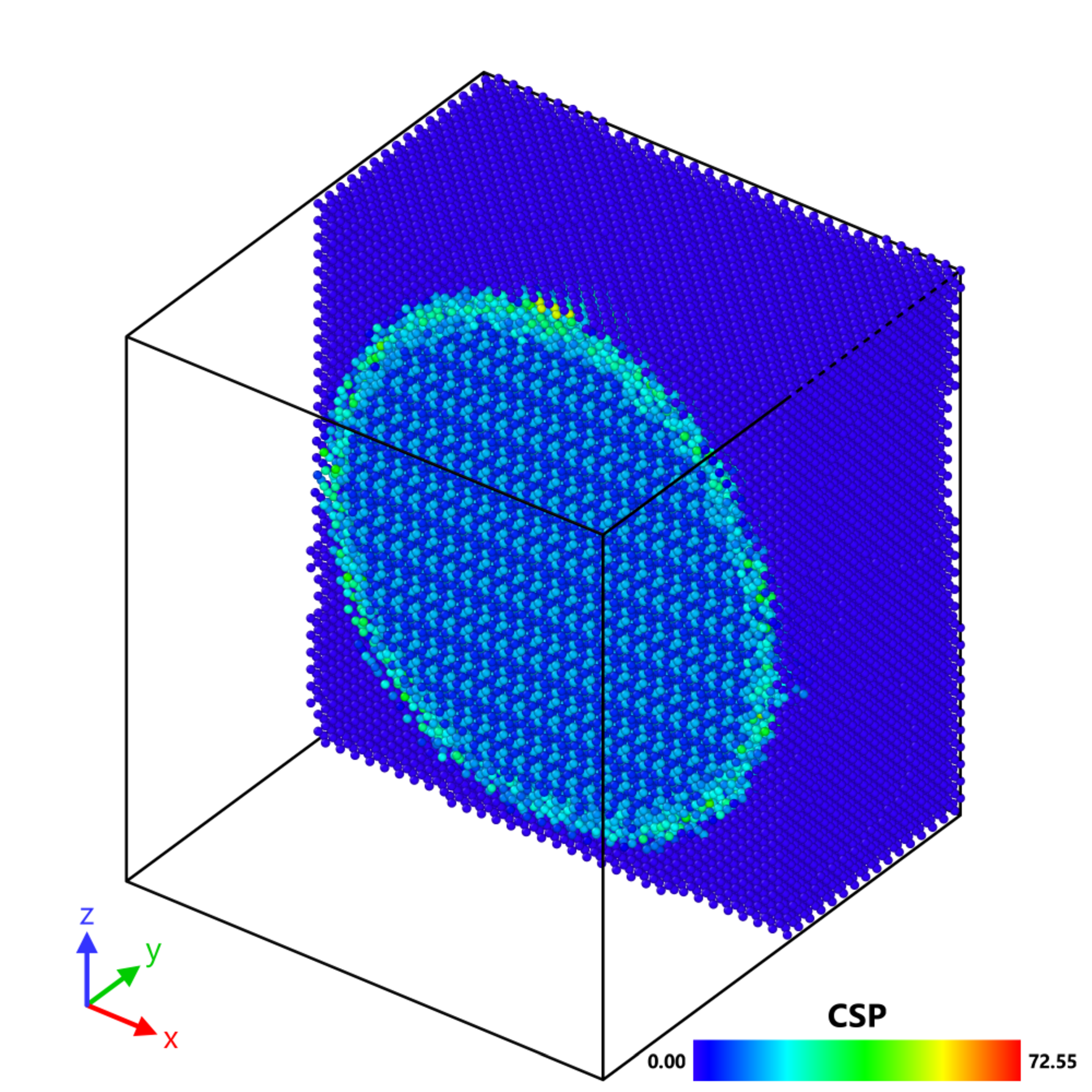} &
			\includegraphics[width=0.310\linewidth]{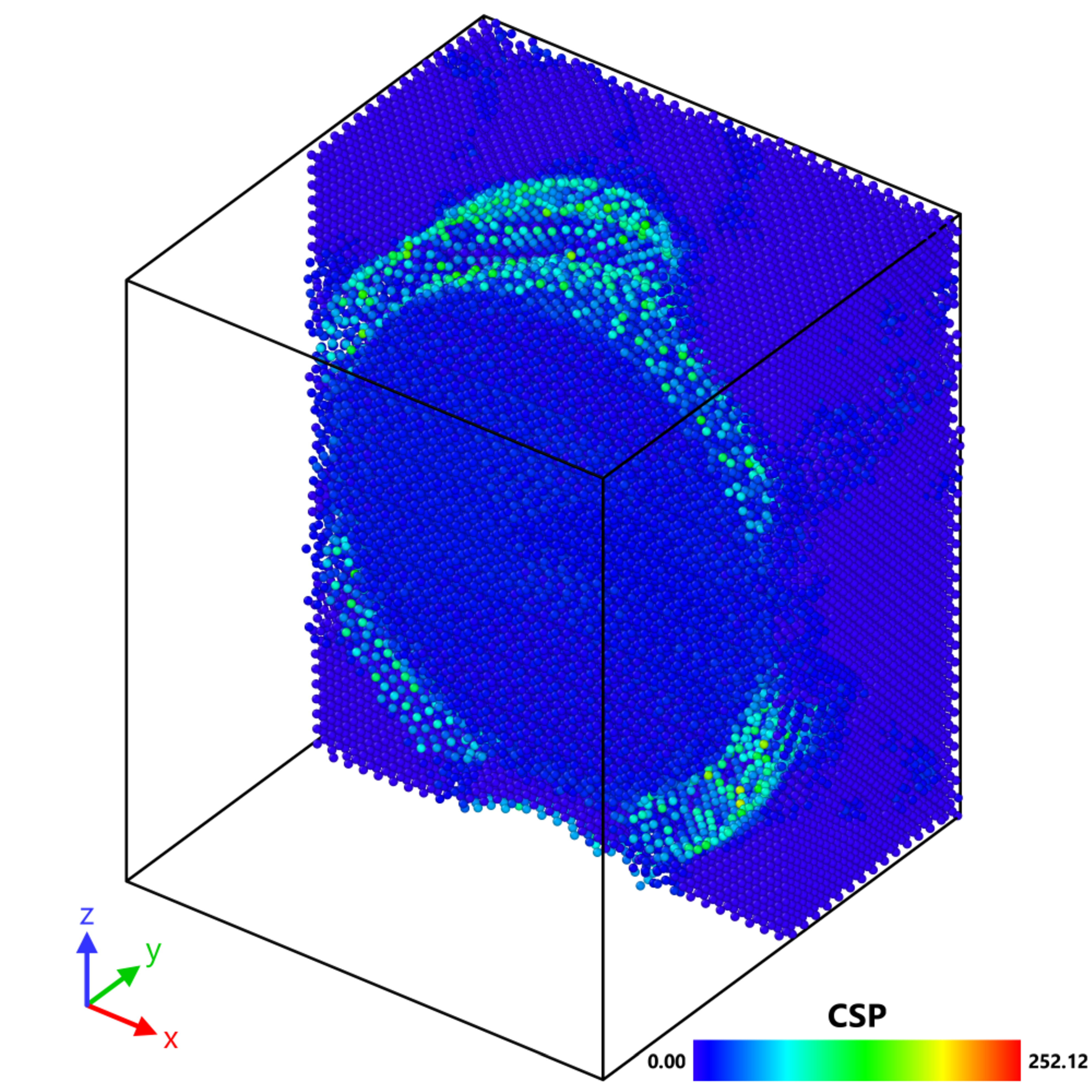}
		\end{tabular}
		\caption{\MM{Cross section of {Al-40\%Al$_2$O$_3$\_32} sample (see Tab.~\ref{tab1}) under uniaxial tensile strain in the Z direction:} \comm{{Al-40\%Al$_2$O$_3$\_32\_usz} (tension) cross section} (a) the cohesive energy per atom (E$_c$/atom) for $\varepsilon_{zz}$=0, (b) E$_c$/atom for $\varepsilon_{zz}$=0.08, (c) E$_c$/atom for $\varepsilon_{zz}$=0.19, (d) the centrosymmetry parameter (CSP) for $\varepsilon_{zz}$=0, (e) CSP for $\varepsilon_{zz}$=0.08 and (f) CSP for $\varepsilon_{zz}$=\MM{0.19.} }
		\label{fig:Al40Al2O3usztens}
	\end{figure}

	Figure~\ref{fig:Al40Al2O3usztens} provides atomistic insight into the deformation and damage evolution of the {Al-Al$_2$O$_3$} using two complementary descriptors: cohesive energy per atom (E$_c$/atom) and the centrosymmetry parameter (CSP). Together, they allow distinguishing elastic response, defect nucleation, and progressive interfacial decohesion.
	
	At the initial state ($\varepsilon_{zz}$\MM{=0.00}, Figs. \ref{fig:Al40Al2O3usztens}a and d), the system is energetically stable and structurally ordered. The cohesive energy field is relatively uniform, with a clear contrast between the Al matrix and the Al$_2$O$_3$
	inclusion due to their different bonding nature. The CSP values remain close to zero throughout most of the domain, confirming a defect-free crystalline configuration. Slightly elevated CSP values may appear at the interface, indicating intrinsic lattice mismatch and local atomic distortion already present in the relaxed configuration.

	At moderate strain ($\varepsilon_{zz}$=0.08, Figs.~\ref{fig:Al40Al2O3usztens}b and e), the onset of inelastic processes becomes evident. The cohesive energy distribution shows localized increases at the matrix–particle interface, forming a non-uniform energy halo around the reinforcement. This indicates stress concentration and progressive weakening of atomic bonds in this region. Simultaneously, CSP maps reveal the nucleation of lattice defects. This confirms that the interface acts as the dominant source of plasticity initiation due to elastic mismatch and stress concentration.
	
	At high strain ($\varepsilon_{zz}$=0.19, Figs.~\ref{fig:Al40Al2O3usztens}c and f), damage localization becomes pronounced. The cohesive energy field exhibits pronounced heterogeneity, with clear regions of elevated energy forming a quasi-continuous band around the particle, indicative of interfacial decohesion or microcrack formation. The CSP distribution shows a significant increase in defective atoms, forming a ring-like damaged zone surrounding the reinforcement. This pattern suggests that failure is governed by interfacial debonding followed by localized plastic deformation in the adjacent matrix. The damage is not uniformly distributed but concentrated along preferred paths, reflecting the evolution of a percolating fracture process zone.

\begin{figure}[!h] 
		\centering
		\begin{tabular}{ccc}
			{\footnotesize (a)} & {\footnotesize (b)} & {\footnotesize (c)} \\
			\includegraphics[width=0.310\linewidth]{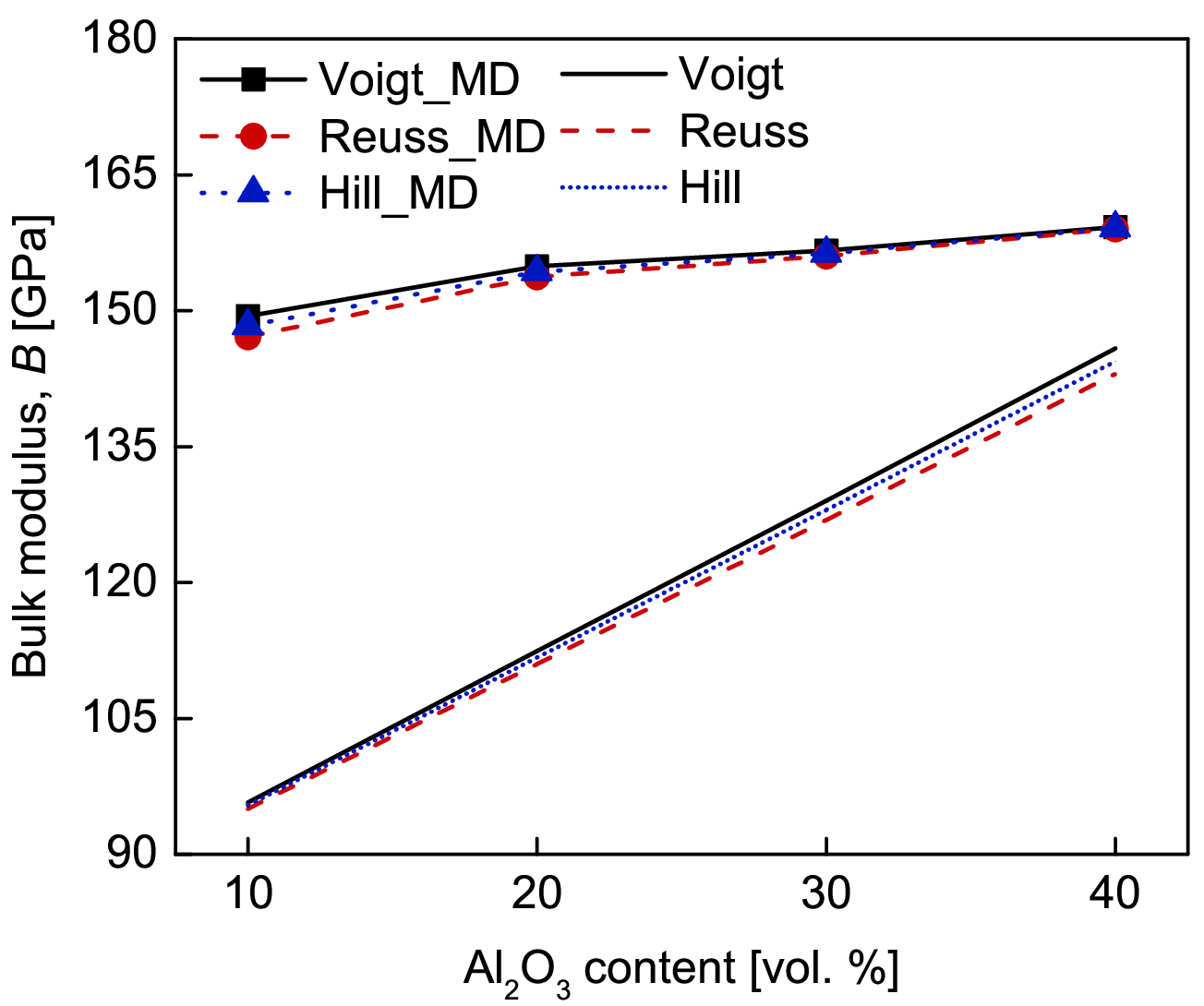} &
			\includegraphics[width=0.310\linewidth]{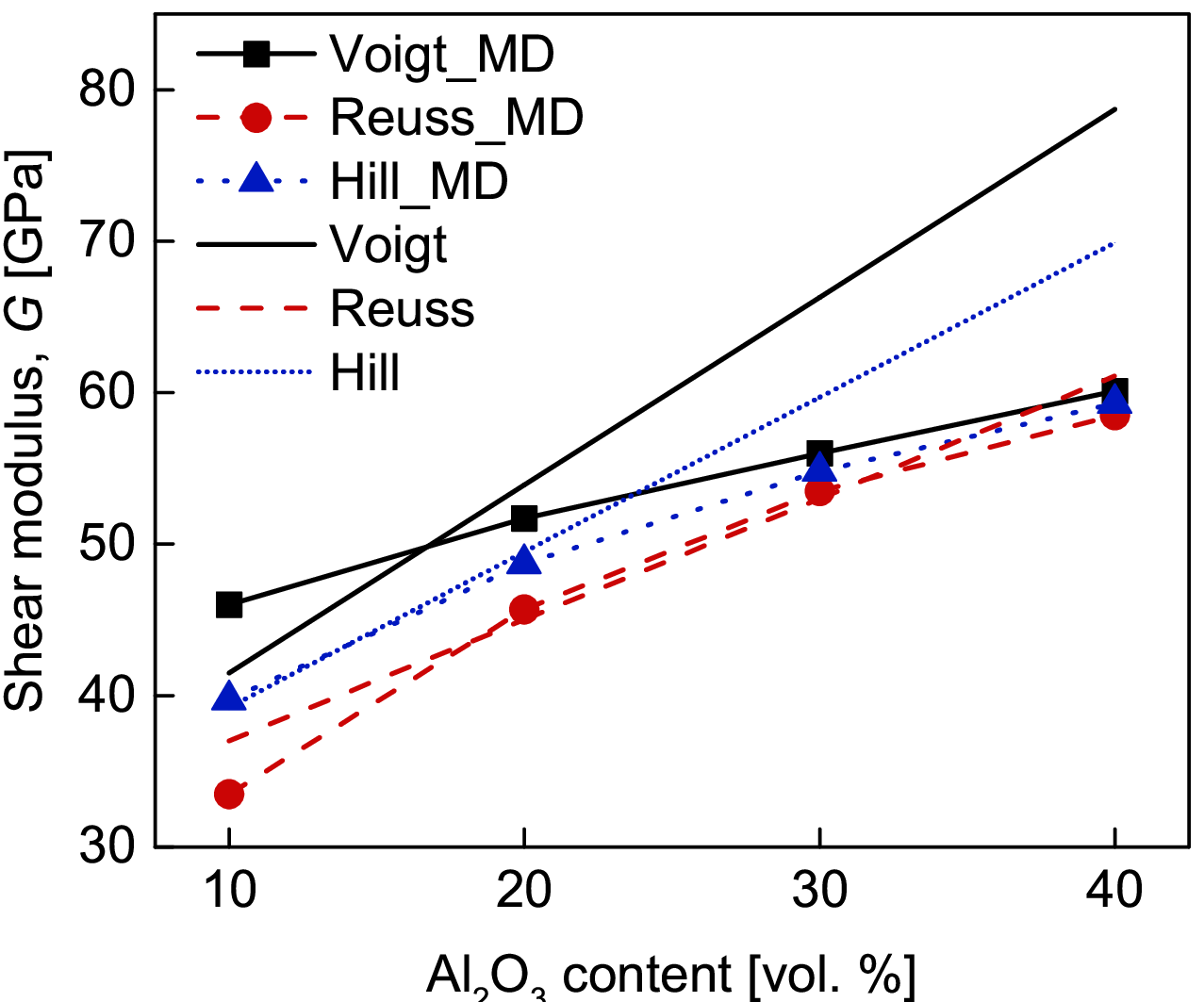} &
			\includegraphics[width=0.310\linewidth]{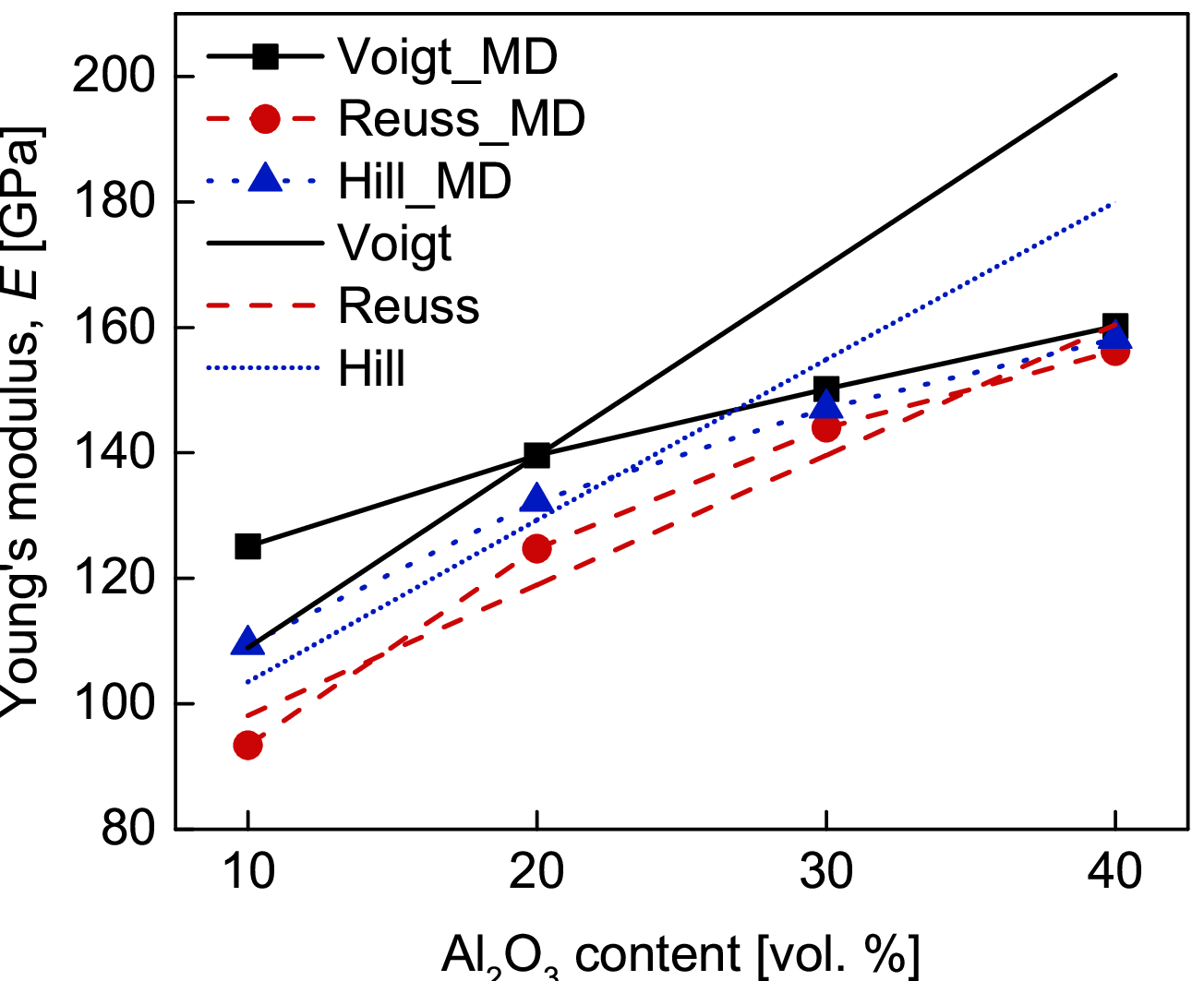}\\
		\end{tabular}
		\caption{Summary of determination of bulk (a), shear (b) and Young's modulus (c)  from MD and analytical estimates.}
		\label{fig:summaryBGE}
	\end{figure}

	Furthermore, the molecular dynamics results of elastic constants have been compared with analytical results. Table~\ref{tab:content_elastic_vm} and Fig.~\ref{fig:summaryBGE} show Voigt, Reuss and Hill bounds for {Al-Al$_2$O$_3$} unit-cell models with various ceramic vol. content together with analogous isotropized stiffness tensors from MD simulations. First, both approaches predict a monotonic increase in elastic stiffness as the Al$_2$O$_3$ volume fraction increases. This is expected due to the ceramic phase's significantly higher stiffness relative to the aluminum matrix. The Young’s modulus obtained from MD (Hill average) increases from 77.6\,GPa for pure Al to 158.2\,GPa at 40 vol.\% and reaches 333.2\,GPa for fully ceramic material, confirming the dominant stiffening effect of Al$_2$O$_3$.
	
	\begin{table}[!h]
		\centering
		\caption{Results of determination of elastic constants of {Al-Al$_2$O$_3$} unit-cell models with various ceramic vol. content using molecular dynamics simulations and analytical estimations based on Voigt, Reuss and Hill bounds.}\vspace{3mm}
		\label{tab:content_elastic_vm}
		\footnotesize
		\setlength{\tabcolsep}{3pt}
		\renewcommand{\arraystretch}{0.8}
		
		\begin{tabular}{c ccccccccc}
			\toprule
			Al$_2$O$_3$  & $B_{Voigt}$  & $G_{Voigt}$ & $E_{Voigt}$ &  $B_{Reuss}$  & $G_{Reuss}$ & $E_{Reuss}$ & $B_{Hill}$  & $G_{Hill}$ & $E_{Hill}$ \\
			(vol.\%) & (GPa) & (GPa) & (GPa)  & (GPa) & (GPa) & (GPa)  & (GPa) & (GPa) & (GPa)\\
			\midrule
			\multicolumn{10}{c}{\textbf{Atomistic results}}\\
			\midrule
			0  & 79.1  & 29.2 & 77.9 & 79.1 &  28.9 & 77.3 & 79.1 & 29.0 & 77.6 \\
			\hdashline 
			10 & 149.4 & 46.0 & 125.1 & 147.1& 33.5 & 93.4 & 148.3 & 39.7 & 109.4\\
			20 & 154.9 & 51.7 & 139.6 & 153.7& 45.7 &124.7 & 154.3 & 48.7 & 132.2\\
			30 & 156.6 & 56.0 & 150.2 & 156.0& 53.5 & 144.0& 156.3 & 54.8 & 147.1\\
			40 & 159.2 & 60.1 & 160.2 & 159.0& 58.5 & 156.2& 159.1 & 59.3 & 158.2\\
			\hdashline
			100& 245.7 & 153.0& 380.0 & 238.6& 109.3& 284.4&242.2 & 131.1& 333.2\\
			\midrule
			\multicolumn{10}{c}{\textbf{Analytical estimates}}\\
			\midrule
			10 &  95.7 & 41.5 & 108.9 & 95.0 & 37.0 & 98.1 & 95.4 & 39.2 & 103.5\\
			20 & 112.4 & 53.9 & 139.5 & 111.0& 45.0 &118.9 &111.7 & 49.5 & 129.3\\
			30 & 129.0 & 66.3 & 169.8 & 126.9& 53.0 & 139.6&128.0 & 59.7 & 154.9\\
			40 & 145.8 & 78.7 & 200.2 & 143.0& 61.1 & 160.4&144.4 & 69.9 & 180.6\\
			\bottomrule
		\end{tabular}					
	\end{table}
	
	Quantitatively, the MD-derived bulk modulus exhibits a systematic deviation from analytical predictions. The largest discrepancy is observed at the lowest ceramic volume fraction. This behavior indicates that atomistic-scale effects—both microstructural (e.g., non-ideal interfacial behavior and lattice anisotropy) and mechanical (e.g., strain localization and constraint effects)—intrinsic to MD simulations are not fully captured by classical homogenization frameworks. As ceramic content increases, the discrepancy between numerical (MD) and analytical estimates progressively diminishes. For the {Al-40\%Al$_2$O$_3$} unit-cell model, the difference falls below 20\,GPa, irrespective of the selected bound. Notably, all three analytical models for the bulk modulus yield nearly identical predictions.
	
	In contrast, the evolution of the shear modulus reveals a markedly different trend. A pronounced separation is observed between the Voigt (upper), Reuss (lower), and Hill (average) bounds in both MD and analytical results. While good agreement between MD and analytical predictions is maintained at lower ceramic contents (10–20 vol.\%), significant deviations emerge at higher volume fractions. Among the models considered, the Reuss estimate consistently remains closest to the MD results across the entire range of compositions.
	
	The trends observed for the Young's modulus closely follow those of the shear modulus, reflecting their functional dependence. Whereas the analytical bounds exhibit approximately linear evolution with increasing ceramic content—leading to a widening spread between the Voigt and Reuss limits—the MD results display a distinctly nonlinear response, with comparatively reduced divergence between the corresponding bounds.
	
\begin{figure}[!h] 
		\centering
		\begin{tabular}{cc}
			{\footnotesize (a)} & {\footnotesize (b)} \\
			\includegraphics[width=0.47\linewidth]{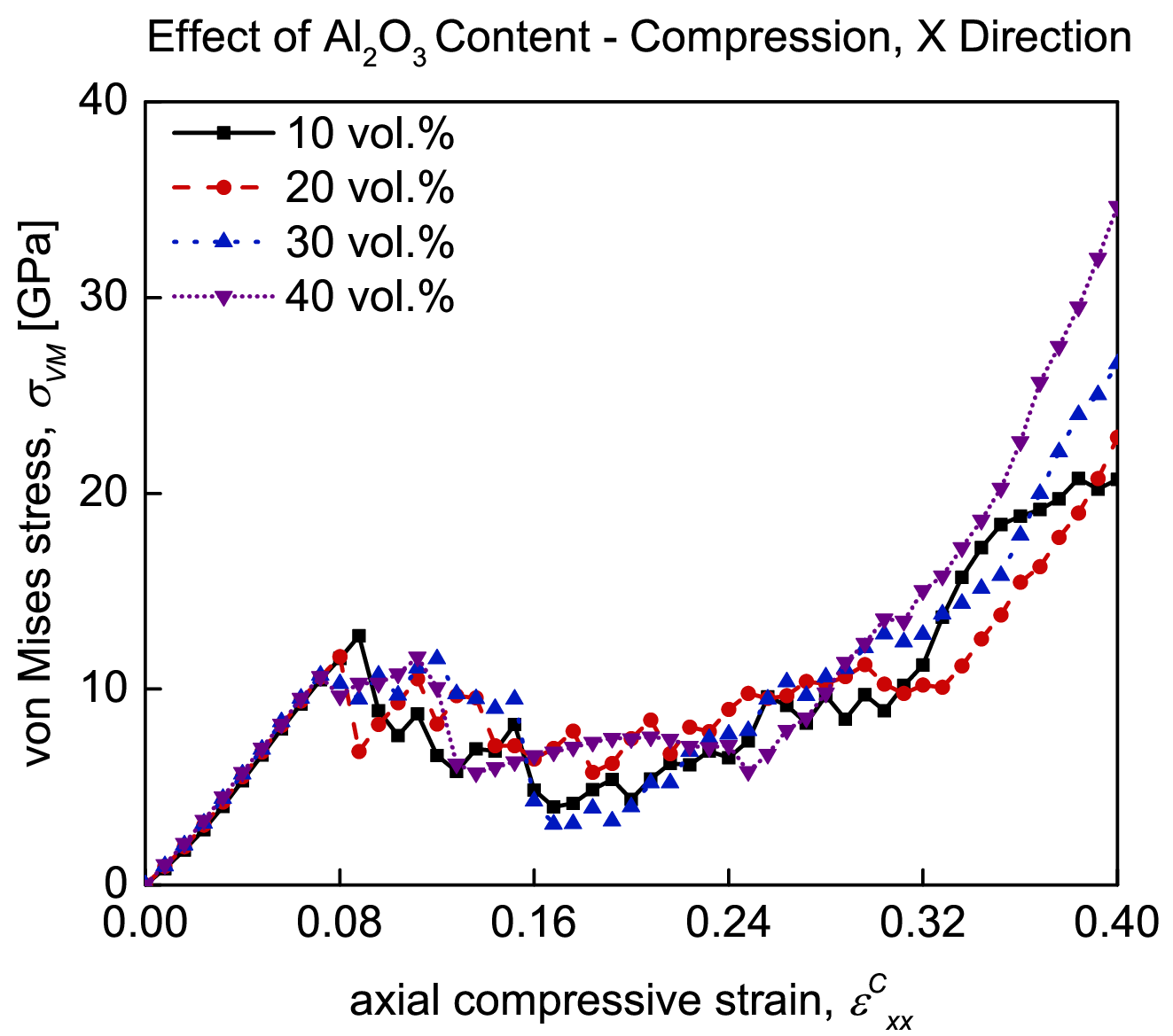} &
			\includegraphics[width=0.47\linewidth]{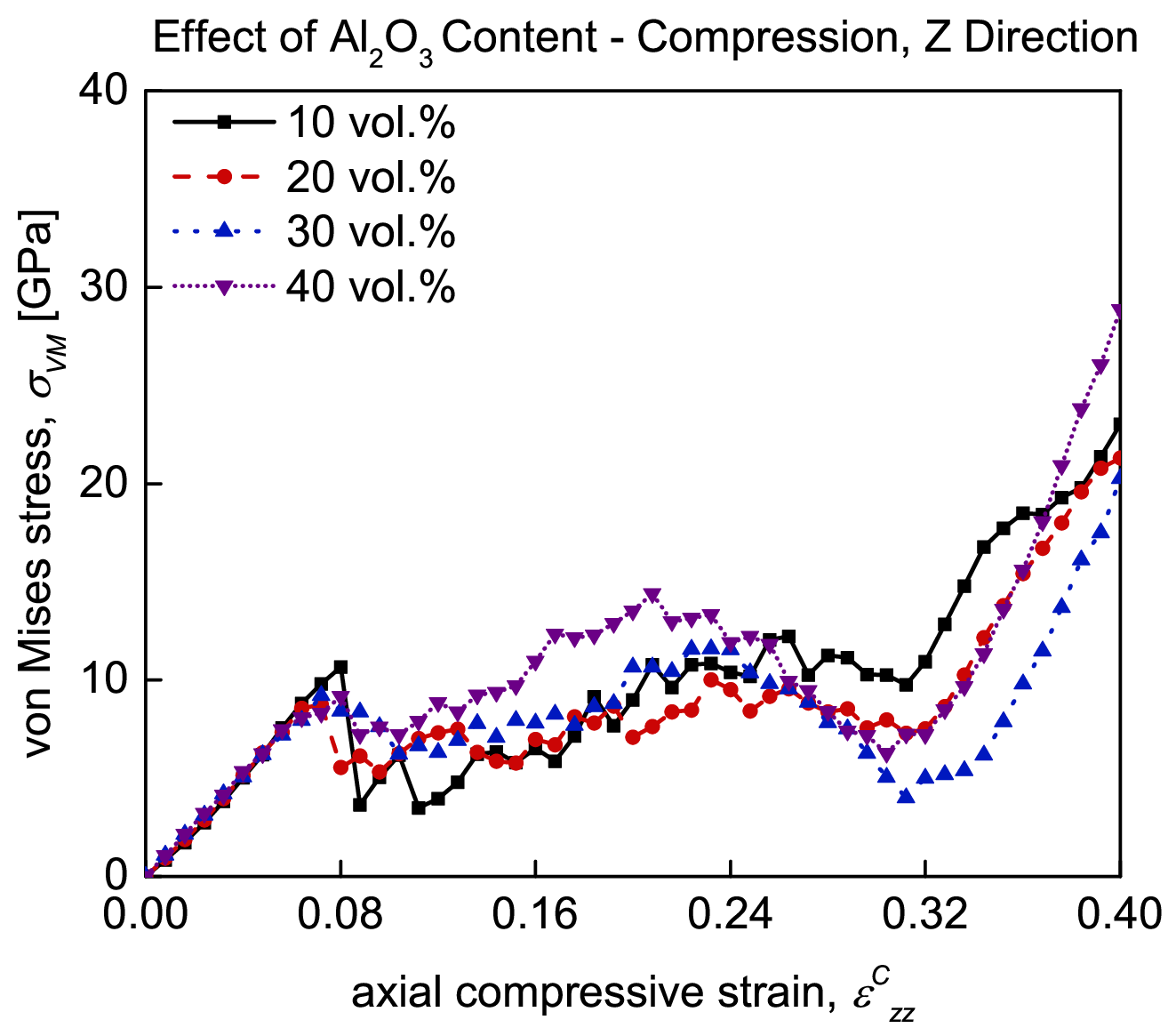}\\
		\end{tabular}
		\caption{The stress-strain evolution of {Al-Al$_2$O$_3$} unit-cell samples with different ceramic content subjected to uniaxial compression in: (a) X and (b) Z direction.}
		\label{fig:Compression_DiffContAlumina_Mono}
	\end{figure}

	Figure~\ref{fig:Compression_DiffContAlumina_Mono} shows the axial compressive stress–strain behavior of Al$_2$O$_3$ unit-cell models with ceramic contents ranging from 10\% to 40\%, loaded along the X (a) and Z (b) directions. As shown in both figures, the unit-cell models exhibit a strongly nonlinear stress–strain response, characterized by a gradual transition from an initially quasi-linear regime ($\epsilon \lesssim$ 0.1) through a gradually softening regime and, finally, to pronounced strain hardening at higher compressive strains. At low strains ($\epsilon \lesssim$ 0.10), differences between ceramic contents are modest, suggesting matrix-dominated deformation. 
	
	An increase in the Al$_2$O$_3$ volume fraction systematically increases the compressive stress at the last part of the strain range. At intermediate and large strains, the curves progressively diverge, with 40\% Al$_2$O$_3$ exhibiting the highest stresses, followed by 30\%, 20\%, and 10\%. At $\epsilon \approx$ 0.40, the maximum compressive stress increases from ~25\,GPa for 10\% Al$_2$O$_3$ to ~30–35\,GPa for 40\% Al$_2$O$_3$, highlighting the dominant role of the ceramic phase in carrying load under large compressive deformation. 
	
\begin{figure}[!h] 
		\centering
		\begin{tabular}{ccc}
			{\footnotesize (a)} & {\footnotesize (b)} & {\footnotesize (c)} \\
			\includegraphics[width=0.310\linewidth]{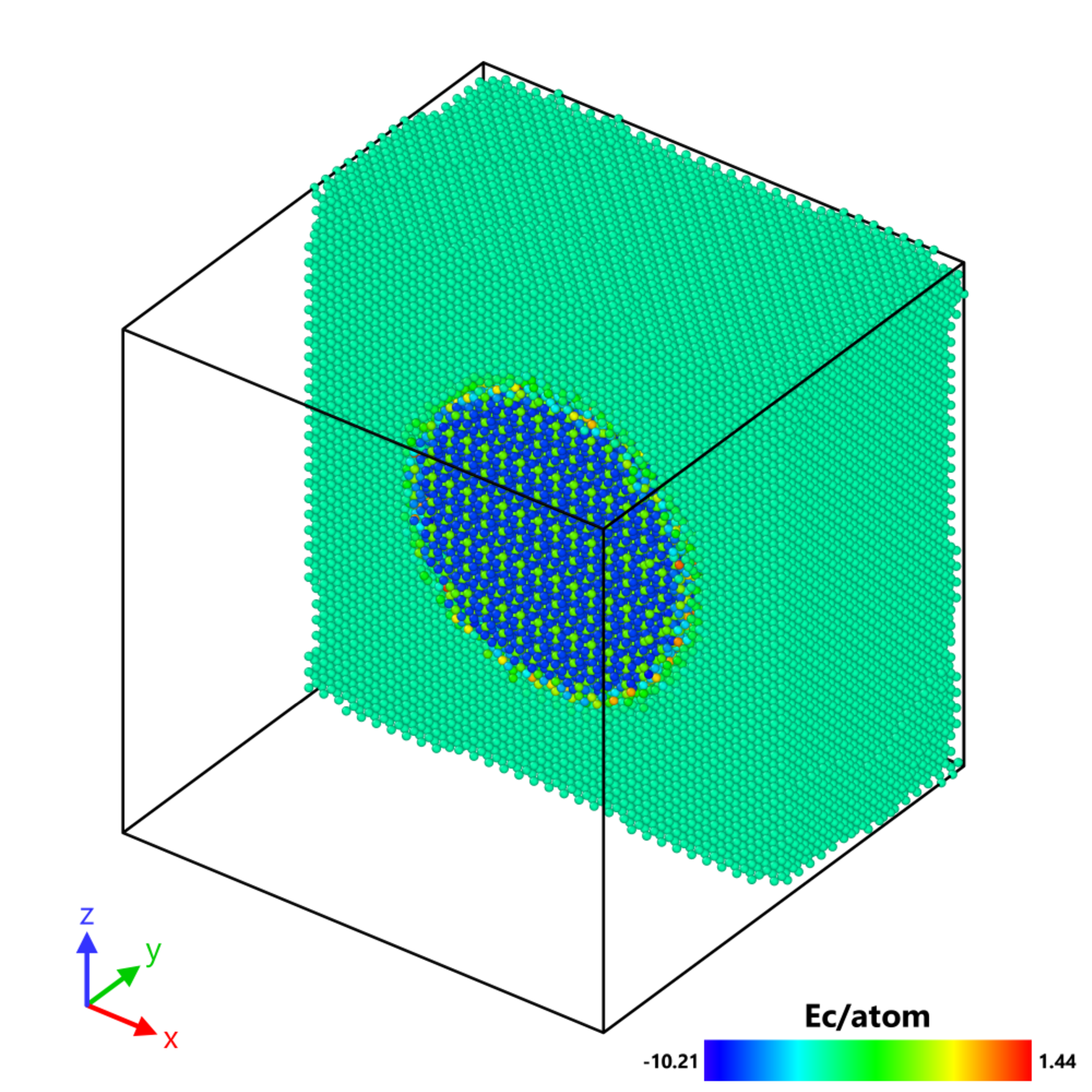} &
			\includegraphics[width=0.310\linewidth]{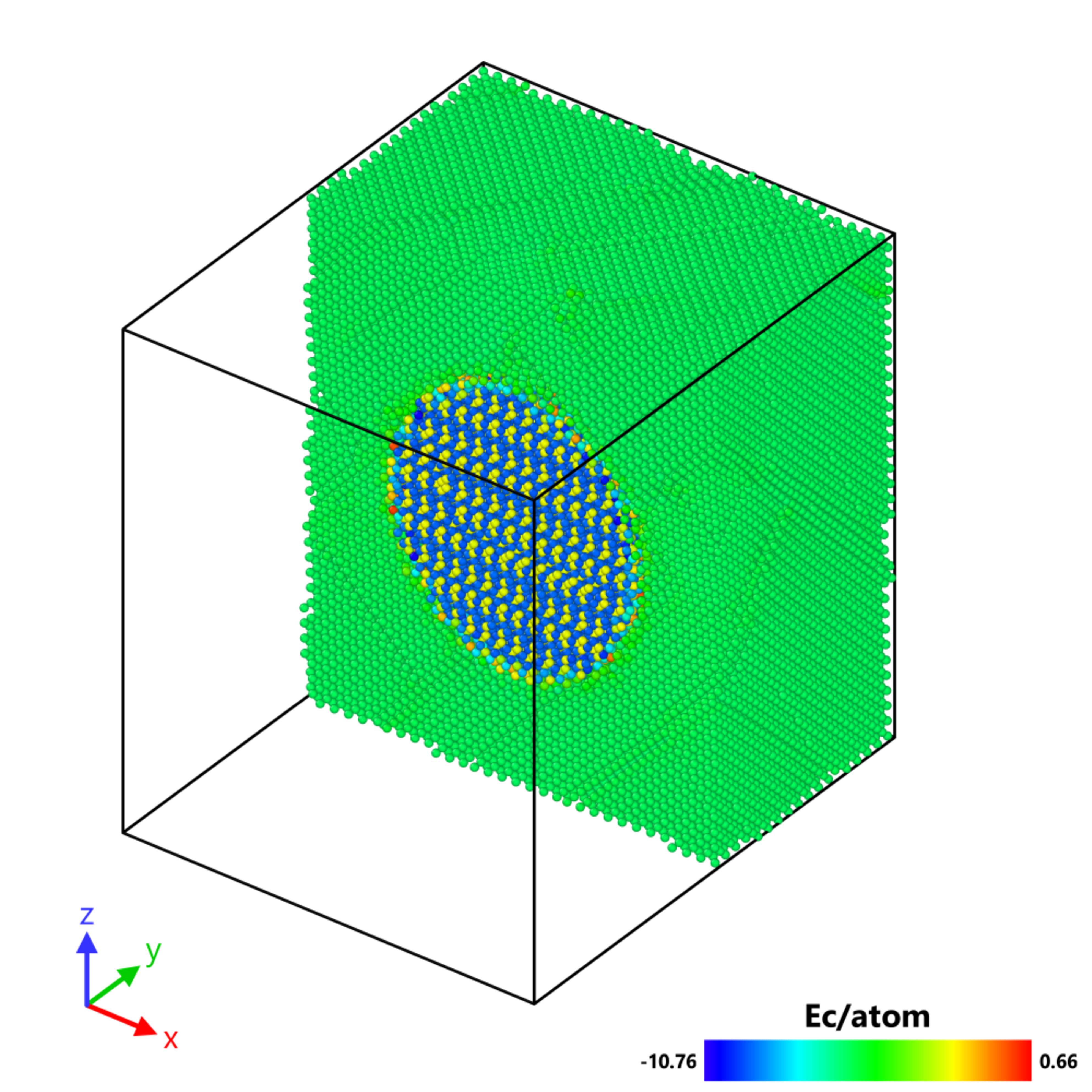} &
			\includegraphics[width=0.310\linewidth]{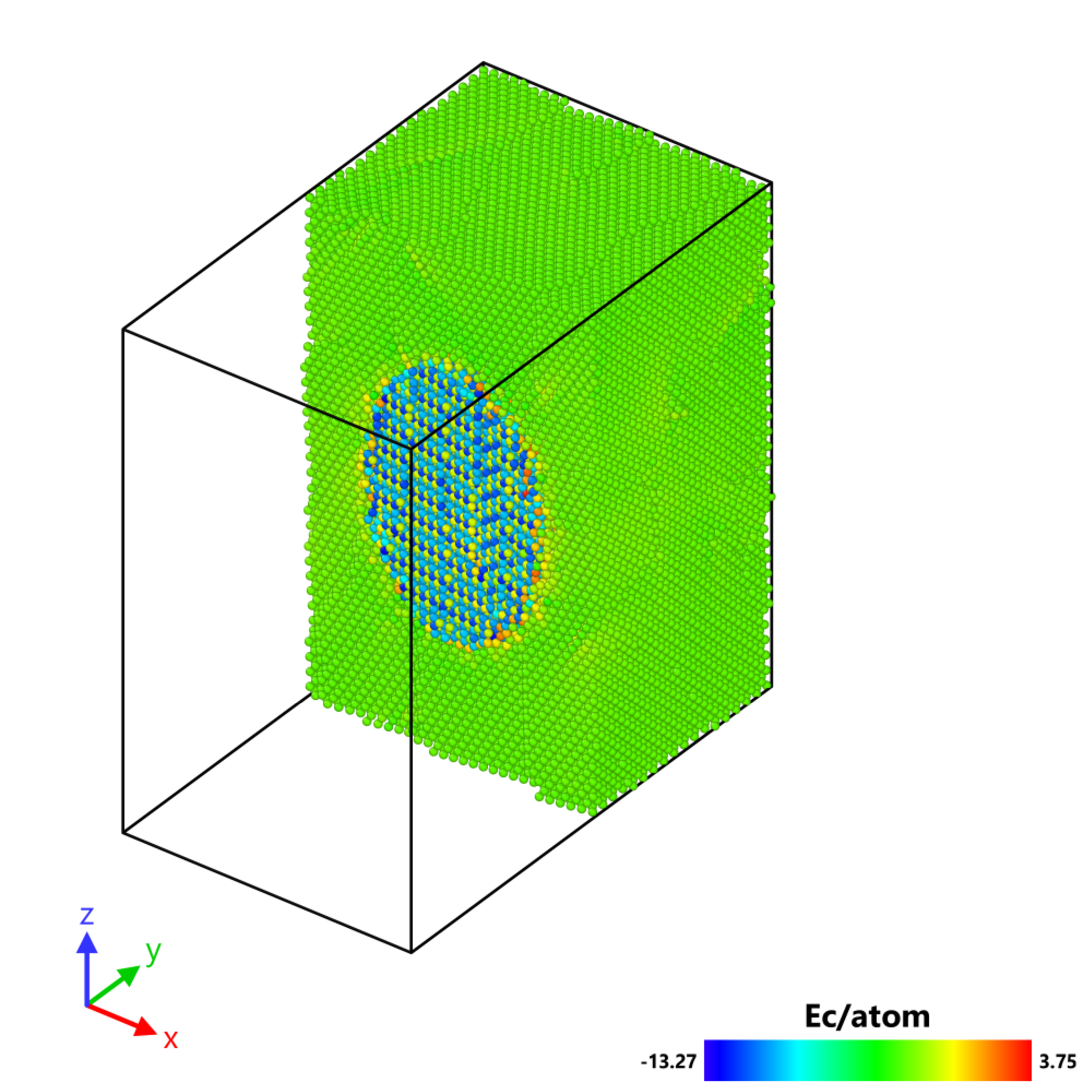}\\
			{\footnotesize (d)} & {\footnotesize (e)} & {\footnotesize (f)} \\
			\includegraphics[width=0.310\linewidth]{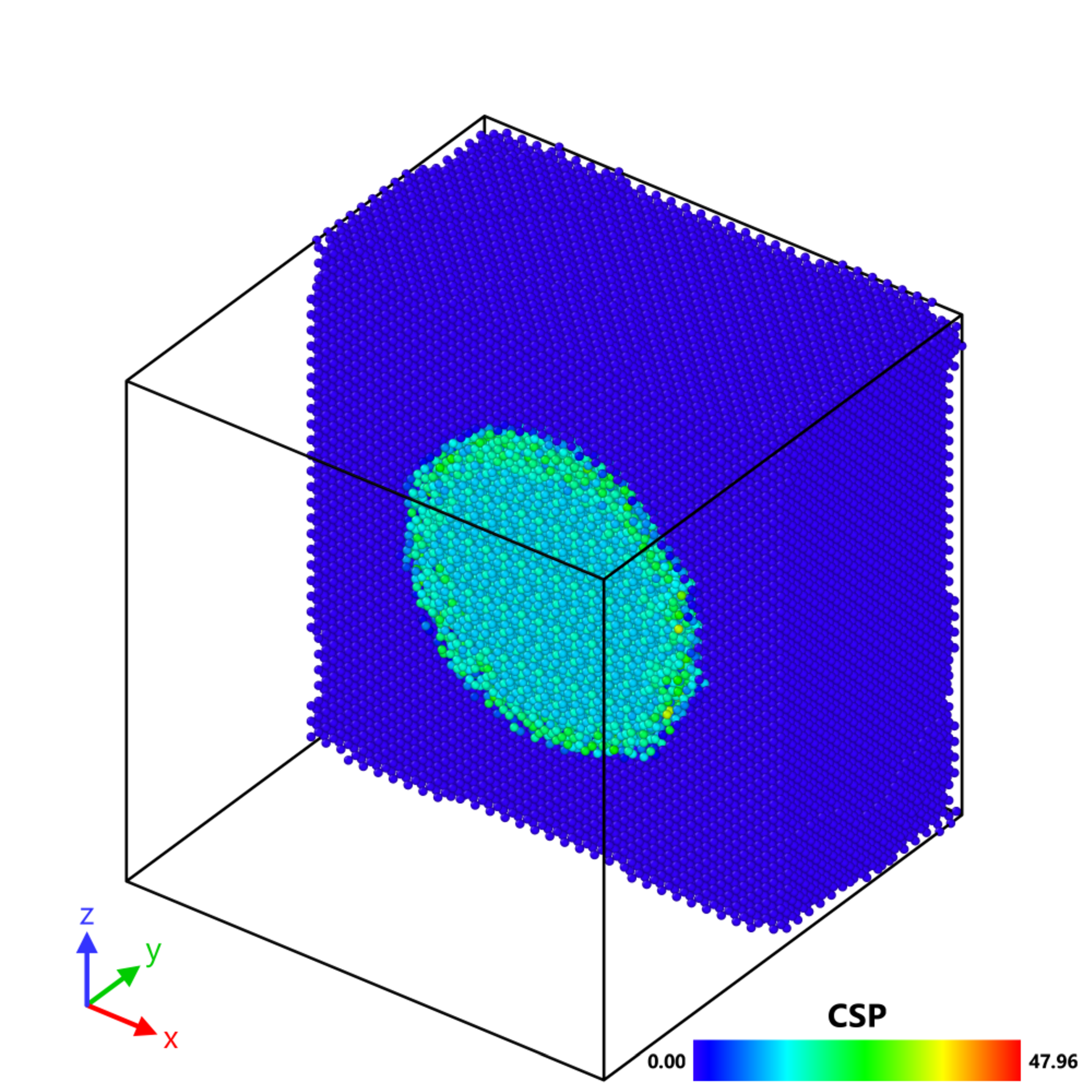} &
			\includegraphics[width=0.310\linewidth]{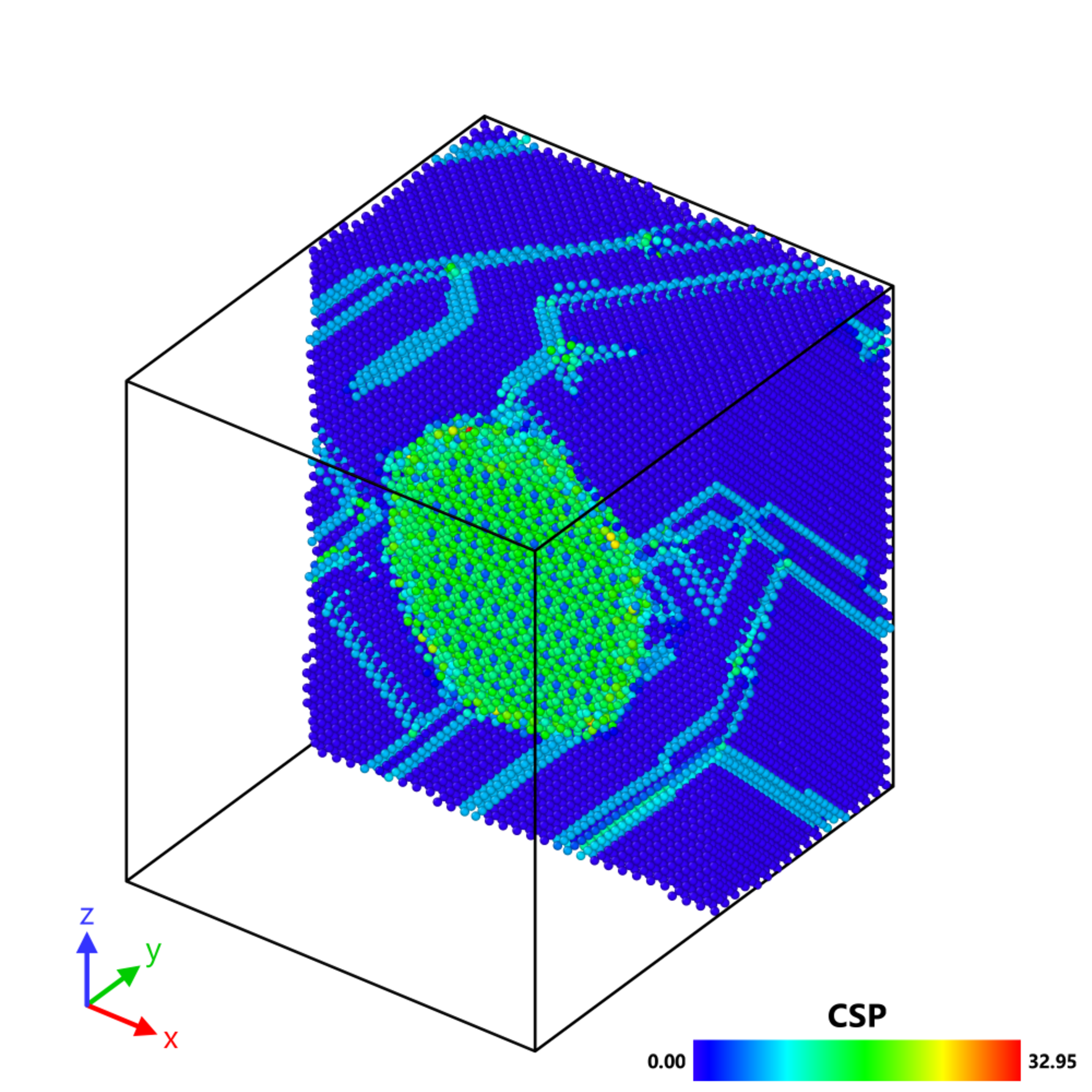} &
			\includegraphics[width=0.310\linewidth]{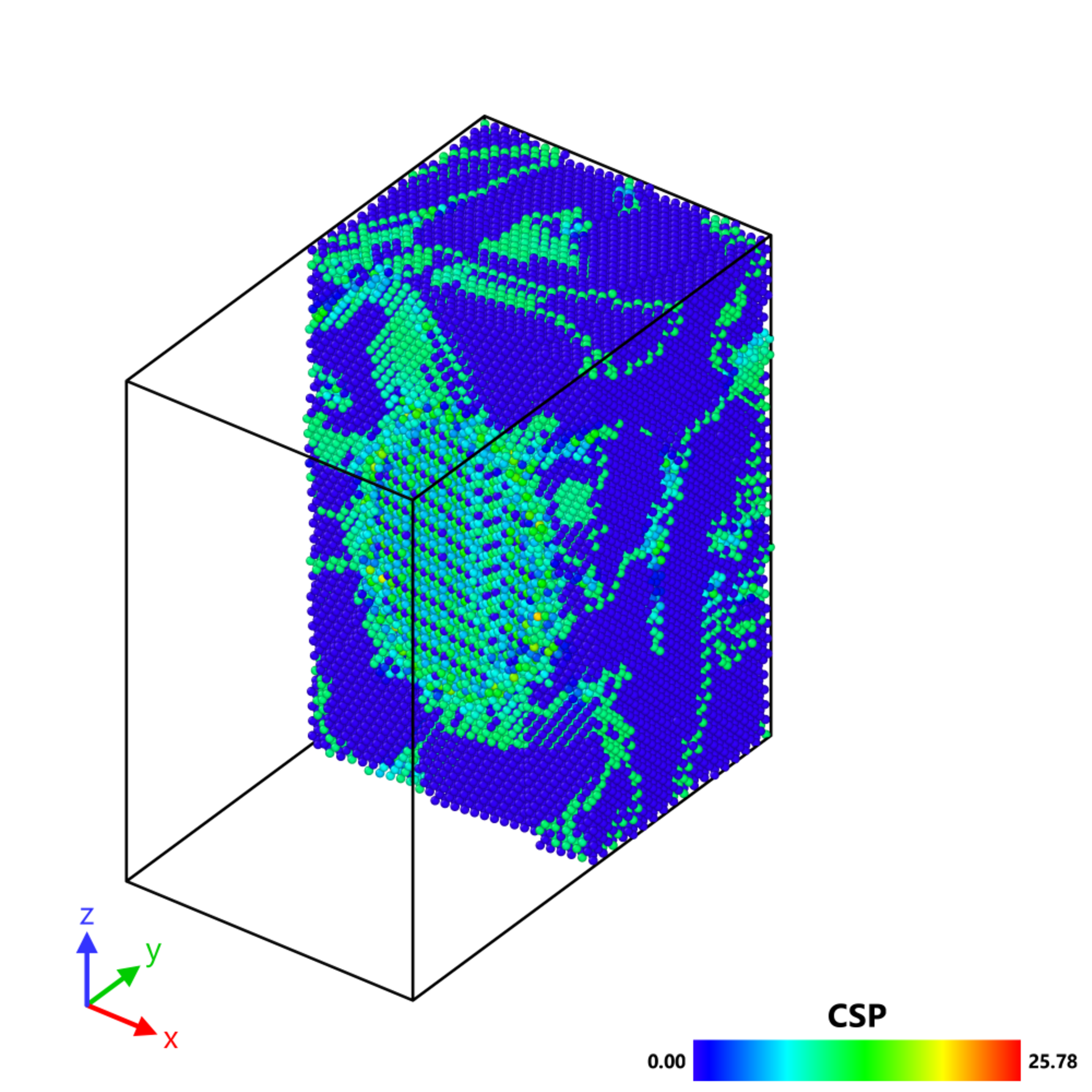}
		\end{tabular}
		\caption{\MM{Cross section of {Al-10\%Al$_2$O$_3$\_32} sample (see Tab.~\ref{tab1}) under uniaxial compressive strain in the X direction:} \comm{{Al-10\%Al$_2$O$_3$\_32\_usx} (compression) cross section} (a) the cohesive energy per atom (E$_c$/atom) for $\varepsilon_{xx}$=0, (b) E$_c$/atom for $\varepsilon_{xx}$=-0.14, (c) E$_c$/atom for $\varepsilon_{xx}$=-0.40, (d) the centrosymmetry parameter (CSP) for $\varepsilon_{xx}$=0, (e) CSP for $\varepsilon_{xx}$=-0.14 and (f) CSP for $\varepsilon_{xx}$=-0.40. }
		\label{fig:Al10Al2O3128gusxcomp}
	\end{figure}

	The occurrence of stress drops and the enhanced strain hardening with increasing ceramic content suggest that plastic flow of the Al matrix is progressively constrained by the rigid Al$_2$O$_3$ particle. Such an effect can be observed in Fig.~\ref{fig:Al10Al2O3128gusxcomp} analyzed through cohesive energy per atom and the centrosymmetry parameter. Compared to the tensile case, the deformation mechanisms are markedly different, with plasticity dominating over decohesion. Here, the deformation mechanism is governed by the activation of plasticity in the Al matrix. The cohesive energy map shows only modest localization, without the pronounced interfacial energy accumulation observed in tension. \SN{In contrast, CSP reveals the formation of extended regions of lattice disorder
that propagate through the Al matrix and around the inclusion. These defect-rich
regions are not confined to the interface but extend into the matrix, indicating
that plastic deformation is distributed rather than localized.}

\SN{At high compression, the microstructure becomes strongly disordered. The CSP
field shows an extended network of defect-rich regions spanning a large part of
the matrix volume, indicating intense and distributed plastic deformation. The
cohesive-energy distribution remains relatively smooth compared with the tensile
case, with no clear evidence of pronounced interfacial decohesion. Instead, the
observed energy variations are associated primarily with local lattice distortion
during deformation.}
	
	\subsection{RVE polycrystal model}
	
\begin{figure}[!h] 
		\centering
		\begin{tabular}{cc}
			{\footnotesize (a)} & {\footnotesize (b)} \\
			\includegraphics[width=0.47\linewidth]{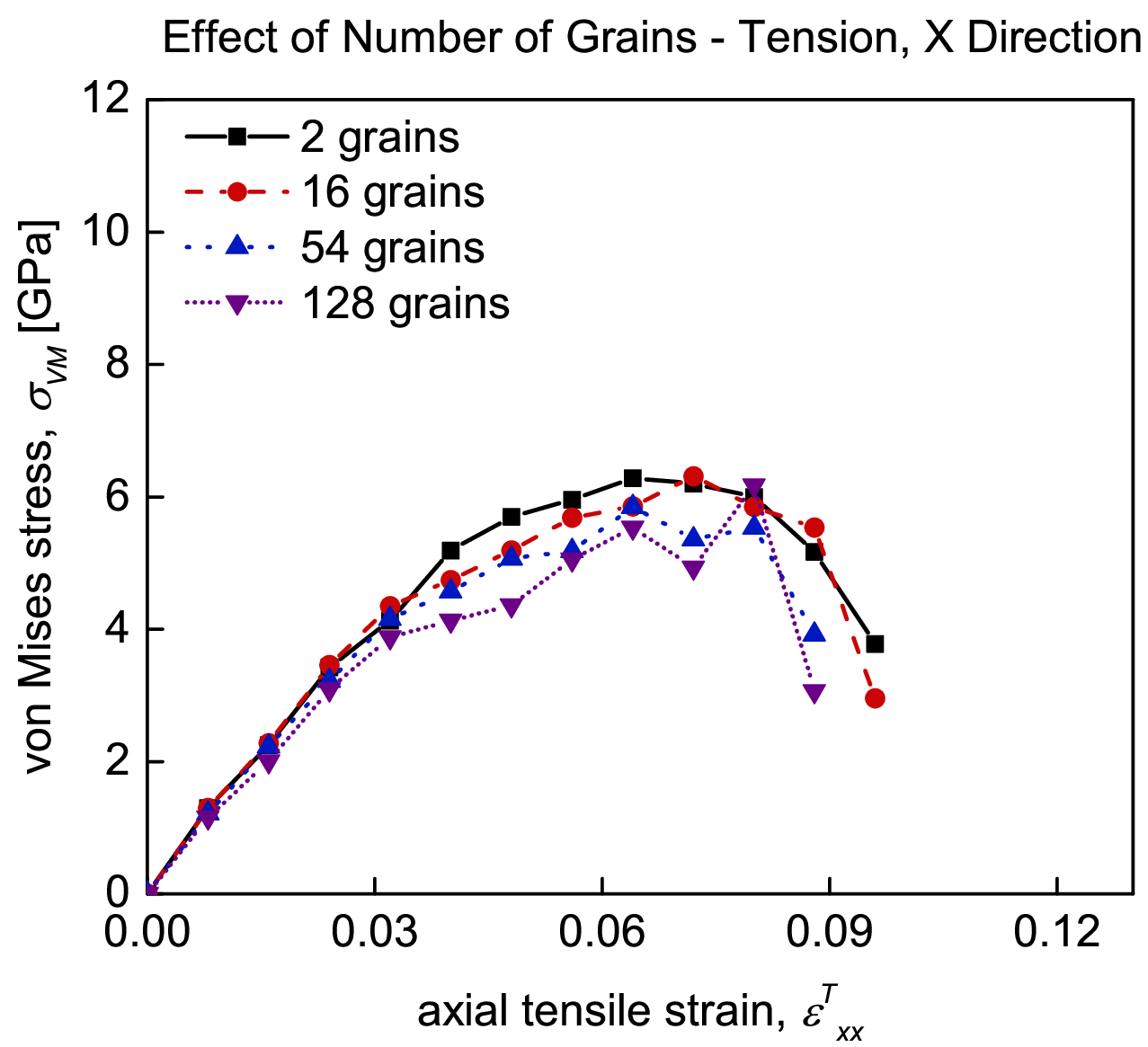} & \hspace{-0.5cm}
			\includegraphics[width=0.47\linewidth]{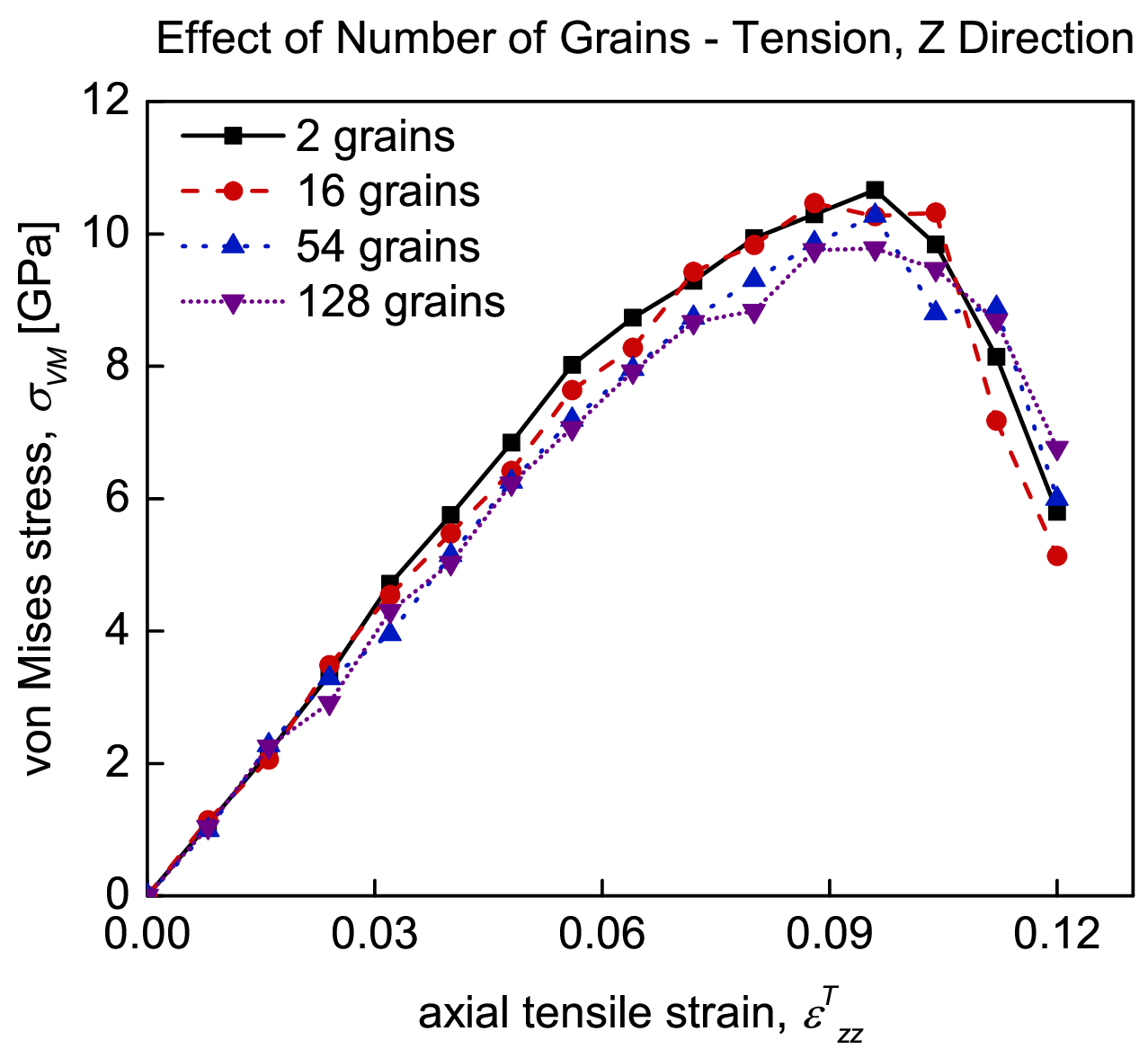}\\
		\end{tabular}
		\caption{The stress-strain evolution of {Al-Al$_2$O$_3$} polycrystal samples consists of different numbers of grains subjected to uniaxial tension in: (a) X and (b) Z directions. }
		\label{fig:Tensile_DiffGrainNumber_Poly}
	\end{figure}

	Figure~\ref{fig:Tensile_DiffGrainNumber_Poly} illustrates the von Mises tensile stress vs. axial strain response of Al–Al$_2$O$_3$ polycrystalline RVE models with a fixed ceramic particle size (32) and ceramic volume fraction (40\%), while varying the number of grains (2, 16, 54, and 128). Results are shown for uniaxial tension applied along the X (a) and Z (b) directions. In both loading directions, all RVEs exhibit nearly identical initial elastic slopes, indicating that the effective elastic response is essentially independent of the number of grains. On the one hand, this confirms that, at small strains, the macroscopic stiffness is governed primarily by phase properties and volume fractions rather than crystallographic discretization. 
	
	On the other hand, the \MM{isotropized MD stress tensor} results \MM{obtained} using the Voigt, Reuss and Hill bounds provide minor correction to the statement. The bulk modulus \textbf{$B$}, regardless of the homogenization scheme (Voigt, Reuss, or Hill), shows only minor variation (on the order of a few percent) across all grain configurations. Since volumetric deformation does not strongly depend on grain orientation or intergranular interactions, even small RVEs provide an adequate approximation of \textbf{$B$}.
	
	In contrast, the shear modulus $G$ decreases significantly as the grain number increases. The Hill-averaged shear modulus decreases from 37.8\,GPa for the 2-grain model to 30.4\,GPa for the 128-grain model, corresponding to a reduction of approximately 20\%. This trend can be attributed to the progressive incorporation of grain orientation variability and the increasing density of grain boundaries, which introduce additional compliance under shear loading. In small RVEs, the limited number of grains leads to artificially aligned or favorable load-bearing paths, resulting in an overestimation of stiffness. As the number of grains increases, the microstructure becomes more statistically representative, and the shear response reflects a more realistic distribution of local strain fields.
	
	\begin{table}[h!]
		\centering
		\caption{Effect of number of grains on elastic properties of RVE model of {Al-40\%Al$_2$O$_3$} composite.} \vspace{3mm}
		\label{tab:grains_elastic_vm}
		\small
		\setlength{\tabcolsep}{2pt}
		\renewcommand{\arraystretch}{0.8}
		
		\begin{tabular}{c ccccccccc}
			\toprule
			Grains number   & $B_{Voigt}$  & $G_{Voigt}$ & $E_{Voigt}$ &  $B_{Reuss}$  & $G_{Reuss}$ & $E_{Reuss}$ & $B_{Hill}$  & $G_{Hill}$ & $E_{Hill}$ \\
			& (GPa) & (GPa) & (GPa) & (GPa) & (GPa) & (GPa) & (GPa) & (GPa) & (GPa)\\
			\midrule
			2   & 149.0 & 45.0 & 122.6 & 145.4& 30.6 & 85.8 & 147.2 & 37.8 & 104.5\\
			16  & 152.5 & 42.4 & 116.5 & 150.9& 24.7 & 70.4 & 151.7 & 33.6 & 93.8\\
			54  & 151.9 & 42.2 & 115.9 & 149.8& 24.0 & 68.4 & 150.9 & 33.1 & 92.5\\
			128 & 153.2 & 39.4 & 108.8 & 151.7& 21.4 & 61.3 & 152.5 & 30.4 & 85.4\\
			\bottomrule
		\end{tabular}
	\end{table}
	
	A similar trend is observed for Young's modulus \textbf{$E$}, which decreases from 104.5\,GPa to 85.4\,GPa with increasing grain number. This behavior is directly linked to the reduction in shear modulus, as \textbf{$E$} is strongly coupled to both \textbf{$B$} and \textbf{$G$}. Since \textbf{$B$} remains nearly constant, the decrease in \textbf{$E$} is primarily driven by the softening of the shear response. This highlights the critical role of microstructural heterogeneity in controlling the composite's effective stiffness.
	
	Furthermore, the gap between Voigt and Reuss bounds becomes more pronounced as the number of grains increases. This widening reflects the growing heterogeneity of the strain and stress fields within the RVE. The Voigt model, assuming uniform strain, increasingly overestimates stiffness, while the Reuss model, based on uniform stress, underestimates it. The Hill average remains a reasonable approximation; however, the increasing spread between bounds indicates that analytical homogenization becomes less accurate as microstructural complexity increases.
	
	As can be shown in Fig.~\ref{fig:Tensile_DiffGrainNumber_Poly} and Table~\ref{tab:grains_strength_vm}, the growth of grain number has a limited influence on damage parameters. The peak tensile stress in the X direction is relatively low ($\approx$ 5.5–6.0\,GPa) for all studied cases, showing only minor sensitivity to grain number. RVEs with 2 and 16 grains attain slightly higher peak stresses, whereas 54- and 128-grain models exhibit marginally reduced maxima. The X direction is more prone to early localization of plastic yielding and reduced load-bearing capacity. In contrast, the Z direction displays significantly higher peak stresses ($\approx$ 10–11\,GPa). The anisotropy persists regardless of grain count, indicating that it originates from the underlying microstructural orientation of the ceramic particle and its load-transfer mechanisms rather than from discretization effects.
	
	\begin{table}[!h]
		\centering
		\caption{Summary of ultimate strength ($\sigma^{US}$), ultimate strain ($\varepsilon^{US}$) and failure strain ($\varepsilon^{FS}$) of {Al-Al$_2$O$_3$} RVE models with various numbers of grains evaluated from MD uniaxial tensile tests in X and Z directions. $\varepsilon^{FS}$ was defined as the strain corresponding to the last data point preceding the first stress drop of at least 25\% between two consecutive simulation points.} \vspace{3mm}
		\label{tab:grains_strength_vm}
		\small
		\setlength{\tabcolsep}{12pt}
		\renewcommand{\arraystretch}{0.8}
		
		\begin{tabular}{c ccc ccc}
			\toprule
			& \multicolumn{3}{c}{x direction} & \multicolumn{3}{c}{z direction}\\
			\midrule
			Grains number &$\sigma^{US}$ & $\varepsilon^{US}_{xx}$ & $\varepsilon^{FS}_{xx}$ &
			$\sigma^{US}$ & $\varepsilon^{US}_{zz}$ & $\varepsilon^{FS}_{zz}$ \\
			\midrule
			2 & 6.28 & 0.064 & 0.088 & 10.66 & 0.096 & 0.112 \\
			16 & 6.31 & 0.072 & 0.088 & 10.46 & 0.088 & 0.104 \\
			54 & 5.85 & 0.064 & 0.080 & 10.27 & 0.096 & 0.112 \\
			128 & 6.17 & 0.080 & 0.080 & 9.78 & 0.096 & 0.112 \\
			\bottomrule
		\end{tabular}
		\vspace{1mm}
	\end{table}
	
	Beyond the peak, all curves exhibit pronounced stress softening associated with strain localization and damage accumulation. However, coarse RVEs (2 and 16 grains) show larger fluctuations, and finer RVEs (54–128 grains) display smoother post-peak evolution and more stable residual stress levels. This behavior reflects improved statistical averaging of grain orientations and stress redistribution in RVEs with a higher grain count.
	
\begin{figure}[!h] 
		\centering
		\begin{tabular}{ccc}
			{\footnotesize (a)} & {\footnotesize (b)} & {\footnotesize (c)} \\
			\includegraphics[width=0.310\linewidth]{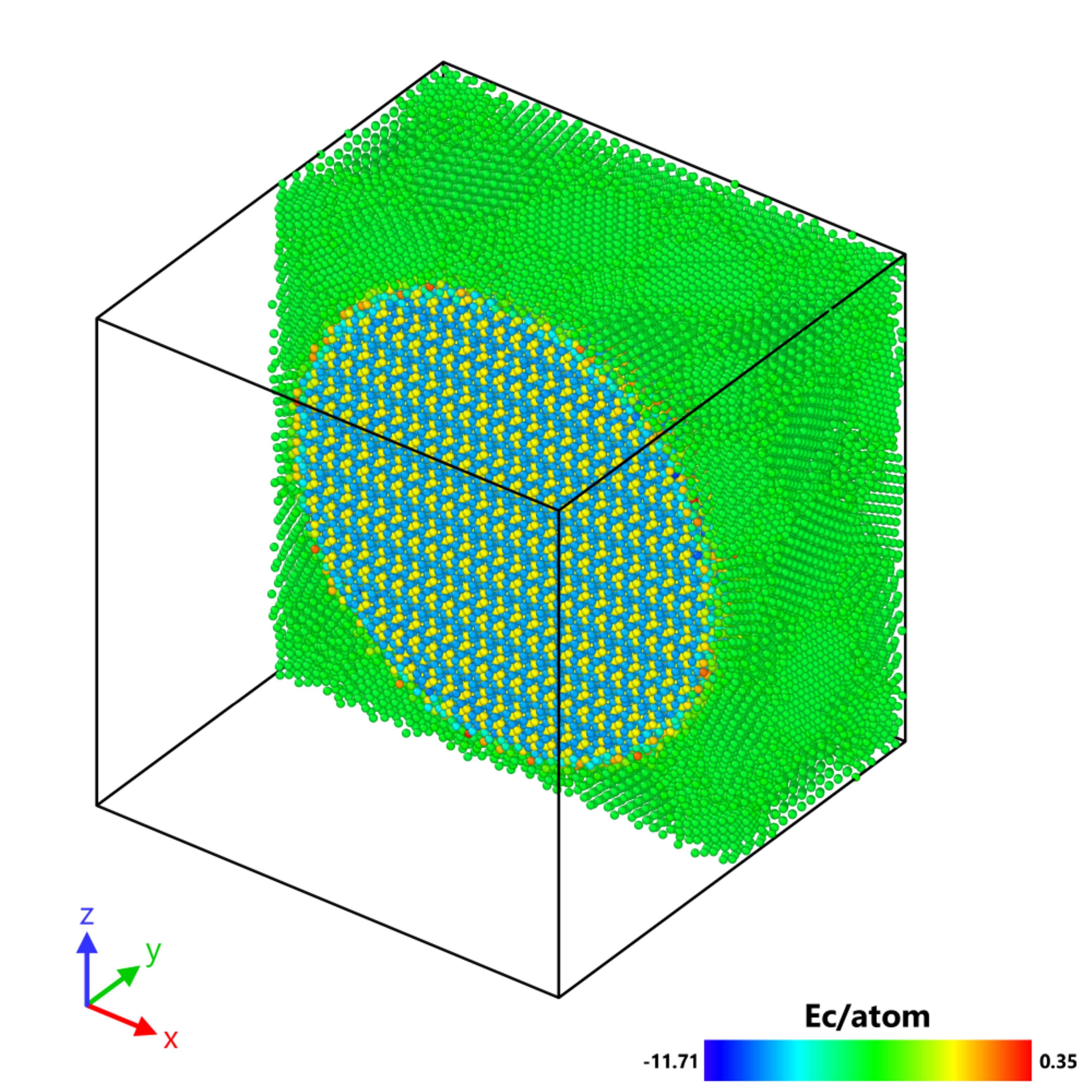} &
			\includegraphics[width=0.310\linewidth]{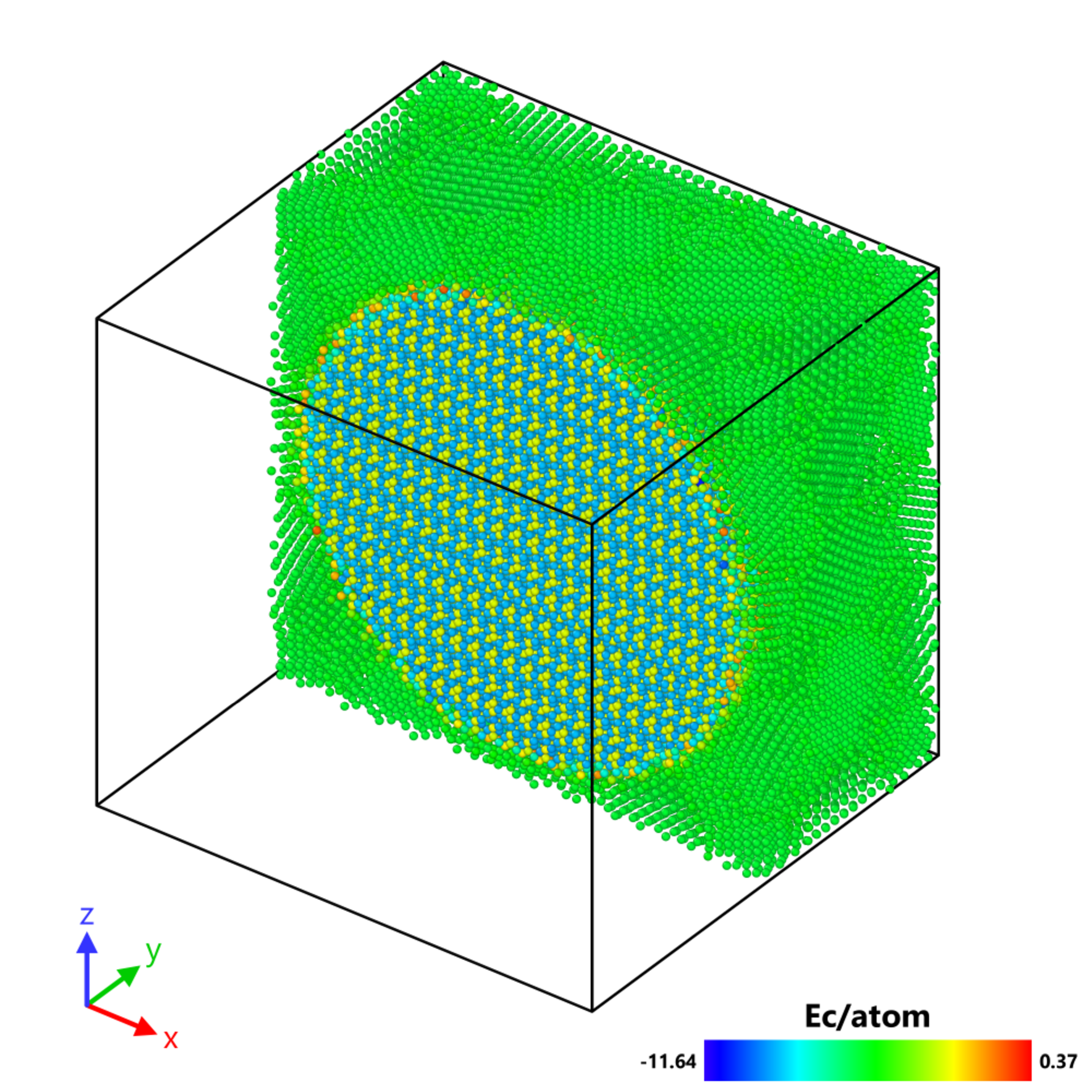} &
			\includegraphics[width=0.310\linewidth]{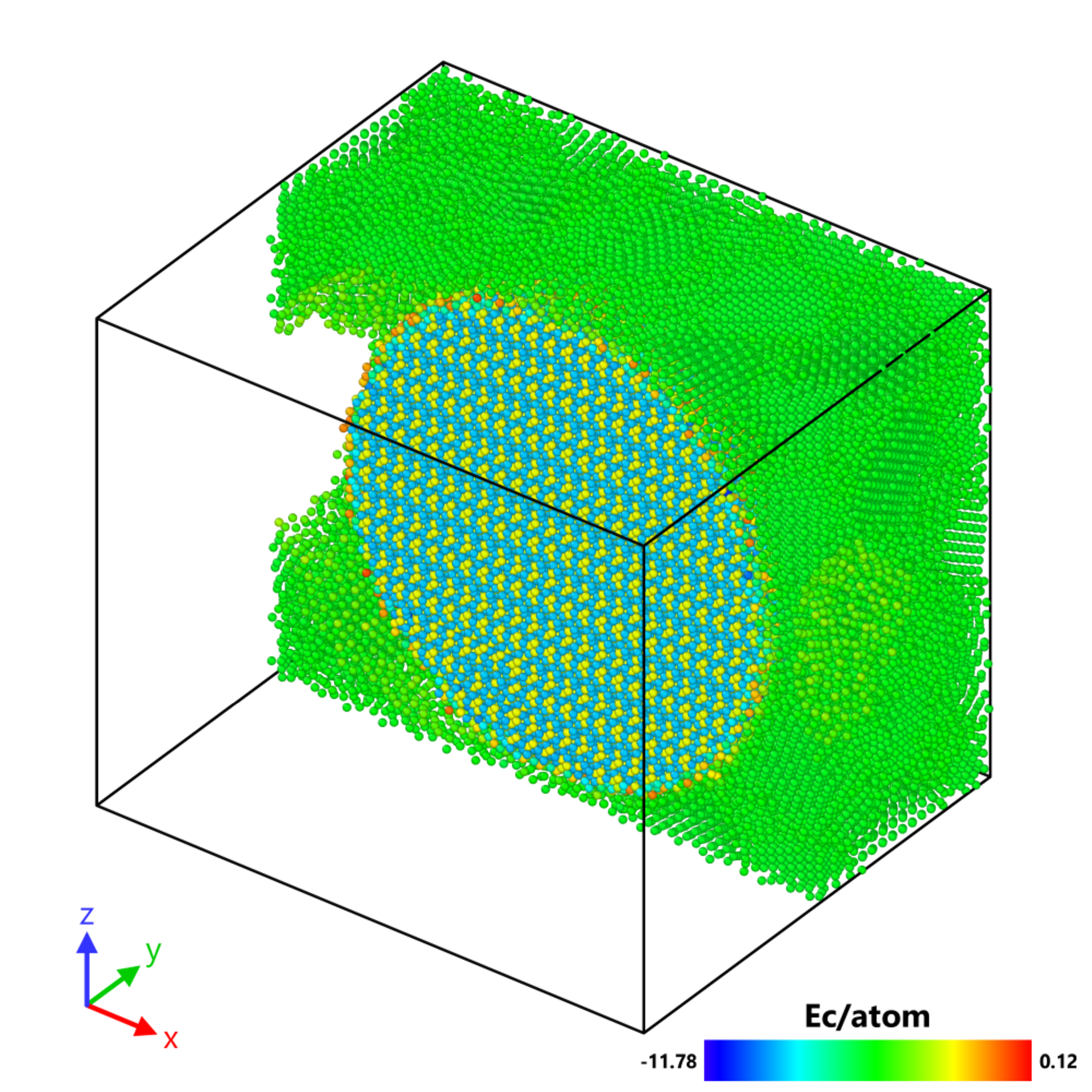}\\
			{\footnotesize (d)} & {\footnotesize (e)} & {\footnotesize (f)} \\
			\includegraphics[width=0.310\linewidth]{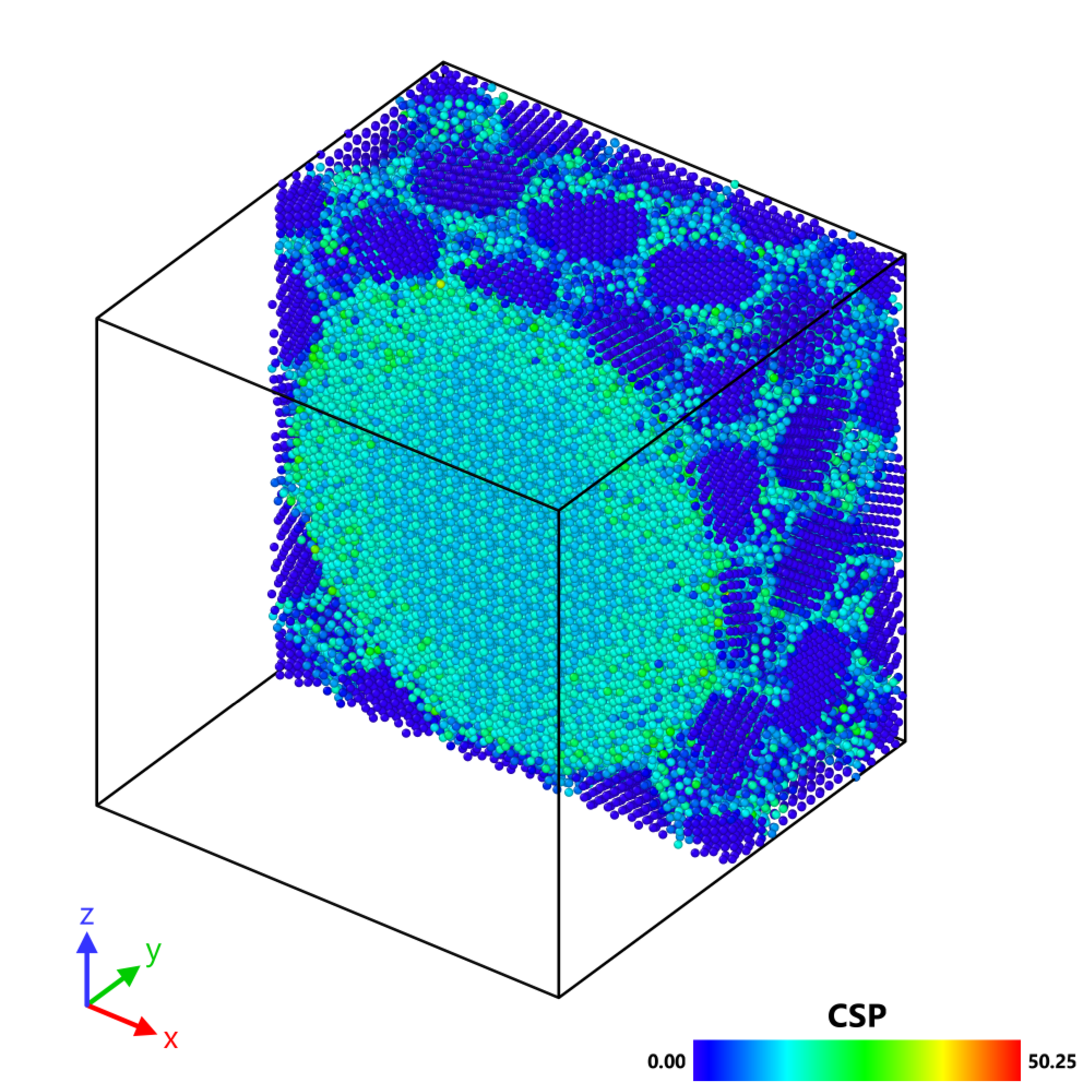} &
			\includegraphics[width=0.310\linewidth]{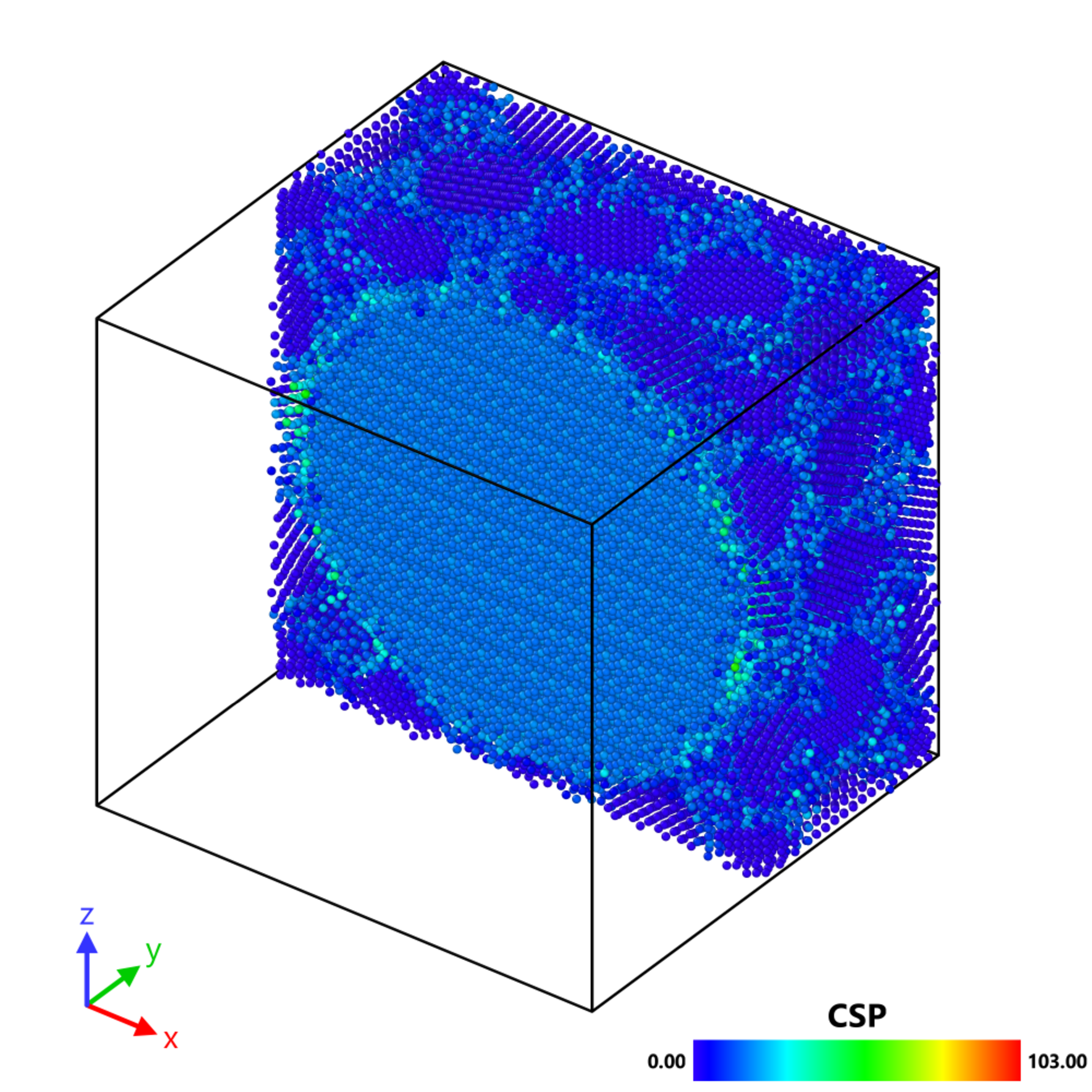} &
			\includegraphics[width=0.310\linewidth]{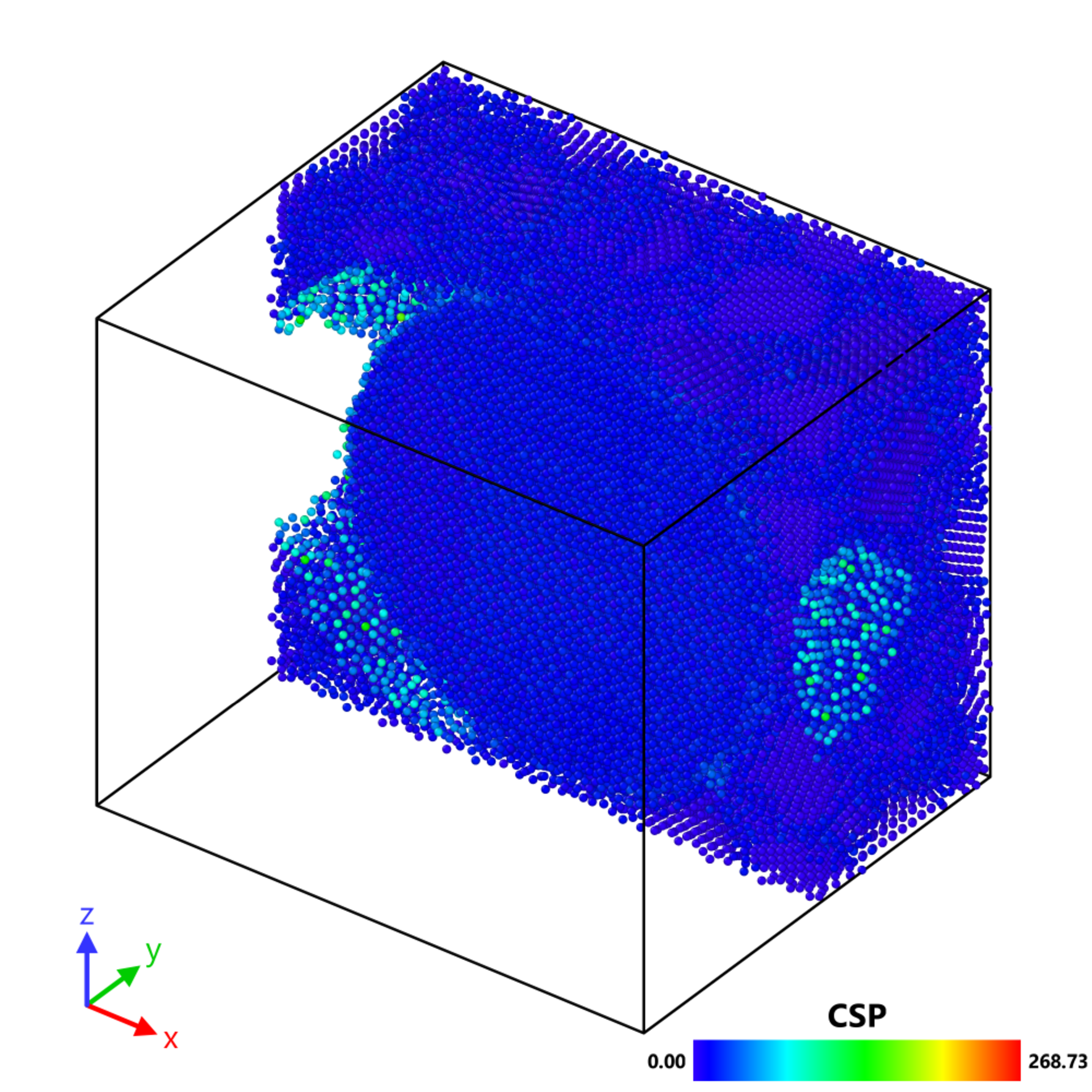}
		\end{tabular}
		  \caption{\MM{Cross section of {Al-40\%Al$_2$O$_3$\_32\_128g} sample (see Tab.~\ref{tab1}) under uniaxial tensile strain in the X direction:} \comm{{Al-40\%Al$_2$O$_3$\_32\_128g\_usx} (tension) cross section:} (a) the cohesive energy per atom (E$_c$/atom) for $\varepsilon_{xx}$=0, (b) E$_c$/atom for $\varepsilon_{xx}$=0.07, (c) E$_c$/atom for $\varepsilon_{xx}$=0.18, (d) the centrosymmetry parameter (CSP) for $\varepsilon_{xx}$=0, (e) CSP for $\varepsilon_{xx}$=0.07 and (f) CSP for $\varepsilon_{xx}$=0.18. }
		\label{fig:Al40Al2O3128gusxtens}
	\end{figure}

	The evolution of deformation and damage mechanisms in the {Al-40\%Al$_2$O$_3$} composite with a 128-grain polycrystalline structure subjected to uniaxial tension has been shown in Fig.~\ref{fig:Al40Al2O3128gusxtens}. Compared to the previously analyzed single-inclusion with monocrystal system, the response here is governed by a significantly more complex and statistically representative microstructure. The onset of plastic deformation is no longer confined to the particle–matrix interface. \SN{Instead, CSP maps reveal widespread development of lattice disorder across
multiple grains. Defect-rich regions develop both near the particle--matrix
interface and at grain boundaries, which act as preferential sites of structural
disturbance and strain localization. The cohesive-energy distribution shows
localized changes around the inclusion and along the grain-boundary network,
indicating that the deformation response involves contributions from both
interface- and grain-boundary-associated mechanisms.}
	
	Importantly, compared to the monocrystal model, damage is no longer dominated solely by interfacial decohesion. Instead, the failure process is governed by a competition between interfacial debonding and grain-boundary-mediated deformation and damage. The presence of a large number of grains introduces multiple pathways for strain accommodation, leading to a more diffuse damage evolution at intermediate strains, followed by localization at higher strains.
	
\begin{figure}[!h] 
		\centering
		\begin{tabular}{cc}
			{\footnotesize (a)} & {\footnotesize (b)} \\
			\includegraphics[width=0.47\linewidth]{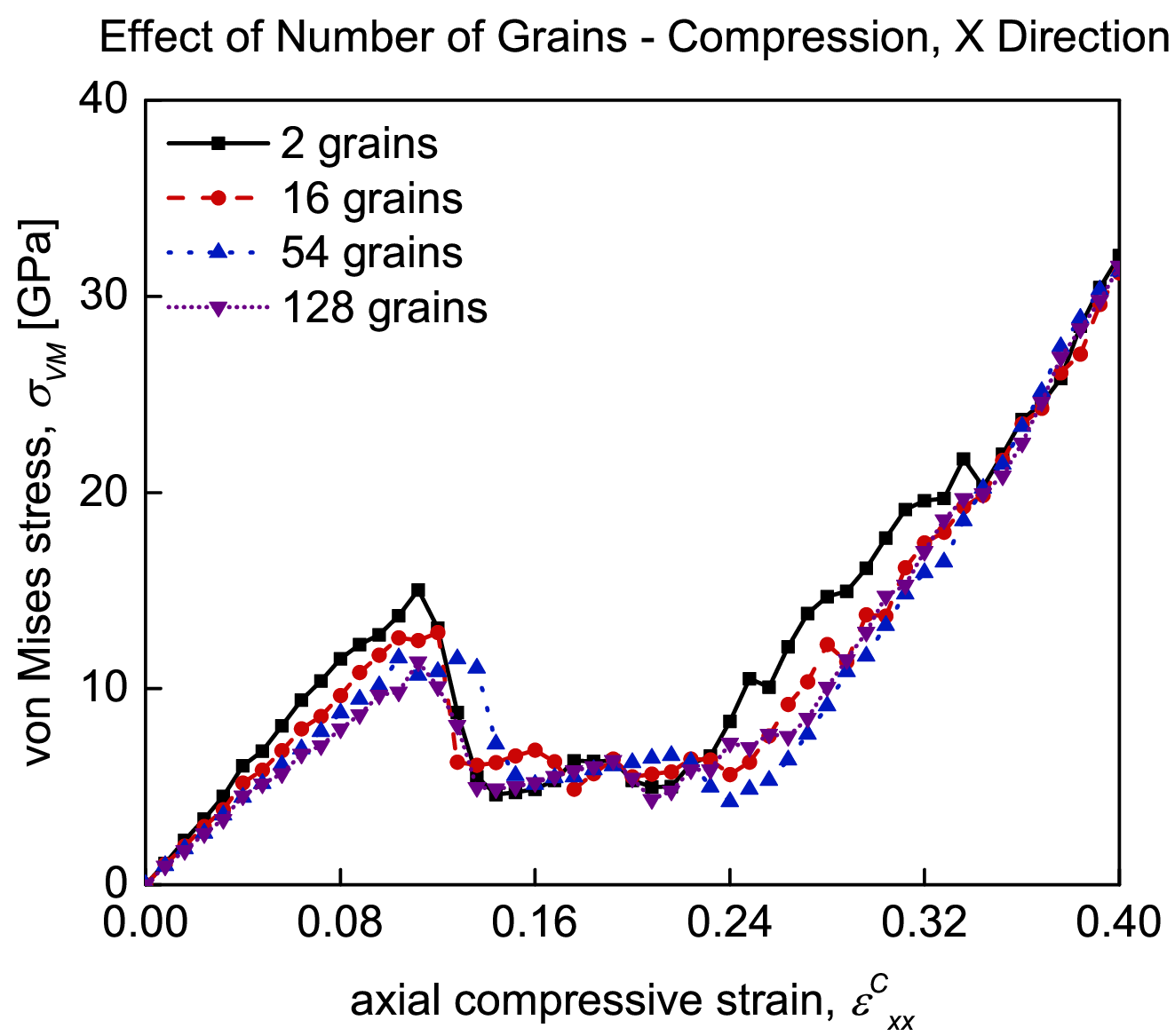} &
			\includegraphics[width=0.47\linewidth]{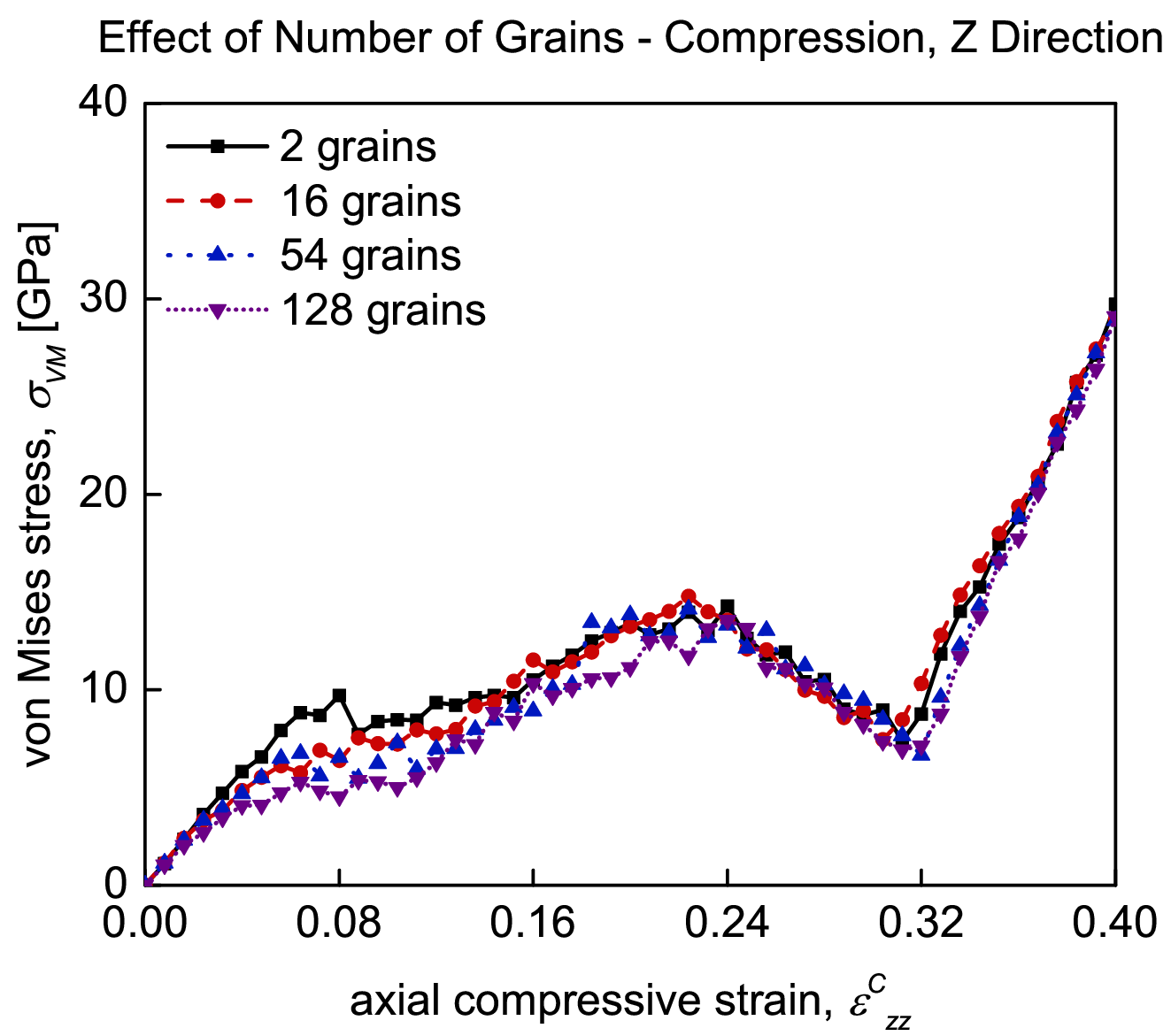}\\
		\end{tabular}
		\caption{The stress-strain evolution of {Al-Al$_2$O$_3$} polycrystal samples consists of different numbers of grains subjected to uniaxial compression in: (a) X and (b) Z directions.}
		\label{fig:Compression_DiffGrainNumber_Poly}
	\end{figure}

	Analogically, compressive stress–strain behavior of the multigrain RVE model was presented in Figure~\ref{fig:Compression_DiffGrainNumber_Poly}. All RVE models exhibit a characteristic three-stage compressive behavior, regardless of loading direction. \MM{The response consists of} (i) an initial quasi-linear elastic regime at small strains, (ii) a transition region with reduced hardening or a mild stress plateau at intermediate strains ($\epsilon \approx $ 0.12–0.25), (iii) pronounced strain hardening at large compressive strains ($\epsilon \gtrsim$ 0.30). Macroscopic softening and loss of load-carrying capacity are observed in the transition between region I and II (X direction) and between region II and III ($\epsilon \approx $ 0.24–0.32) in Z direction. Comparing the X and Z loading directions shows pronounced directional anisotropy. Loading along X produces higher stresses at small strains but an early stress drop and extended plateau, whereas loading along Z exhibits sustained hardening to a higher mid-strain maximum and a smoother transition into large-strain densification.
	
\begin{figure}[!h] 
		\centering
		\begin{tabular}{ccc}
			{\footnotesize (a)} & {\footnotesize (b)} & {\footnotesize (c)} \\
			\includegraphics[width=0.310\linewidth]{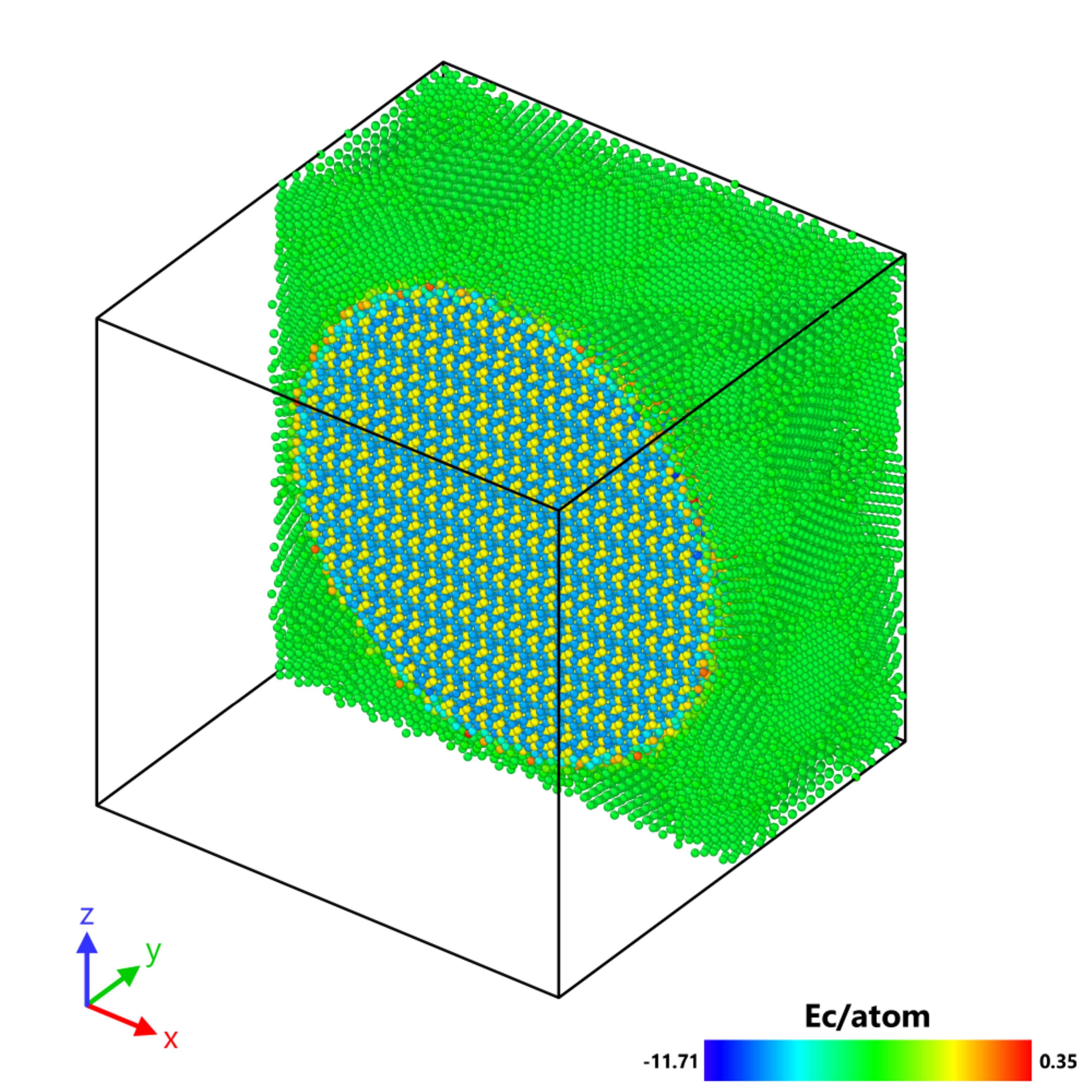} &
			\includegraphics[width=0.310\linewidth]{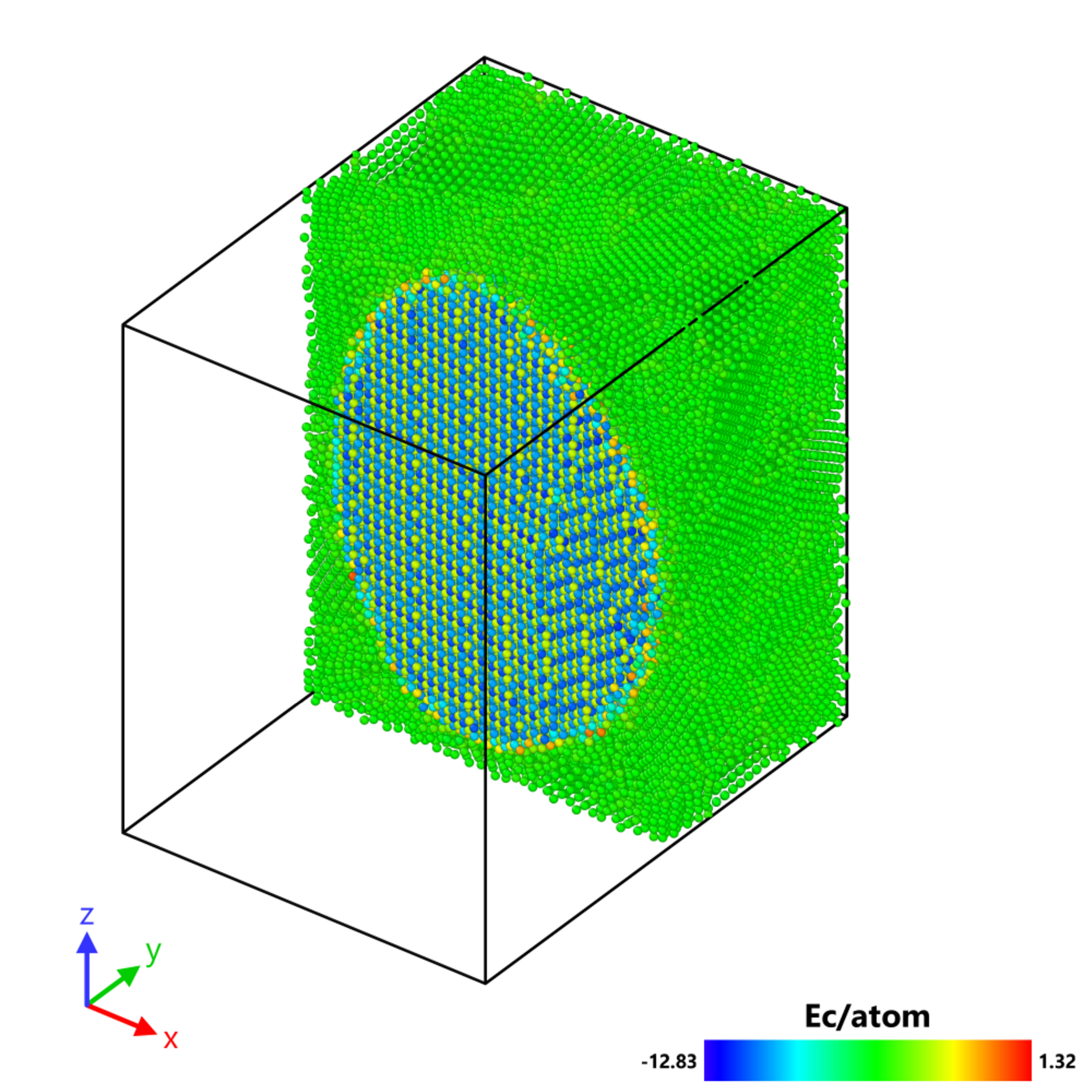} &
			\includegraphics[width=0.310\linewidth]{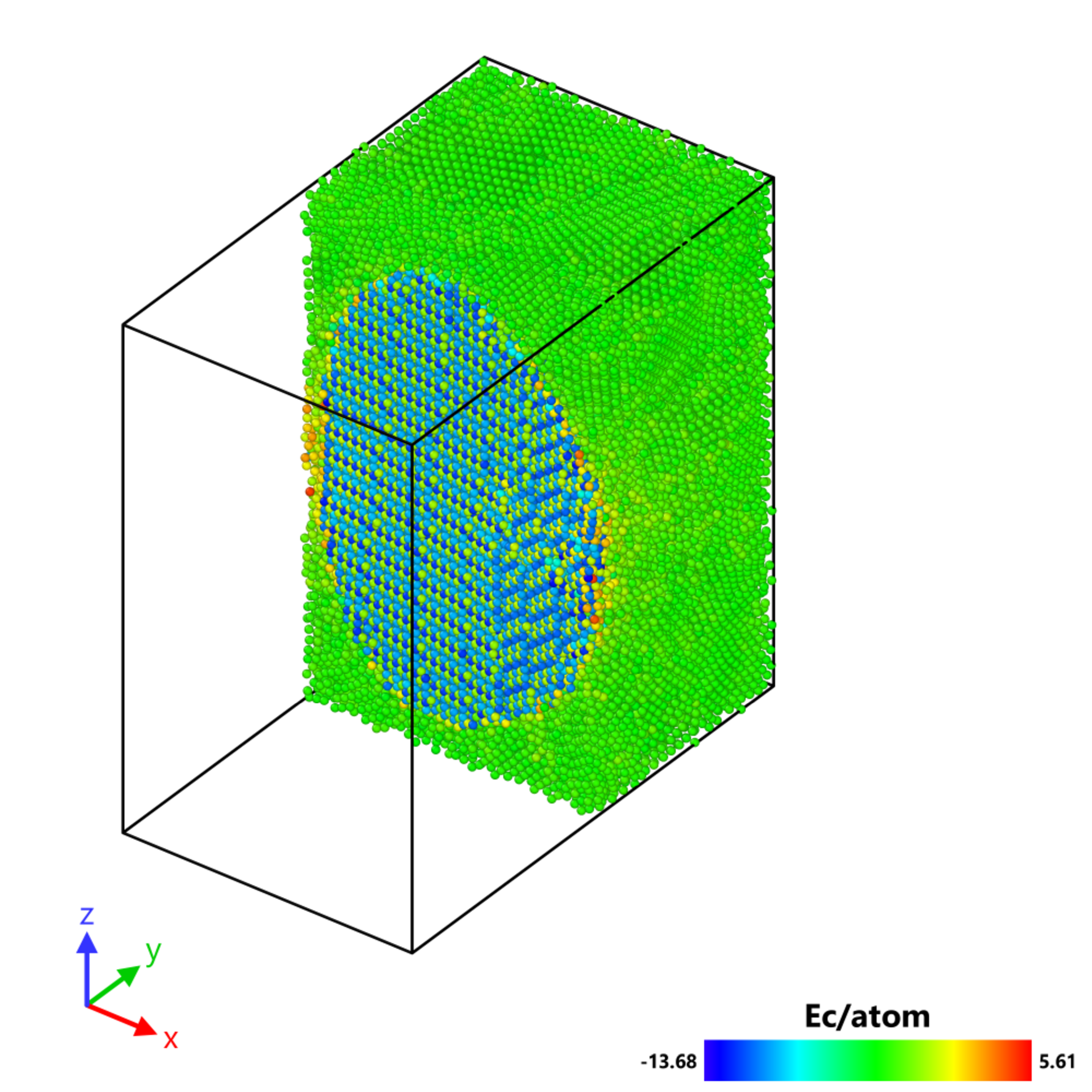}\\
			{\footnotesize (d)} & {\footnotesize (e)} & {\footnotesize (f)} \\
			\includegraphics[width=0.310\linewidth]{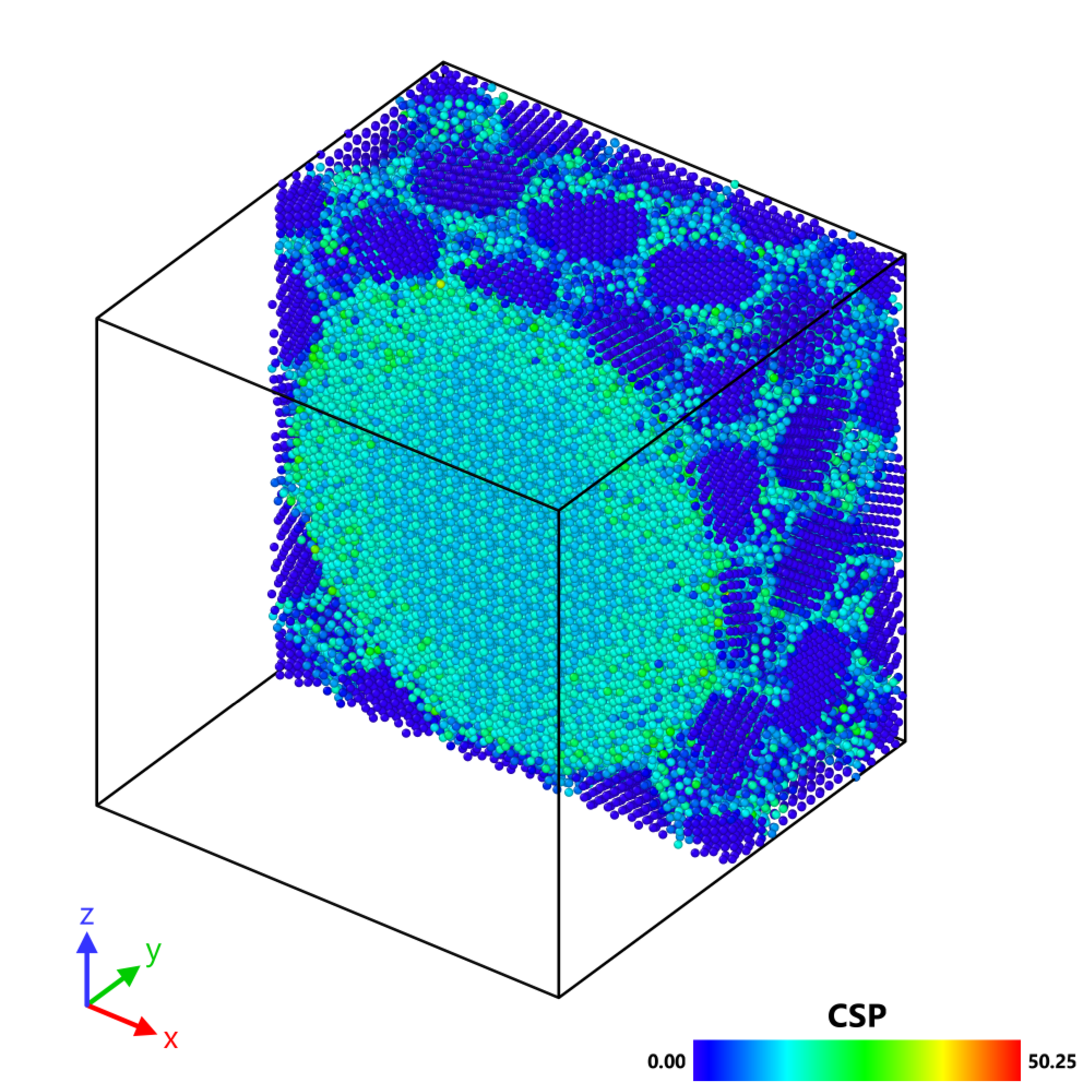} &
			\includegraphics[width=0.310\linewidth]{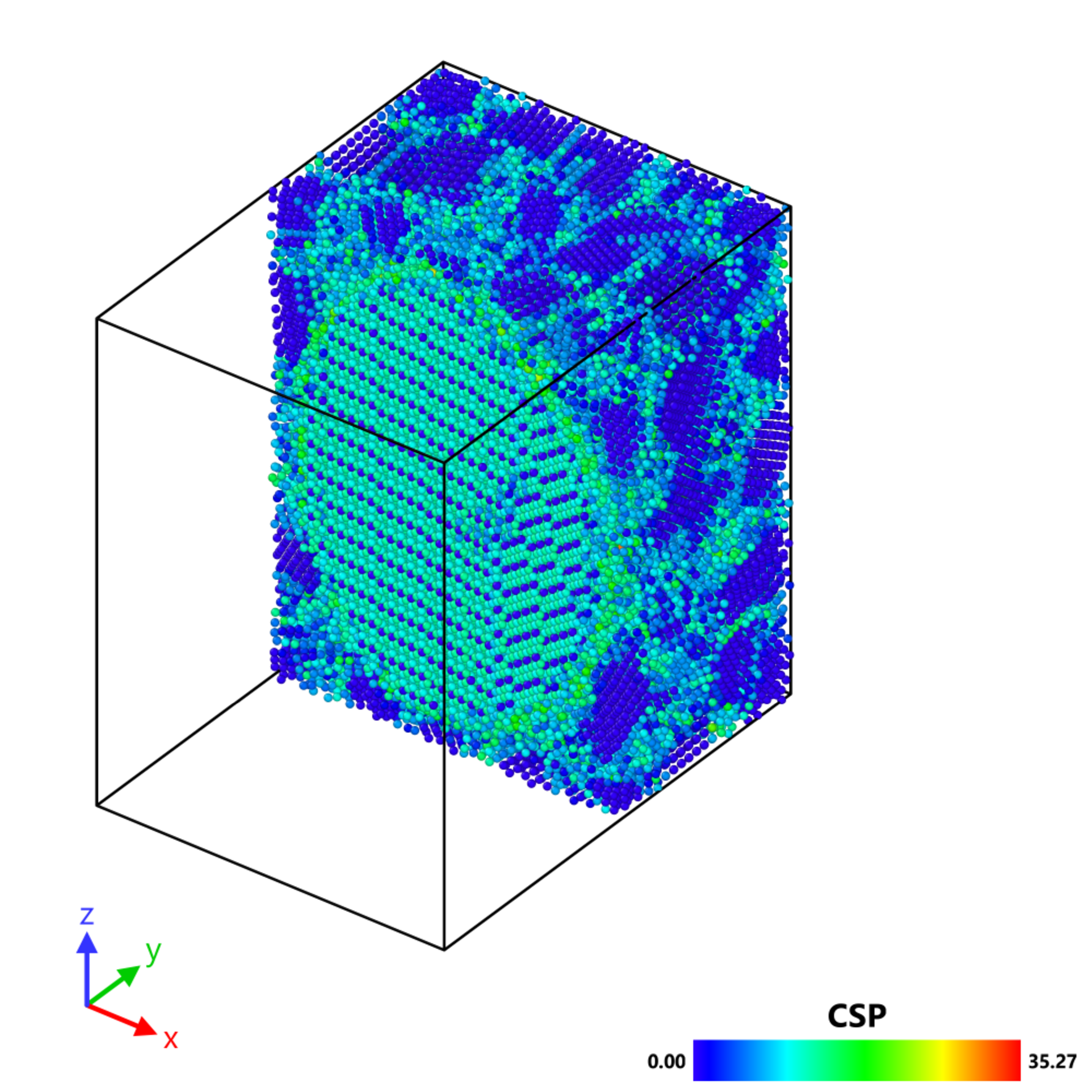} &
			\includegraphics[width=0.310\linewidth]{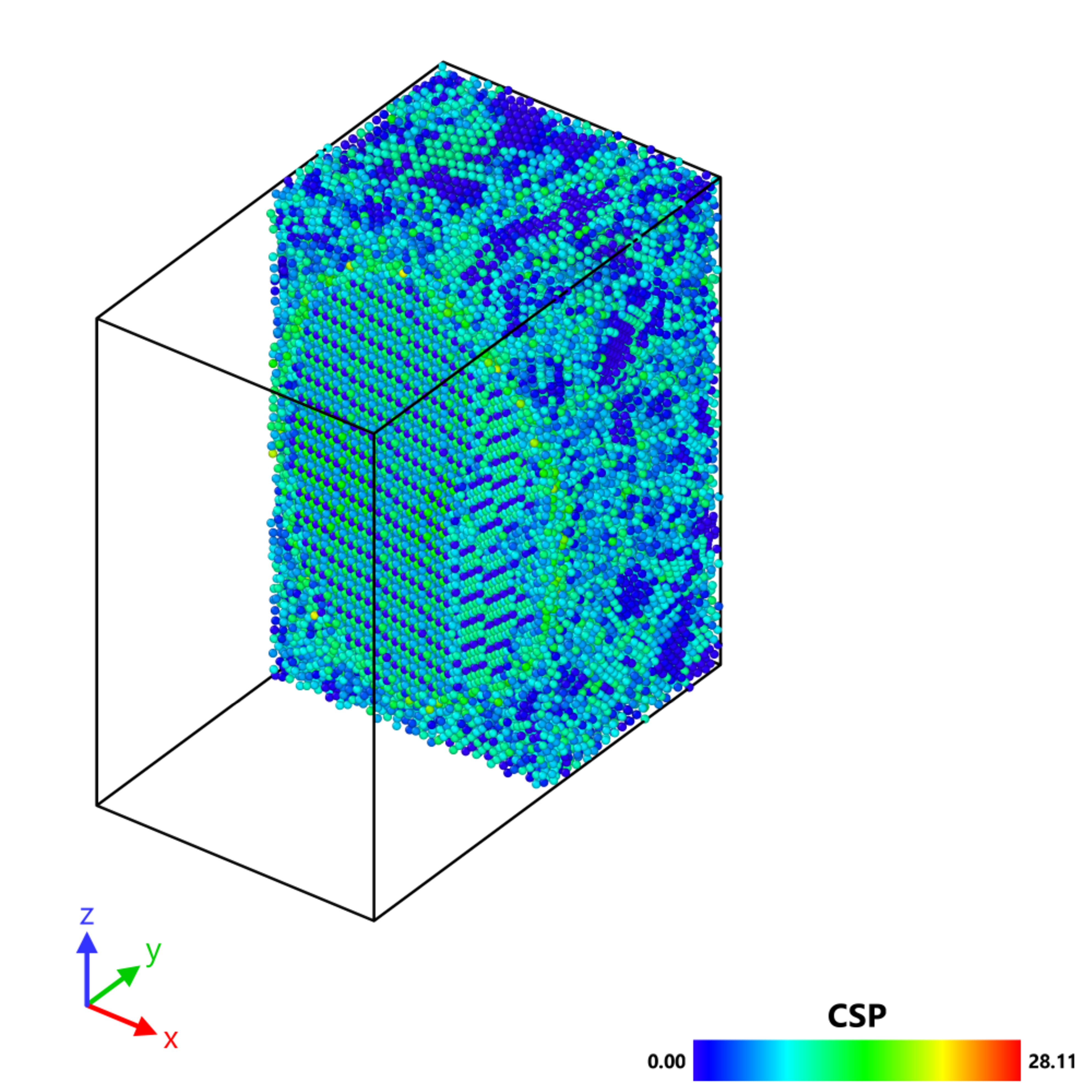}
		\end{tabular}
		\caption{\MM{Cross section of {Al-40\%Al$_2$O$_3$\_32\_128g} sample (see Tab.~\ref{tab1}) under uniaxial compressive strain in the X direction:} \comm{{Al-40\%Al$_2$O$_3$\_32\_128g\_usx} (compression) cross section:} (a) the cohesive energy per atom (E$_c$/atom) for $\varepsilon_{xx}$=0, (b) E$_c$/atom for $\varepsilon_{xx}$=-0.25, (c) E$_c$/atom for $\varepsilon_{xx}$=-0.40, (d) the centrosymmetry parameter (CSP) for $\varepsilon_{xx}$=0, (e) CSP for $\varepsilon_{xx}$=-0.25 and (f) CSP for $\varepsilon_{xx}$=-0.40. }
		\label{fig:Al40Al2O3128gusxcomp}
	\end{figure}

	The stress–strain curves for different grain counts are closely clustered across the entire strain range, indicating that the macroscopic compressive response is only weakly sensitive to grain count. RVEs with 2 grains exhibit slightly higher stresses at small to intermediate strains, likely due to limited crystallographic averaging and more favorable grain orientations. Increasing the number of grains to 16, 54, and 128 progressively smooths the response, reducing fluctuations and improving the stability of the stress evolution. At large strains, all RVEs converge toward similar stress levels (especially in the Z direction), with differences typically below a few percent. This convergence indicates that grain-scale heterogeneity has a limited influence on compressive load-bearing capacity once plastic flow and constraint effects dominate, as shown in Fig.~\ref{fig:Al40Al2O3128gusxcomp}.

	\SN{As the deformation proceeds, lattice disorder becomes widespread, with
defect-rich regions developing at both grain boundaries and the
particle--matrix interface and extending across multiple grains. The matrix
exhibits a dense network of structurally disturbed regions, indicating intense
and distributed plastic deformation. Unlike under tension, no pronounced
interfacial decohesion is observed; instead, deformation is distributed through
the matrix and grain-boundary regions. The inclusion remains intact and acts
as a rigid constraint, promoting heterogeneous but non-localized deformation.}
	
	\section{Discussion}
    \label{sec:Disc}
	
	The mechanical response of the {Al–Al$_2$O$_3$} nanostructures — unit-cell and RVE models — under tensile and compressive loading depended on both the volume fraction of the reinforcing phase and the loading direction, highlighting a complex interplay among strength, stiffness, and the dominant deformation mechanisms.
	
	A clear trend, particularly pronounced under tensile loading along the Z direction, is observed: the initial slope of the stress–strain curves increases with increasing Al$_2$O$_3$ content, indicating enhanced effective stiffness. This behavior arises from the growing fraction of the stiff reinforcing phase, which is characterized by significantly higher elastic moduli (\textbf{$B$}, \MM{\textbf{$G$}} and \textbf{$E$}) compared to the Al matrix \cite{Chawla2001}. 
	
\SN{It is readily apparent from the comparison of the atomistic predictions of
elastic constants with the analytical Voigt, Reuss, and Hill estimates that
both approaches predict an increase in stiffness with increasing
Al$_2$O$_3$ content (Fig.~\ref{fig:summaryBGE}). Quantitatively, the Reuss estimate provides the
closest agreement with the atomistically calculated Young's and shear moduli
over the investigated range of ceramic volume fractions. For Young's modulus,
the difference between the atomistic Reuss values and the corresponding
analytical estimates remains within approximately 5\%. This numerical
agreement should not, however, be interpreted as evidence that the atomistic
models satisfy an isostress condition. The Voigt and Reuss formulations
represent ideal upper and lower bounds based on uniform-strain and
uniform-stress assumptions, respectively, whereas the investigated atomistic
configurations exhibit heterogeneous local stress and strain fields associated
with the pronounced stiffness contrast between the Al matrix and the
Al$_2$O$_3$ inclusion and with the explicitly resolved matrix--particle
interface. }

	Secondly, a high level of disagreement between MD and analytical estimates is observed in the determination of the bulk modulus. This discrepancy indicates that several nanoscale mechanisms, inherently captured in atomistic simulations, are not accounted for in classical homogenization schemes such as the Voigt–Reuss–Hill \cite{Man2010}. In particular, MD explicitly resolves the atomic structure and chemistry of the Al$_2$O$_3$ interface, which leads to non-ideal interfacial behavior characterized by finite stiffness, strong chemical bonding, and local lattice coherency or semi-coherency \cite{Pan2023}. Another important contribution arises from interfacial excess volume and atomic-scale stress fields, including residual stresses generated during equilibration due to lattice misfit \cite{DEHOSSON19994077}. These internal stresses modify the effective elastic response but are not included in analytical estimates \cite{Reshetniak2025}. These features enable more efficient stress transfer than is assumed in analytical bounds, which treat phases as perfectly homogeneous and interfaces as mathematically ideal (i.e., perfectly bonded, with no thickness or compliance). Moreover, MD captures strain localization \cite{DELOGU2008904, ijms251910777, Falk_Shi_2002} and constraint effects at the nanoscale, where the metallic matrix is confined between stiff ceramic inclusions. This geometric confinement restricts plastic or elastic deformation modes, thereby effectively increasing the composite's apparent stiffness. Such constraint-driven stiffening is absent in VRH models, which assume uniform strain (Voigt) or stress (Reuss) fields and do not resolve local strain or stress gradients. Finally, the discrete nature of MD allows capturing anisotropy at the local scale \cite{MAZDZIARZ2024109953} and its statistical averaging, whereas analytical bounds assume fully isotropic phases and neglect orientation-dependent stiffness contributions unless explicitly modeled. \SN{In addition, the discrepancy may be influenced by the finite dimensions of the
simulated systems, lattice mismatch and local coherency conditions at the
interface, and residual atomic-scale stress fields generated during structural
relaxation. Since the individual contributions of these effects were not
separated quantitatively in the present study, the deviation from the analytical
estimates should be interpreted as the combined consequence of several
atomistic, interfacial, and finite-size effects rather than being attributed
solely to nanoscale interface phenomena.}
	
	As a result, MD tends to predict higher effective stiffness-particularly at low to intermediate reinforcement levels-where interfacial and confinement effects are most pronounced and cannot be adequately represented by classical homogenization theories.
	In general, the increase in composite stiffness observed in this study is consistent with previous computational and experimental studies reported in Al (or Al-based alloy)–Al$_2$O$_3$ systems \cite{Luster1993} \cite{Gudlur2012} as well as in other ceramic–metal composites \cite{Rezaei2018}.
	
	The present results reveal a clear directional sensitivity of the mechanical response. This is reflected in the ultimate strength, which varies with loading direction. Under tensile loading along the X direction, increasing the reinforcement fraction does not enhance strength, with the highest value observed at 10\% Al$_2$O$_3$, followed by a slight reduction at higher contents. In contrast, for loading along the Z direction, the ultimate strength increases monotonically with increasing Al$_2$O$_3$ fraction. These results indicate that load transfer in the composite is also governed by direction-dependent mechanisms and is strongly influenced by the reinforcement's intrinsic anisotropy. \MM{This directional dependence is also consistent with previous atomistic results showing that {Al–Al$_2$O$_3$} interfacial strength is sensitive to the crystallographic orientation relationship \cite{SAZGAR2015200}.}  Atomistic simulations of single-crystal $\alpha$-Al$_2$O$_3$ have demonstrated a pronounced orientation dependence of mechanical behavior, with significant variations in strength and deformation mechanisms depending on the loading direction \cite{Xu2023}. Although these studies were conducted under compressive loading, they clearly highlight that the load-bearing capacity of alumina is inherently anisotropic at the atomic scale.

	In experimental {Al–Al$_2$O$_3$} systems reinforced with nanometric alumina particles, tensile strength has been reported to increase with reinforcement content only up to a limited fraction. \MM{Above this level,} the strengthening effect tends to level off due to particle clustering and reduced effectiveness of further reinforcement additions \cite{Kang2004}. It should be noted, however, that such clustering effects are absent in our model, which considers an idealized system with a single reinforcement particle. \MM{Numerical studies further indicate that the strengthening effect depends not only on reinforcement content but also on particle geometry and interfacial characteristics \cite{Paknia2016,Huo2018}.}
    
	Generally, MD simulations of unit-cell and RVE models of {Al-Al$_2$O$_3$} nanocomposites reveal significantly different deformation and damage characteristics under tension and compression; see Figs.~\ref{fig:Tensile_DiffContAlumina_Mono} and \ref{fig:Compression_DiffContAlumina_Mono}. Deformation under compression is dominated by distributed plasticity rather than localized interfacial failure. The absence of significant decohesion, even at large strains, highlights the fundamental asymmetry between tensile and compressive loading \cite{HONG1999433}. While tension promotes interface-driven damage and crack initiation \cite{app152011173}, compression is characterized by more broadly distributed matrix deformation.
This interpretation is supported by the CSP and cohesive-energy distributions
observed in the present simulations, where compression leads to the formation
of extended defect-rich regions distributed throughout the matrix rather than
localized at the interface. In contrast, tensile loading is characterized by
pronounced localization of both CSP and cohesive-energy changes near the
matrix--particle interface, reflecting the dominant role of interfacial
processes in the early stages of deformation.
	
	The role of the interface remains critical for damage initiation, as demonstrated in atomistic studies of Al-matrix composites \cite{Pan2023}, in which misfit dislocation networks serve as preferential sites for dislocation emission and stress concentration. In such cases, failure is governed by void nucleation and coalescence near the interface, which contrasts with the compression-dominated behavior observed here but aligns with the tensile response of the present system.
    
	This perspective can be further extended by considering the combined contribution of interface- and grain-boundary-mediated plasticity, as reported by Vu et al. \cite{Vu2023} for Al-based composites reinforced with SiC particles. In such systems, both interfaces and grain boundaries act as primary sources of dislocation activity, while increasing particle size or content constrains plastic flow and promotes heterogeneous deformation. A similar tendency is observed in the present simulations, where CSP-based analysis reveals that defect nucleation and evolution are distributed across both interfacial regions and grain boundaries, reflecting a coupled contribution of these microstructural features.
	
	Alternative deformation mechanisms at the nanoscale have been reported in metallic nanoparticles \cite{Pizzagalli2024}, where plasticity occurs at high strains and stresses via a mechanism combining diffusive and displacive atomic displacements, rather than classical dislocation activity, leading to progressive structural disordering and partial structural recovery upon unloading. Notably, for strain levels of $\varepsilon \approx 0.07$–$0.10$, the reported stress values for pure Al nanoparticles ($\approx 7.5$–$12$\,GPa) are comparable to those obtained in this study for {Al–Al$_2$O$_3$} composites under compression.

 \SN{Finally, the peak tensile stresses obtained in the present atomistic simulations
(approximately 6--12\,GPa) are considerably higher than the strengths typically
measured for macroscopic {Al-Al$_2$O$_3$} composites. However, GPa-level strength
is commonly reported in atomistic simulations of Al and Al-based nanoscale
systems. Mei et al.~\cite{Mei2013} reported strengths of approximately 6.5--6.8\,GPa for
{Al-Al$_2$O$_3$} models, while values of approximately 2--7\,GPa were reported for
atomistically modeled Al by Mahata et al.~\cite{Mahata2024Bayesian}, Zeng et al.~\cite{Zeng2021SLMAl}, and Yadav et
al.~\cite{Yadav2024HallPetch}. Atomistic Al$_2$O$_3$ models may exhibit even higher fracture stresses
\cite{Chang2023, Streitz1996Alumina}. Such values reflect the very small, highly controlled, and nearly
defect-free simulated volumes and should therefore be interpreted as local
atomistic-scale strength parameters.}

\section{Conclusions}
\label{sec:Concl}

\SN{
The present study employed atomistic simulations to determine the effective elastic and damage-related response of particle-reinforced {Al-Al$_2$O$_3$} nanocomposite configurations. The results demonstrate that an increase in the Al$_2$O$_3$ volume fraction generally enhances the effective stiffness of the composite, while reducing the strain associated with peak stress and subsequent failure. The influence of reinforcement content on strength is strongly dependent on the loading direction. This directional dependence reflects the crystallographic anisotropy of the constituent phases and the specific matrix--particle orientation relationship.

The simulations further reveal a pronounced difference between tensile and compressive deformation. Under tensile loading, damage preferentially localizes in the vicinity of the {Al-Al$_2$O$_3$} interface, and progressive interfacial decohesion becomes an important contribution to failure. Under compression, the deformation is more broadly distributed throughout the Al matrix, with extended regions of lattice disorder and defect-mediated plastic deformation rather than pronounced interfacial separation. 

Introducing a polycrystalline Al matrix changes the effective response by adding grain-boundary-mediated deformation and a broader distribution of local crystallographic orientations. The grain structure has a particularly noticeable effect on the shear-related elastic response, whereas the calculated peak tensile stresses and failure strains show a comparatively weaker dependence on the investigated grain count. These results indicate that grain boundaries and matrix polycrystallinity primarily modify the redistribution of local stresses and strains rather than producing a proportional change in all effective mechanical parameters. 

Comparison with the Voigt--Reuss--Hill estimates confirms that classical homogenization provides a useful reference for the effective elastic response, but cannot explicitly represent the local matrix--particle interface, crystallographic anisotropy, grain-boundary effects, or strain localization resolved by the atomistic models. The calculated parameters should therefore be interpreted as effective atomistic-scale properties of the investigated configurations rather than as direct macroscopic material properties. Within these limitations, the present results provide an atomistically informed basis for subsequent scale-bridging procedures, in which the identified elastic and damage-related parameters may be incorporated into higher-scale FEM or DEM models after appropriate constitutive mapping and validation.
}

	\section*{CRediT authorship contribution statement}
	\textbf{Bartosz Bucholc:} Conceptualization, Methodology, Visualization, Writing – original draft, Writing – review \& editing.\\ 
	\textbf{Marcin Ma\'zdziarz:} Conceptualization, Methodology, Resources, Software, Visualization, Writing – original draft.\\
	\textbf{Szymon Nosewicz:} Conceptualization, Methodology, Visualization, Writing – original draft, Writing – review \& editing, Funding acquisition.
	
	\section*{Declaration of generative AI and AI-assisted technologies in the writing process}
	During the preparation of this work, the authors used Grammarly and ChatGPT by OpenAI in order to correct the spelling and English grammar. After using this tool/service, the authors reviewed and edited the content as needed and take full responsibility for the content of the publication.
	
	\section*{Declaration of Competing Interest}
	The authors declare that they have no known competing financial interests or personal relationships that could have appeared to influence the work reported in this paper.
	
	\section*{ACKNOWLEDGMENTS}
	The authors would like to acknowledge the financial support of the National Science Centre, Poland, under Grant Agreement No. OPUS 2023/49/B/ST8/04228 for the project “Tailoring mechanical
	properties of metal matrix composites by controlling reinforcing particles distribution".
	Additional support for the work was provided by the computing cluster GRAFEN at Biocentrum Ochota, the Interdisciplinary Centre for Mathematical and Computational Modelling of Warsaw University (ICM UW) and Pozna\'n Supercomputing and Networking Center (PSNC).
		\appendix
		
		\section{Supplementary material}
		\label{app:Sm}
		
		The following is available online at \href{run:./SI.pdf}{Supplementary data}
	
	\bibliography{References}

	
	
	
	
\end{document}